\documentclass[11pt,document,nofootinbib,superscriptaddress,onecolumn,preprintnumbers,balancelastpage]{article}
\pdfoutput=1

\usepackage{jheppub_1}

\usepackage{mathtools}
\usepackage{multirow}
\usepackage{caption}
\usepackage{subcaption}
\usepackage[countmax]{subfloat}
\usepackage{framed}
\usepackage{mdframed}
\usepackage{graphicx}
\usepackage{tikz-feynman}
\usepackage{scalerel}
\usepackage[countmax]{subfloat}
\usepackage{dcolumn}
\usepackage{cancel}
\usepackage{slashed}
\usepackage{booktabs} 
\usepackage{array}

\makeatletter  
\newcommand{\thickhline}{%
    \noalign {\ifnum 0=`}\fi \hrule height 1pt
    \futurelet \reserved@a \@xhline
}

\newcolumntype{"}{@{\hskip\tabcolsep\vrule width 1pt\hskip\tabcolsep}}

\makeatother

\usepackage{latexsym}
\usepackage{braket}
\usepackage{amssymb}
\usepackage{epsfig,amsmath,graphics}
\usepackage{listings}
\usepackage{color}
\usepackage{dcolumn}
\usepackage{slashed}
\usepackage{cancel}
\usepackage{comment,latexsym} 
\usepackage{bm}
\usepackage{verbatim}
\usepackage{tabularx}
\usepackage{dcolumn}
\definecolor{Mahogany}{rgb}{0.62,0.24,0.15}
\definecolor{colorLink}{rgb}{0.7,0,0}
\definecolor{colorCite}{rgb}{0,.7,0}
\definecolor{colorURL}{rgb}{0,0,0.7}
\usepackage[colorlinks=true,linktocpage=true,linkcolor=black,citecolor=colorCite,urlcolor=colorURL]{hyperref}
\usepackage{xspace}
\usepackage{multirow}
\usepackage{titlesec}
\usepackage[bbgreekl]{mathbbol}
\usepackage{bm}
\usepackage{float}
\usepackage{placeins}
\usepackage{afterpage}
\usepackage{footmisc}
\usepackage{breakcites}
\usepackage[most]{tcolorbox}
\tcbset{highlight math style={enhanced,colframe=green!30!black,colback=white,arc=5pt,boxrule=1.5pt}}

\usepackage[toc,page]{appendix}

\usepackage{etoolbox}
\appto\appendix{\addtocontents{toc}{\protect\setcounter{tocdepth}{1}}}

\usepackage{scalerel}

\usepackage{arydshln}

\allowdisplaybreaks

\expandafter\def\expandafter\normalsize\expandafter{%
    \normalsize
    \setlength\abovedisplayskip{8pt}
    \setlength\belowdisplayskip{8pt}
    \setlength\abovedisplayshortskip{8pt}
    \setlength\belowdisplayshortskip{8pt}
}

\newcommand{\diff}{ \mathrm{d}}
\newcommand{\bino}{{\tilde{B}}}
\newcommand{\squark}{{\tilde{q}}}
\newcommand{\me}{\mathrm{e}}
\newcommand{\lambdapp}{\lambda^{\prime\prime}}
\newcommand{\lambdapps}{\lambda^{\prime\prime *}}
\newcommand\ptwiddle[1]{\mathord{\mathop{#1}\limits^{\scaleto{(\sim)}{4pt}}}}
\newcommand{\MET}{E_T^{\mathrm{miss}}}

\DeclareRobustCommand{\Eq}[1]{Eq.~(\ref{#1})}

\newcommand{\be}{\begin{equation}}
\newcommand{\ee}{\end{equation}}
\newcommand{\bea}{\begin{eqnarray}}
\newcommand{\eea}{\end{eqnarray}}

\newcommand{\bphi}{\boldsymbol{\phi}}

\newcommand{\bS}{ \mathbf S}
\newcommand{\bT}{ \mathbf T}
\newcommand{\bO}{ \mathbf O}
\newcommand{\bW}{ \mathbf W }
\newcommand{\bQ}{ \mathbf Q}
\newcommand{\bH}{ \mathbf H}
\newcommand{\bU}{ \mathbf U}
\newcommand{\bD}{ \mathbf D}
\newcommand{\bL}{ \mathbf L}
\newcommand{\bE}{ \mathbf E}
\newcommand{\bR}{ \mathbf R}
\newcommand{\Br}{\text{Br} \left( B^0 \to \mathcal{B} + X \right)}

\usepackage{latexsym}
\usepackage{amssymb}
\usepackage{epsfig,amsmath,graphics}
\usepackage{listings}
\usepackage{color}
\usepackage{dcolumn}
\usepackage{slashed}
\usepackage{cancel}
\usepackage{comment,latexsym} 
\usepackage{bm}
\usepackage{verbatim}
\usepackage{tabularx}
\usepackage{dcolumn}
\definecolor{Mahogany}{rgb}{0.62,0.24,0.15}
\definecolor{colorLink}{rgb}{0.7,0,0}
\definecolor{colorCite}{rgb}{0,.7,0}
\definecolor{colorURL}{rgb}{0,0,0.7}
\usepackage[colorlinks=true,linktocpage=true,linkcolor=black,citecolor=colorCite,urlcolor=colorURL]{hyperref}
\usepackage{setspace}
\usepackage{xspace}
\usepackage{multirow}
\usepackage{titlesec}
\usepackage[bbgreekl]{mathbbol}
\usepackage{bm}
\usepackage{float}
\usepackage{placeins}
\usepackage{afterpage}
\usepackage{footmisc}
\usepackage{breakcites}
\usepackage{tikz-feynman}
\usepackage{scalerel}
\usepackage{braket}

\newcommand{\symfootnote}[1]{%
\let\oldthefootnote=\thefootnote%
\stepcounter{mpfootnote}%
\addtocounter{footnote}{-1}%
\renewcommand{\thefootnote}{\fnsymbol{mpfootnote}}%
\footnote{#1}%
\let\thefootnote=\oldthefootnote%
}

\title{
A Supersymmetric Theory of Baryogenesis and Sterile Sneutrino Dark Matter from B Mesons
}

\author[a]{Gonzalo Alonso-\'Alvarez,}
\author[b]{Gilly Elor,}
\author[b]{Ann E. Nelson\symfootnote{Ann Nelson passed away after this manuscript was written. Her contribution made this work possible, particle physics a richer field and the whole world a little bit brighter. We are forever grateful for her kindness and inspiration.},}
\author[b]{and Huangyu Xiao\,}

\affiliation[a]{\footnotesize Institut f\"ur Theoretische Physik, Universit\"at Heidelberg,
Philosophenweg 16, 69120 Heidelberg, Germany}
\affiliation[b]{\footnotesize Department of Physics, Box 1560, University of Washington, Seattle, WA 98195, U.S.A.}

\emailAdd{alonso@thphys.uni-heidelberg.de}
\emailAdd{gelor@uw.edu}
\emailAdd{huangyu@uw.edu}

\abstract{
Low-scale baryogenesis and dark matter generation can occur via the production of  neutral $B$ mesons at MeV temperatures in the early Universe, which undergo CP-violating oscillations and subsequently decay into a dark sector. In this work, we discuss the consequences of realizing this mechanism in a supersymmetric model with an unbroken $U(1)_R$ symmetry which is identified with baryon number.
$B$ mesons decay into a dark sector through a baryon number conserving operator mediated by TeV scale squarks and a GeV scale Dirac bino. The dark sector particles can be identified with sterile neutrinos and their superpartners in a type-I seesaw framework for neutrino masses. The sterile sneutrinos are sufficiently long lived and constitute the dark matter. 
The produced matter-antimatter asymmetry is directly related to observables measurable at $B$ factories and hadron colliders, the most relevant of which are the semileptonic-leptonic asymmetries in neutral $B$ meson systems and the inclusive branching fraction of $B$ mesons into hadrons and missing energy. 
We discuss model independent constraints on these experimental observables before quoting predictions made in the supersymmetric context. Constraints from astrophysics, neutrino physics and flavor observables are studied, as are potential LHC signals with a focus on novel long lived particle searches which are directly linked to properties of the dark sector. 
}

\begin{document} 
\maketitle
\setcounter{page}{2}
\begin{spacing}{1.2}

\pagebreak

\section{Introduction}
\label{sec:intro}
\noindent   
The nature and genesis of dark matter (DM) and the dynamical origin of the matter-antimatter asymmetry of the Universe are two outstanding mysteries that the Standard Model of Particle Physics (SM) cannot explain. Meanwhile, the SM also suffers from aesthetic or fine-tuning problems, especially the gauge hierarchy problem. The hope of solving these puzzles has driven theorists and experimentalists alike to search for new physics (NP).

Many particle physics models have been proposed to explain the origin of the DM relic abundance, measured to be $\Omega_{\rm DM} h^2 = 0.1200 \pm 0.0012$~\cite{Aghanim:2018eyx}, \emph{i.e.} roughly 26\% of the critical energy density of the Universe~\cite{Ade:2015xua,Aghanim:2018eyx}. However, searches for DM at colliders and at designated direct detection experiments together with studies of the possible indirect effects of DM in astrophysical observations have yet to shed light on its nature, thereby severely constraining many scenarios.

The origin of the primordial matter-antimatter asymmetry $Y_B \equiv (n_{B}-n_{\bar{B}})/s = \left( 8.718 \pm 0.004 \right) \times 10^{-11}$, inferred from measurements of the Cosmic Microwave Background (CMB)~\cite{Ade:2015xua,Aghanim:2018eyx} and Big Bang Nucleosynthesis (BBN)~\cite{Cyburt:2015mya,Tanabashi:2018oca}, requires a dynamical explanation: {\it baryogenesis}. Many mechanisms  of baryogenesis (satisfying the three Sakharov conditions \cite{sakharov}:  C and CP Violation (CPV), baryon number violation, and departure from thermal equilibrium) have been proposed, but remain experimentally challenging to verify due to the inaccessibly high scales and very massive particles involved.

A new mechanism for low-scale baryogenesis and DM production was introduced in  \cite{Elor:2018twp}\,\footnote{Many models and mechanisms that simultaneously generate a baryon asymmetry and produce the DM abundance in the early Universe  have been proposed. For instance, in models of  Asymmetric Dark Matter \cite{Nussinov:1985xr,Dodelson:1989cq,Barr:1990ca,Kaplan:1991ah,Farrar:2005zd,Kaplan:2009ag}, DM carries a conserved charge just as baryons do. }. 
Contrary to the standard lore that baryogenesis is difficult to test experimentally, the mechanism of~\cite{Elor:2018twp} predicts distinct signals at $B$ factories and hadron colliders. In particular, a large positive enhancement of the charge/semileptonic-leptonic asymmetry in neutral $B$ meson oscillation systems is required, as are new decay modes of charged and neutral $B$ mesons into baryons and missing energy together with exotic decays of $b$-flavored baryons. 

Weak scale supersymmetry (SUSY) offers a theoretically well motivated solution to the gauge hierarchy problem with a  plethora of implications for LHC searches. However, the lack of discoveries at the LHC has pushed the Minimal Supersymmetric Standard Model (MSSM) to ever more tuned corners of parameter space \cite{Cabrera:2011bi}, prompting theorists to introduce new variations of weak scale SUSY where experimental constraints may be evaded or relaxed. For instance, supersymmetric models with Dirac gauginos~\cite{Fayet:1974pd, Fayet:1975yi} and an exact $U(1)_R$ symmetry ($R$-SUSY)~\cite{Hall:1990hq, Randall:1992cq} are phenomenologically appealing~\cite{Fox:2002bu}. Exact $U(1)_R$ symmetries \textemdash under which particles within the same supersymmetric multiplet carry different charges\textemdash~ are motivated by top down constructions. By forbidding Majorana gaugino masses and supersymmetric $a$-terms, the majority of collider constraints are circumvented in models of $R$-SUSY.  Therefore, such extensions of the MSSM are intriguing in their own right, but are arguably even more motivated if the deviation from minimality accommodates a solution to other outstanding problems of the SM, such as DM and baryogenesis.

In the present work, we realize the mechanism of baryogenesis and DM production introduced in~\cite{Elor:2018twp} in an $R$-SUSY setup where an exact $U(1)_R$ symmetry is identified with $U(1)_B$ baryon number\footnote{\footnotesize{$R$-SUSY models can accommodate diverse baryogenesis scenarios~\cite{Fok:2012fb, Ipek:2016bpf}, sometimes at the cost of breaking the $R$-symmetry~\cite{Beauchesne:2017jou}. In this regard, realizing the mechanism of~\cite{Elor:2018twp} is a novel application of such ingredients and, as will be discussed, motivates a rather unstudied region of parameter space: a light (GeV mass scale) Dirac bino.}}. In this baryogenesis scenario, neutral $B$ mesons are produced at MeV temperatures in the early Universe, undergo CP-violating oscillations and subsequently decay to a Dirac gaugino (which is charged under the $R$-symmetry and therefore baryon number) through a four-fermion interaction obtained by integrating out heavy squarks. 
Successful baryogenesis requires at least one squark to have a mass of $\mathcal{O}(1\  \text{TeV})$, and critically a Dirac gaugino with mass of a few GeV (typically smaller than 4 GeV in order for $B$ meson decay to be kinematically allowed). Current constraints do not allow the existence of so light gluinos or charginos, but the bounds do not apply to the color neutral bino~\cite{Cabrera:2011bi}. 

The collider phenomenology of various $R$-SUSY setups has been greatly studied~\cite{Frugiuele:2012pe, Kalinowski:2015eca, Beauchesne:2017jou, Diessner:2017ske, Alvarado:2018rfl, Fox:2019ube, Chalons:2018gez}, along with constraints from low energy flavor observables  and electric dipole moments \cite{Kribs:2007ac}. However, previous work on $R$-SUSY models focused on scenarios with heavy Dirac gauginos (for instance \cite{Fox:2002bu} considers flavor constraints on a model where all the gauginos share the same mass scale of order $500\ \text{GeV} - \text{TeV}$). Realizing the mechanism of \cite{Elor:2018twp} motivates us to study the phenomenology of a complementary region of $R$-SUSY parameter space \textemdash one where multiple gaugino mass scales are present. This allows the bino to be light enough for baryogenesis purposes while the other gauge superpartners are heavier and avoid collider constraints. This scenario can be realized, for instance, if different SUSY breaking scales are assumed \cite{Cohen:2016kuf}. 

In the present work we further extend the $R$-SUSY setup to include a sterile neutrino and sneutrino with GeV scale masses, and discuss their role in providing a mass for SM neutrinos via a type-I seesaw~\cite{Mohapatra:1979ia,GellMann:1980vs,Minkowski:1977sc,Schechter:1980gr}. In this scenario, the bino decays predominantly into a sterile neutrino-sneutrino pair. The sterile neutrino decays are generically long lived and may be searched for at SHiP~\cite{Anelli:2015pba}, FASER~\cite{Feng:2017uoz}, CODEX-b~\cite{Gligorov:2017nwh}, MATHUSLA~\cite{Lubatti:2019vkf} and in the ATLAS muon tracker~\cite{Aaboud:2018aqj}, while the sneutrino is stable on cosmological time scales and constitutes the DM. This opens up a variety of astrophysical constraints and signals which are discussed and assessed.

While the existence of an $R$-symmetry is theoretically well motivated, it is generically expected that this symmetry is not realized exactly, as supergravity considerations lead to a gravitino whose mass is not $U(1)_R$ symmetric. Supergravity breaking of the $U(1)_R$ symmetry introduces, for instance, Majorana sparticle masses generated from the conformal anomaly along with $a$terms and small soft squark masses, resulting in additional constraints (see \cite{Aitken:2017wie} for a detailed discussion). We quantify the degree to which the $R$-symmetry can be broken while still being able to accommodate the baryogenesis and DM production mechanism from $B$ mesons.

This paper is organized as follows: in Sec.~\ref{sec:mechanism} we review the key ingredients and results of  \cite{Elor:2018twp}, \emph{i.e.} the generic model and parameter space needed to achieve baryogenesis and DM from $B$ mesons. In Sec.~\ref{sec:ModelIndep} we present new model independent constraints on the experimental observables within our framework. Next, Sec.~\ref{sec:Model} introduces the model of R-SUSY with Dirac gauginos and right-handed neutrino multiplets, in addition to discussing the degree to which we require the $R$-symmetry to be exact. In Sec.~\ref{sec:Predictions} we present our results for flavor observables within the SUSY model \textemdash namely the contributions to the semileptonic-leptonic asymmetries in neutral $B$ meson oscillations and various other flavor constraints. Next, in Sec.~\ref{sec:BandDMconst} we discuss what region of parameter space is able to generate the observed matter-antimatter asymmetry and the DM relic abundance. This is followed by a discussion of additional constraints on the model parameters arising from neutrino physics, DM stability, supersymmetry breaking and neutron stars. In Sec.~\ref{sec:Signals} various signals at colliders and $B$ factories that can probe this model are discussed. Finally, we conclude in Sec.~\ref{sec:Outlook} by summarizing the allowed parameter space of the model that realizes baryogenesis. Some further  details of the computations are given in the appendices.

\section{Baryogenesis and Dark Matter from $B$ Mesons}
\label{sec:mechanism}
Here we review the mechanism of low scale baryogenesis and DM production from the oscillations and subsequent decay of neutral $B$ mesons, which was proposed in~\cite{Elor:2018twp}. An inflaton-like\footnote{\footnotesize{The details of an inflation model are beyond the scope of this work. This role can also be played by a modulus or any other late-time decaying particle.}} field $\Phi$ with mass $m_\Phi$ decays out of thermal equilibrium in the early Universe, at temperatures of order $T_R \sim 1-100\, \mathrm{MeV}$. The main decay products are assumed to be quarks and antiquarks that hadronize producing neutral $B^0_{q = s, d} = |\bar{b} \, q \rangle$ mesons and antimesons $\bar{B}^0$, which then undergo CP-violating oscillations before decaying partially into a dark sector\footnote{\footnotesize{Additional mesons may also be produced, but are uninteresting for the current setup as only $B$ mesons can give rise to baryogenesis in this way due to kinematic considerations.}}. 

A minimal set of four new particles is required to generate the necessary interactions between the visible and dark sectors (see~\cite{Elor:2018twp,Aitken:2017wie} for details). First consider an electrically charged ($Q_e=-1/3$), baryon number carrying ($B=-2/3$), color triplet scalar which couples to SM quarks \textemdash foreshadowing, we denote this particle by $\tilde{q}_R$. Such a particle couples to the SM quarks through the fully $SU(3)_c$ antisymmetric term $\tilde{q}_R^* \, \bar{u} \, b^c $, and various other flavor combinations.  Additionally, we introduce a Dirac fermion $\psi$ carrying baryon number $-1$, so that couplings of the form  $\tilde{q}_R \bar{\psi} \,  s^c$ are allowed. 
To avoid collider constraints, the new colored scalar is assumed to have a mass around the TeV scale (see Sec.~\ref{sec:LHC_colored_scalars} for a more detailed discussion), allowing us to integrate it out arriving at an effective Hamiltonian valid a low energies:
\begin{align}
\label{eq:Lag_psi}
\mathcal{H}_{eff} \,\, = \,\, \frac{\lambda_{\rm eff}}{m_{\tilde{d}}^2} u \, s\,  b\, \psi \, .
\end{align}
To simplify the notation, we have chosen a particular combination of light quark flavors to showcase the mechanism, and have parametrized the effective coupling by $
\lambda_{\rm eff}$. It will however be true that any other combination of couplings that yields an effective four-fermion interaction involving $b$ and $\psi$ and two other light quarks\footnote{More precisely, any up- plus down-type pair of first or second generation quarks, in order to preserve hypercharge and for the $B$ meson to be able to decay through the effective operator. We refer to Table III in~\cite{Elor:2018twp} for an exhaustive list of options.} is equally suitable.

For a sufficiently light $\psi$, with mass smaller than about $4$ GeV, the effective operator in Eq~\eqref{eq:Lag_psi} allows the $\bar{b}$ quark within the $B$ meson to  decay;  $B^0_q \rightarrow \psi + \text{Baryon} + X$, where $X$ parametrizes mesons or other additional SM particles. Since $\psi$ carries baryon number, this decay is baryon number conserving. However, an asymmetry $n_\psi - \bar{n}_\psi$ can be generated due to the CP-violating nature of the neutral meson oscillations.
If $\psi$ is allowed to decay back into visible sector particles, the generated asymmetry will be erased. However, the asymmetry may be preserved if $\psi$ dominantly decays into stable DM particles. This is made possible by the existence of a dark scalar baryon $\phi$ with baryon number $-1$ and a dark Majorana fermion $\xi$. If a discrete $\mathbb{Z}_2$ symmetry is assumed under which the dark sector particles transform as $\psi \rightarrow \psi$, $\phi\rightarrow-\phi$ and $\xi\rightarrow-\xi$, we can write a Yukawa operator 
\begin{align}
\label{eq:DarkYukawa}
\mathcal{L} \,\, \supset \,\,  -\lambda_N \, \bar{\psi}\, \phi \, \xi\ ,
\end{align}
mediating the decay of $\psi$ into the dark sector, where an overall negative baryon number is stored. A positive baryon asymmetry is thus produced in the visible sector without globally breaking baryon number. Note that the scalar $\phi$ carries baryon number and as such its mass is constrained to be greater than $\sim \, 1.2$ GeV \cite{McKeen:2018xwc}\footnote{\footnotesize{The authors of~\cite{McKeen:2018xwc} showed that the observation of neutron stars with masses greater than $2 M_{\odot}$ constrains the mass of a dark particle carrying baryon number to be greater than the neutron chemical potential inside the star, which is about $ 1.2 \, \text{GeV}$. Otherwise, if a light dark baryon did exist, a process such as $n+n\rightarrow \text{DM} + \text{DM}$ could occur within the star. As a consequence, neutrons would be replaced by DM particles and the neutron Fermi pressure would diminish. This would destabilize the neutron star, leading to gravitational collapse. }}.
In this paper we consider a scenario in which the scalar constitutes the DM (the fermion will turn out to be long lived but unstable on cosmological time scales). The combination of kinematics, proton decay, dark matter stability, and neutron star stability leads us to consider a parameter space where
\begin{align}
    \label{eq:MassParamRange}
m_{B}-m_n   > m_\psi > m_\phi + m_\xi > 1.5 \, \text{GeV} \, \quad \text{and} \quad m_\xi+m_n > m_\phi > 1.2 \, \text{GeV} \, ,
\end{align}
where $m_n$ is the neutron mass and $m_B$ is the $B$ meson mass.

The asymmetry and DM abundance can be determined by solving a system of coupled Boltzmann equations \cite{Elor:2018twp, Nelson:2019fln}. Doing this, the comoving baryon asymmetry is found to scale as
\begin{align}
\label{eq:YB}
Y_B   \,\, \propto \,\, \Br \sum_{q = s, d} \alpha_q \left( T_R, \Delta m_{B_q}  \right)  \times a_{\ell \ell}^q    \,.
\end{align}
Here, $ \text{Br}(B^0 \to \mathcal{B} + X)$ is the branching fraction of the $B^0$ meson to a baryon $\mathcal{B}$ and $X$, which stands for the dark sector particles plus potentially other mesons or SM particles. 
The ``reheat temperature" $T_R$ is taken to be the temperature at which the $\Phi$ decays become significant, that is, when $3 H(T_R) = \Gamma_\Phi$, where $\Gamma_\Phi$ is the decay width of $\Phi$. 
Furthermore, $a_{\ell \ell}^q$ is the leptonic charge asymmetry (which in this set-up is effectively equivalent to the semileptonic-leptonic asymmetry $a_{\mathrm{sl}}^q$), an experimental observable which parametrizes CP violation in the $B_{s, \, d}^0$ systems. An important point is that, as in neutrino physics, neutral $B$ meson oscillations can only occur in a coherent system. Additional interactions with the mesons can act to decohere the oscillations~\cite{Cirelli:2011ac,Tulin:2012re} by ``measuring" the system, thereby suppressing the CP violation and diminishing the potential to generate an asymmetry. In the early Universe, decoherence is caused by the scattering of $e^\pm$ off $B^0$ mesons due to their charge asymmetry. For CP violation to effectively happen, $B$ mesons must oscillate at a rate similar to or faster than the $e^{\pm} B^0_q \rightarrow e^{\pm} B^0_q$ scattering in the plasma. The factor of $\alpha_q \left( T_R, \Delta m_{B_q}  \right)$ in Eq. \eqref{eq:YB} is a measure of decoherence at the temperature $T_R$ for each of the $q = s, d$ systems. 

The kinematically allowed and currently experimentally unconstrained parameter space that accommodates the measured baryon asymmetry and DM relic abundance was mapped out in \cite{Elor:2018twp}. In order to reproduce the observed value of  $Y_B \sim 8.7 \times 10^{-11}$, the experimental observables in Eq.~\eqref{eq:YB} must, after scanning over the $(m_\Phi, \Gamma_\Phi)$ parameter space, lie within the ranges
\begin{align}
\label{eq:YBprediction}
\text{Br}(B^0 \to \mathcal{B} + X) \sim 2\times10^{-4}-10^{-1} \, , \quad \text{and} \quad  \sum_{q = s, d}  a_{\mathrm{sl}}^q \sim 10^{-5}-10^{-3} > 0  \,.
\end{align} 
If we compare with the SM predictions~\cite{Artuso:2015swg} 
\begin{align}
a^{s,\,\mathrm{SM}}_{\mathrm{sl}} = \left( 2.22 \pm 0.27 \right)\times 10^{-5}\quad \text{and} \quad a^{d,\,\mathrm{SM}}_{\mathrm{sl}} = \left( -4.7 \pm 0.6 \right)\times 10^{-4}\,,
\end{align}
we conclude that the amount of CP violation predicted by the SM is typically not enough for successful baryogenesis. However, given the current experimental constraints~\cite{Tanabashi:2018oca}
\begin{align}\label{eq:experimental_semileptonic_asymmetries}
a^{s,\,\mathrm{exp}}_{\mathrm{sl}} = \left( -0.6 \pm 2.8 \right)\times 10^{-3}\quad \text{and} \quad a^{d,\,\mathrm{exp}}_{\mathrm{sl}} = \left( -2.1 \pm 1.7 \right)\times 10^{-3}\,,
\end{align}
we see that there is room for new physics to enhance the CP violation and accommodate the required values in Eq.~\eqref{eq:YBprediction}. We will explore this in more detail in Sec.~\ref{sec:ModelIndep}. The other relevant quantity is the branching ratio of $B$ mesons into a baryon and dark sector states which would contribute to missing energy in an experiment. To the best of our knowledge, to the date an inclusive search for such a decay has not been performed. Given the experimental constraints on similar processes like $B\rightarrow K\nu\nu$, which are in the $\mathcal{O}(10^{-5})$ level, we expect that existing and future data from BaBar, Belle (-II) and LHCb may be in position to fully test the parameter range that allows for successful baryogenesis.

The symmetric component of the DM is generally overproduced in this set-up. Therefore, additional interactions allowing the DM to annihilate are required in order to deplete its abundance. The results of \cite{Elor:2018twp} show that an annihilation cross section
\begin{align}
\label{eq:DMprediction}
\left<\sigma v \right>_{\rm dark} 
&=\, (6-20) \times 10^{-25} \,\text{cm}^3/\text{s}
\end{align}
is needed in order to reproduce the observed DM density $\Omega_{\rm DM} h^2 \sim 0.12$.

One of the main appeals of this mechanism is its susceptibility to be tested in terrestrial experiments such as flavor facilities and accelerators. Indeed, experimental constraints on $a_{\mathrm{sl}}$ can directly constrain the mechanism. We explore this model-independent bounds in Sec.~\ref{sec:ModelIndep}. Next, we construct a realization of this mechanism, \emph{i.e.} \eqref{eq:Lag_psi} and \eqref{eq:DarkYukawa}, in a supersymmetric framework. Having a full model will allow us to consistently calculate the new physics contributions to $a_{\mathrm{sl}}$ and the branching fraction $\text{Br}(B^0 \to \mathcal{B} + X)$, together with any other related flavor observables that the new particles and couplings may modify. In doing so, we will be able to identify the phenomenologically viable region of parameter space where baryogenesis can successfully be achieved, and develop the best approach to experimentally test the mechanism in high and low energy experiments.

\newpage

\section{Model Independent Constraints}
\label{sec:ModelIndep}
We now conduct a model independent study to further quantify the required new physics (NP) contributions to realize baryogenesis and DM from $B$ mesons. This will set the stage for successfully embedding this mechanism in a full model. As seen in  \eqref{eq:YB}, the baryon asymmetry depends critically on the semileptonic asymmetres $a_{\mathrm{sl}}^q$ and the branching fraction $\Br$ of neutral $B$ mesons to baryons and dark sector states. As discussed above, the SM prediction for the  semileptonic asymmetries is not enough to generate the observed baryon asymmetry required for baryogenesis, and in this section we study the ability for general NP contributions to enhance this asymmetry, taking into account all the relevant experimental constraints. The amount of NP introduced is constrained by measurements of different $B^0_q$ oscillation parameters, in turn limiting the size of the enhancement of the semileptonic-leptonic asymmetry which is consistent with experimental data. As a consequence of these bounds, we obtain a lower limit on the required branching fraction in order to generate the measured baryon asymmetry of the Universe. In this section we remain agnostic about the origin of the NP contributions and perform the study in a data driven and model independent way.

\subsection{The Semileptonic Asymmetries} 
\label{sec:model_indep_asl}
Experimental measurements of the oscillation parameters in neutral $B$ meson mixing constrain the size of NP effects that introduce new CP violation into these systems. We now quantify such constraints. Recall that the evolution of the neutral $B_q^0 - \bar{B}_q^0$ system is described by
\begin{equation}
i\frac{\diff}{\diff t} \begin{pmatrix}
\ket{B^0_q(t)}\\
\ket{\bar{B}^0_q(t)}
\end{pmatrix}
= \begin{pmatrix}
\mathcal{H}^q_{\rm osc}
\end{pmatrix}
\begin{pmatrix}
\ket{B^0_q(t)}\\
\ket{\bar{B}^0_q(t)}
\end{pmatrix} \,,
\end{equation}
where the Hamiltonian $2 \times 2$ matrix $\mathcal{H}^q_{\rm osc} = M^q - i \Gamma^q$ contains off diagonal terms which result in CP violation in the neutral meson mixing. Writing the off-diagonal elements as
\begin{equation}
M^q_{12} = \left| M^q_{12} \right| \me^{i\phi^q_M} \quad\text{and}\quad \Gamma^q_{12} = \left| \Gamma^q_{12} \right| \me^{i\phi^q_\Gamma} \,,
\end{equation}
the mass and decay rate differences between the two propagation eigenstates are given by
\begin{align}\label{eq:deltaM_deltaGamma}
\Delta M^q \simeq 2 \left| M^q_{12} \right|\quad\text{and}\quad\Delta \Gamma^q \simeq 2 \left| \Gamma^q_{12} \right| \cos{\phi^q_{12}} \,,
\end{align}
where we have defined
\begin{equation}
\phi^q_{12} = \mathrm{arg}\left( -\frac{M^q_{12}}{\Gamma^q_{12}} \right) = \pi + \phi_M - \phi_\Gamma \, .
\end{equation}
To get to Eq.~\eqref{eq:deltaM_deltaGamma} we have assumed that $\left| \Gamma^q_{12} \right| \ll \left| M^q_{12} \right|$, which is well established experimentally. Note that although both $M^q_{12}$ and $\Gamma^q_{12}$ can be complex, only the phase difference between the two is physical. This phase is the main quantity controlling the CP violation in the system. In the SM, where the off-diagonal elements of the oscillation Hamiltonian are generated by box diagrams, nonzero $\phi^q_M$ and $\phi^q_\Gamma$ arise from the phase in the CKM matrix~\cite{Lenz:2011ti,Lenz:2012mb,Botella:2014qya,Gershon:2016fda}.

The NP contributions that are relevant for baryogenesis only modify $M^q$ and leave $\Gamma^q$ unchanged from its SM value. In particular, there is no extra source of \emph{direct CP violation}, which measures differences in the tree-level decay of $B^0\rightarrow f$ with respect to the CP conjugate $\bar{B}^0\rightarrow \bar{f}$. In addition, no new common final state to which both $B^0$ and $\bar{B^0}$ can decay is introduced. Therefore, the CP-violating observables that may be modified by NP contributions are:
\begin{itemize}
\item \emph{CP violation in mixing}, which is described by the mixing phase $\phi^q_{12}$. CP violation in mixing can be measured via the \emph{semileptonic CP asymmetries}, also known as CP asymmetries in flavor-specific decays. The asymmetry is defined as
\begin{equation}
a^q_{\mathrm{sl}} = \frac{\Gamma\left( \bar{B}^0_q \rightarrow f \right) - \Gamma\left( B^0_q \rightarrow \bar{f} \right)}{\Gamma\left( \bar{B}^0_q \rightarrow f \right) + \Gamma\left( B^0_q \rightarrow \bar{f} \right)} = \mathrm{Im}\left( \frac{\Gamma^q_{12}}{M^q_{12}} \right) = \left| \frac{\Gamma^q_{12}}{M^q_{12}} \right| \sin\phi^q_{12},
\end{equation}
where the final state is such that the decays $B^0_q \rightarrow f$ and $\bar{B}^0_q \rightarrow \bar{f}$ are forbidden.
\item \emph{Mixing-induced CP violation}, also known as \emph{CP violation in interference}, which arises from interference between mixing and decay. Here, we consider a final state $f$ to which in principle both $B^0$ and $\bar{B^0}$ can decay, and define
\begin{equation}
A^q_{CP,f}(t) = \frac{\Gamma\left( \bar{B}^0_q \rightarrow f \right) - \Gamma\left( B^0_q \rightarrow f \right)}{\Gamma\left( \bar{B}^0_q \rightarrow f \right) + \Gamma\left( B^0_q \rightarrow f \right)}.
\end{equation}
This form of CP violation is controlled by the phase $\phi^q$, which is defined as
\begin{equation}
\phi^q = -\pi + \phi^q_M + \mathrm{arg}\left( \frac{\bar{\mathcal{A}}_f}{\mathcal{A}_f} \right),
\end{equation}
where $\mathcal{A}_f$ ($\bar{\mathcal{A}}_f$) denotes the decay amplitude of $B^0$ ($\bar{B}^0$) to $f$.
\end{itemize}
In order to assess the extent to which NP can impact these observables, it is customary to define~\cite{Artuso:2015swg}
\begin{align}
M^q_{12} = M^{q,\mathrm{SM}}_{12}\cdot \Delta^q  \equiv M^{q,\mathrm{SM}}_{12}\cdot \left| \Delta^q \right| \me^{i\phi^q_\Delta},\, \quad \Gamma^q_{12} = \Gamma^{q,\mathrm{SM}}_{12}.
\end{align}
We reiterate that we are only allowing modifications in $M^q_{12}$, but not $\Gamma^q_{12}$, which is fixed to its SM value. The complex parameter $\Delta^q$ is a measure of the effects of NP. This parametrization has the advantage that each of $\Delta^q$ and $\phi^q_\Delta$ are individually related to the observables $\Delta M^q$ and $\phi^q$. It also allows to write a simple expression for the semileptonic asymmetries in the $B^0_{s,d}$ systems. For each $q=s, d$ we have
\begin{equation}
a^q_{\mathrm{sl}} = \frac{\left| \Gamma_{12}^{q,\mathrm{SM}} \right|}{\left| M_{12}^{q,\mathrm{SM}} \right|} \frac{\sin\left( \phi_{12}^{q,\mathrm{SM}} + \phi^q_\Delta \right)}{\left| \Delta^q \right|}.
\end{equation}
It is however more convenient for us to parametrize $M^{q,\mathrm{NP}}_{12}$ in a way that makes the relation with the parameters of the NP model more transparent. For that purpose, we define
\begin{equation}
M^{q,\mathrm{NP}}_{12} = M^{q,\mathrm{SM}}_{12}\cdot \alpha^q = M^{q,\mathrm{SM}}_{12}\cdot \left| \alpha^q \right| \me^{i\phi^q_\alpha}.
\end{equation}
Note that $\Delta = 1 + \alpha$. The NP effects are thus normalized with respect the SM values as
\begin{equation}\label{eq:alpha_phi_alpha}
\alpha^q = \frac{\left| M_{12}^{q,\mathrm{NP}} \right|}{\left| M_{12}^{q,\mathrm{SM}} \right|}\,,\qquad \phi^q_\alpha = \phi_M^{q,\mathrm{NP}} - \phi_M^{q,\mathrm{SM}}.
\end{equation}
The sign of the contribution to $a^q_{\rm sl}$ depends on the relative alignment of the NP phase with respect to the SM one (recall that a positive contribution is required for successful baryogenesis, see Sec.~\ref{sec:mechanism}). In the SM, the CP violation can be traced back to the single physical complex phase in the CKM matrix. Using the Wolfenstein parametrization, we obtain the SM predictions for the phases of the mass splittings,
\begin{align}
\phi_M^{d, \, \rm SM} \simeq \arctan\frac{2\eta(1-\rho)}{(1-\rho)^2-\eta^2} \simeq 0.768\,\mathrm{rad},\quad \text{and} \quad
\phi_M^{s, \, \rm SM} \simeq \arctan\frac{-2\lambda^2\eta}{1-\lambda^4\eta^2} \simeq -0.036\,\mathrm{rad}.
\end{align}

The experimental measurements of the oscillation period of neutral $B$ mesons (which constraints $\Delta M^q$) and the CP violation in interference (which constrains $\phi^q$) can be turned into exclusion limits on the size of NP contributions. We use the most up-to-date averages by the HFLAV group~\cite{HFLAV16} to derive our constraints. A key ingredient for the analysis are the SM predictions for the values of the relevant quantities $\Delta M^{q,\mathrm{SM}}$, $\Delta\Gamma^{q,\mathrm{SM}}$, $\phi_{12}^{q,\mathrm{SM}}$, $\phi^{q,\mathrm{SM}}$ and $a_{\rm sl}^{q,\mathrm{SM}}$, which we obtain from~\cite{Artuso:2015swg}.

\begin{figure*}[t!]
\centering
\includegraphics[width=0.99\textwidth]{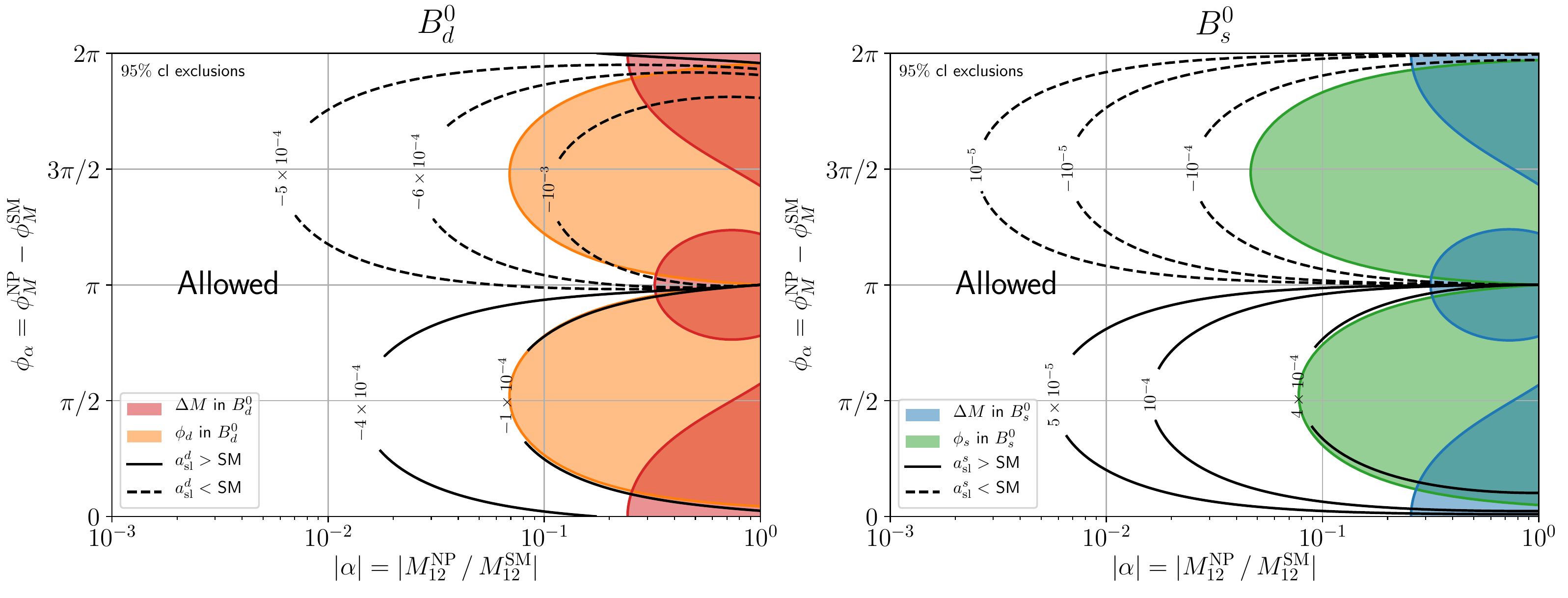}
\caption{\footnotesize{Bounds on the NP contributions to the absolute value (horizontal axis) and the phase (vertical axis) of $M_{12}$, assuming $\Gamma_{12}$ is given by the SM value. The left panel corresponds to the $B^0_d$ system, and the right panel to $B^0_s$. Colored contours are excluded by measurements of $\Delta M$ and mixing-induced CP violation. The contour levels give the values of the semileptonic asymmetry, with solid (dashed) lines corresponding to values larger (smaller) than the SM ones.}}
\label{fig:semileptonic_asymmetry_constraints}
\end{figure*}

The results of our analysis are shown in Fig.~\ref{fig:semileptonic_asymmetry_constraints} for the $B_d$ and $B_s$ systems. The main conclusion is that the combination of measurements pushes the absolute value of any NP contributions to be roughly one order of magnitude smaller than the SM values. This statement is slightly dependent on the phase of the NP contributions. Additionally, for small enough values of $\left| M^{q,\mathrm{NP}}_{12} \right|$, the phase $\phi^{q,\mathrm{NP}}_{M}$ is essentially unconstrained. In Fig.~\ref{fig:semileptonic_asymmetry_constraints}, we have also overlaid contours showing the value of the $a_{\mathrm{sl}}^q$ arising from the NP parameters. As expected, phases in the $(0,\pi)$ range enhance $a_{\mathrm{sl}}^q$ with respect to its SM value, whereas phases in the $(\pi,2\pi)$ range give a negative contribution to the asymmetry. It is important to note that in the $B^0_d$ system the asymmetry is bound to be negative, while in the $B^0_s$ system it can be positive. More precisely, we obtain the following $95\%$ cl ranges for the semileptonic asymmetries\footnote{Our limits differ by up to a factor of $2$ from the ones derived in~\cite{Bona:2007vi} (see \url{http://www.utfit.org} for up-to-date results), which are extracted from a global fit of all SM+NP parameters including all relevant constraints.}:
\begin{align}\label{eq:fit_semileptonic_asymmetries}
a^d_{\mathrm{sl}} \in \left[-8.9\times 10^{-4},\,-9.0\times 10^{-5}\right], \quad \text{and} \quad a^s_{\mathrm{sl}} \in \left[-2.1\times 10^{-4},\,4.1\times 10^{-4}\right].
\end{align}
This is a much tighter region than the one quoted in Eq.~\eqref{eq:experimental_semileptonic_asymmetries}, which is an average of only direct measurements of the asymmetries. 
Of course, the one we have derived here is reliant on the extra assumptions mentioned above regarding the NP contributions. 

\subsection{The Branching Fraction $\text{Br}(B^0 \to \mathcal{B} + X)$} 
\label{sec:model_indep_br}
\begin{figure}[t!]
  \begin{subfigure}[b]{.49\textwidth}
    \centering
    \includegraphics[width=0.885\textwidth]{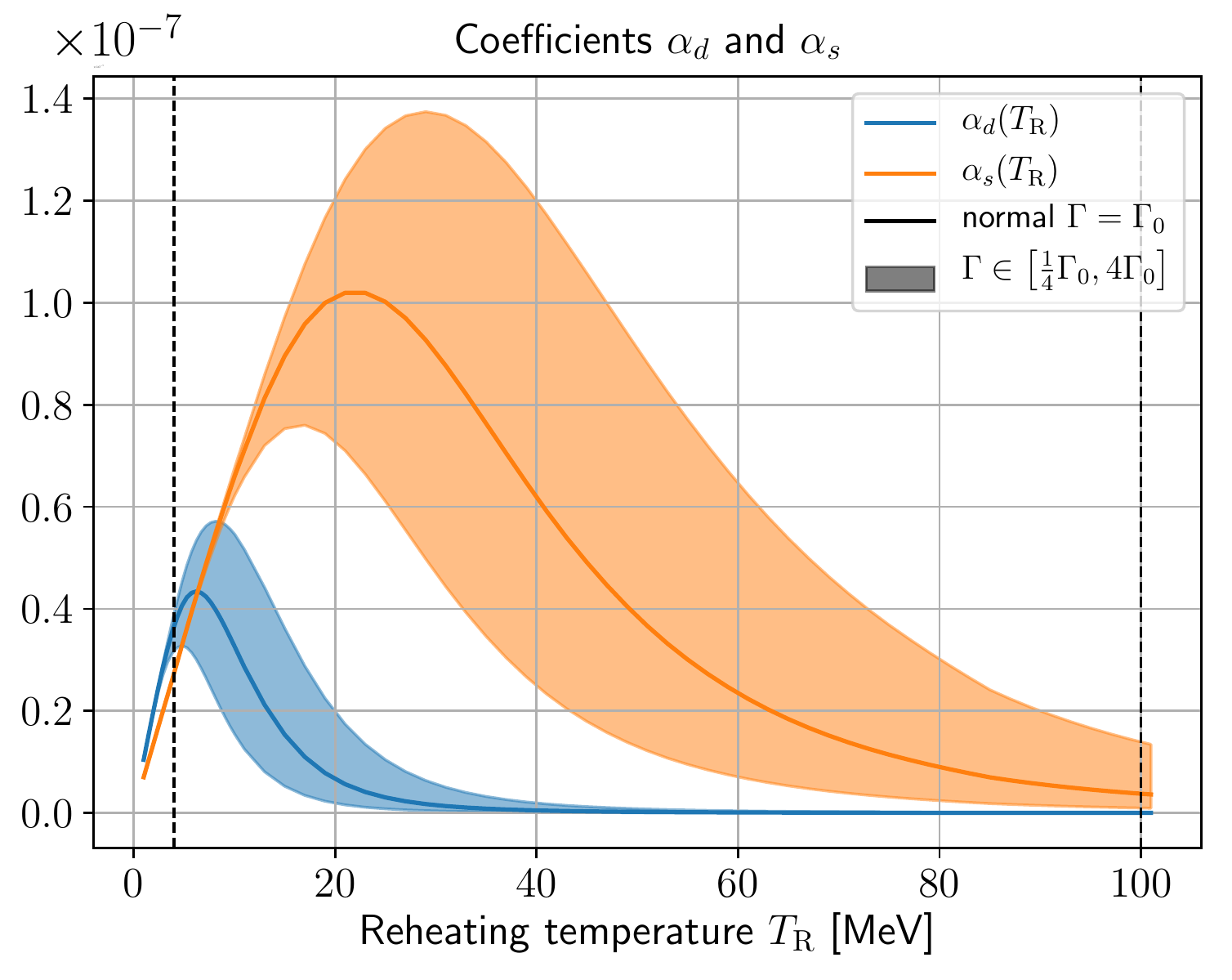}
    \label{fig:alpha_beta}
  \end{subfigure}
  \begin{subfigure}[b]{.49\textwidth}
    \centering
    \includegraphics[width=0.95\textwidth]{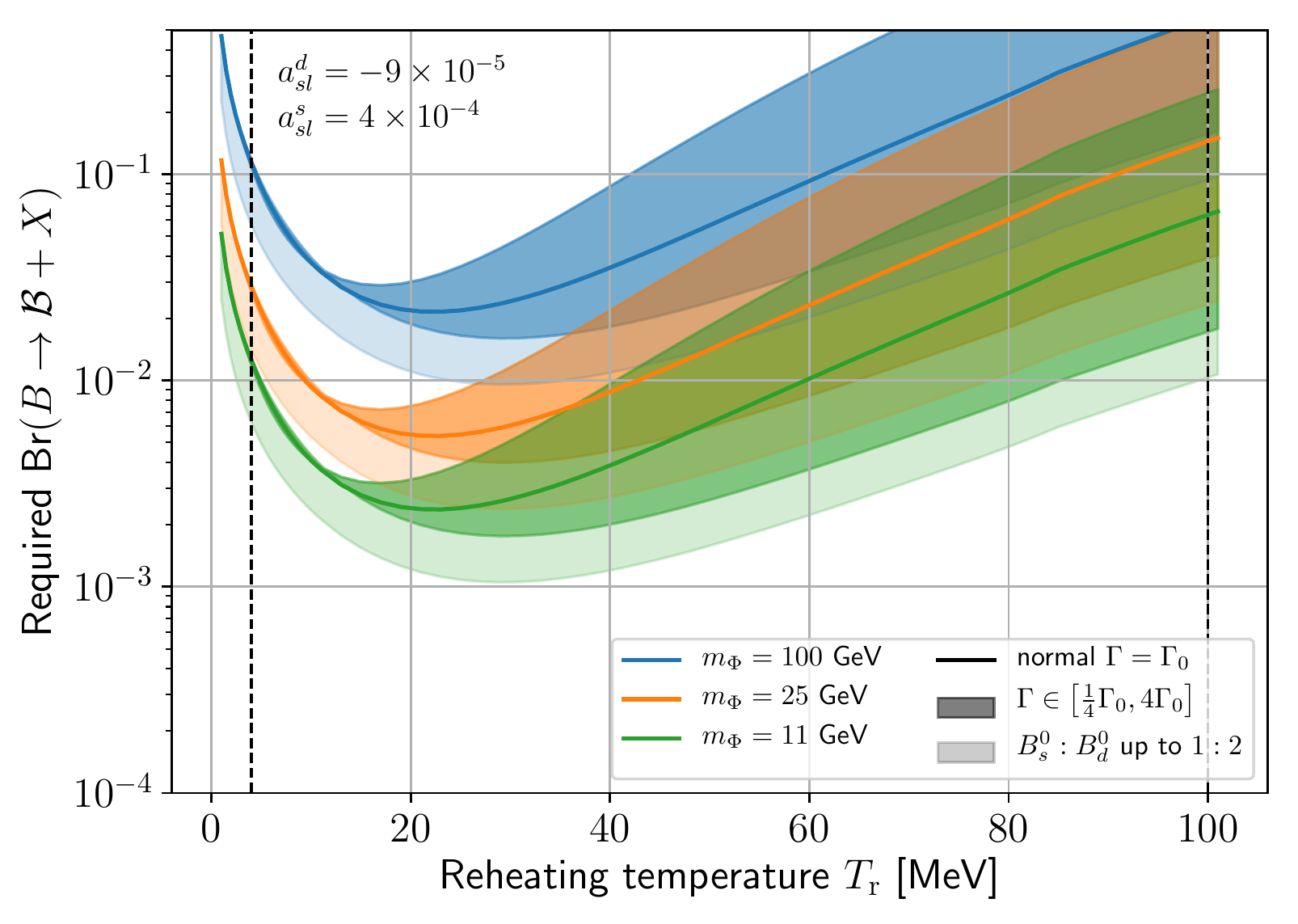}
    \label{fig:branching_ratio}
  \end{subfigure}
  \caption{\footnotesize{\textit{Left:}  Values of the numerically determined functions $\alpha_d$ and $\alpha_s$ as a function of the reheating temperature over the range of interest $4-100$~MeV, computed following the prescription of~\cite{Nelson:2019fln}. In order to account for the uncertainty in the calculation coming from the lack of precise knowledge of the scattering rate $\Gamma$ of $e^\pm$ off $B$ mesons in the primordial plasma, we conservatively allow for up to a factor of $4$ uncertainty in $\Gamma$. This uncertainty mostly stems to the lack of knowledge about the charge radius of the neutral $B$ mesons. \textit{Right:} Branching ratio of $B$ mesons to visible baryons and dark antibaryons necessary to reproduce the baryon asymmetry of the Universe, also as a function of the reheating temperature. We use the largest experimentally allowed values for the semileptonic asymmetries and give the results for 3 different values of the mass of  $\Phi$. We use different shades of the colors to highlight uncertainties coming from the scattering rate as explained above and, additionally, from potential variations in the fragmentation ratio of $b$ quarks to $B^0_s$ and $B^0_d$ mesons~\cite{Tanabashi:2018oca} (we use a benchmark ratio of $1:4$ but to be conservative allow it to be as large as $1:2$). In this range of parameters, branching ratios in the $0.001-0.1$ range can lead to successful baryogenesis.}}
  \label{fig:alpha_beta_branching_ratio}
\end{figure}
\end{spacing}

Taking the values in Eq.~\eqref{eq:fit_semileptonic_asymmetries} as reference for the maximum positive asymmetry that can be accommodated by NP contributions, we may use \eqref{eq:YB} to set a limit on the possible branching fraction of neutral $B$ mesons to baryons and dark sector states $\Br$ that is required to reproduce the observed baryon asymmetry of the Universe. 

The baryon asymmetry is solved for by evolving a set of coupled Boltzmann equations \cite{Aitken:2017wie,Elor:2018twp,Nelson:2019fln} governing the evolution of the number densities of the relevant particles $\Phi$, $B$, $\bar{B}$ and the DM. The solution may be found numerically, and factorizes into the following form:
\begin{equation}
\label{eq:Y3p2}
Y_B = \left( \frac{\Br}{10^{-2}} \right) \left( \frac{100\,\mathrm{GeV}}{m_\Phi} \right) \left( \alpha_d(T_{R})\,a^d_{\mathrm{sl}} + \alpha_s(T_{R})\,a^s_{\mathrm{sl}} \right) \,.
\end{equation}
In addition to the branching fraction and the semileptonic asymmetry, the asymmetry depends on  $m_\Phi$ and $\Gamma_\Phi$. The $\Phi$ width may be related to the late era reheating temperature in the usual way $\Gamma_\Phi = 3 H( T_R )$. Following the prescription of ~\cite{Nelson:2019fln}, the value of the asymmetry is simply inversely proportional to $m_\Phi$ and depends on the decay width through a numerically determined polynomial denoted by $\alpha_{q}(T_R)$ for the $B_{q = s, d}^0$ systems, whose dependence on $T_R$ is depicted in Fig.~\ref{fig:alpha_beta_branching_ratio}. Roughly speaking, $\alpha_q$ acts to parametrize the decoherence effects of the system. 

In analogy with neutrino oscillations, additional interactions with the $B$ mesons can ``measure" the state of the meson and decohere the oscillating system through the Zeno effect \cite{Venugopalan:1995je}. 
Such decoherence will spoil the oscillations and therefore prevent a baryon asymmetry from being generated. For the setup at hand, the only possible source of dechorence is the $e^{\pm}$ scattering off the neutral $B$ meson in the early Universe plasma. To a good approximation, decoherence becomes important when the scattering rate exceeds the oscillation lengths of the $B_s^0$ or $B_d^0$ system, which are measured to be $\Delta M_{s}  = 1.17 \times 10^{-11}\, \mathrm{GeV}$ and  $\Delta M_{d} = 3.34 \times 10^{-13}\, \mathrm{GeV}$. The temperature-dependent scattering rate is roughly $\Gamma_{e^{\pm} B \to e^{\pm} B} \sim 10^{-11}\, \mathrm{GeV} \left( T/ 20  \, \text{MeV} \right)^5$ up to uncertainties in the charge radius of $B^0_{s,d}$ mesons. Therefore, decoherence becomes more relevant at high temperatures. More specifically, the oscillations decohere for $T  \gtrsim ( 15, \, 30 ) \, \text{MeV}$ for the $B_d$ and $B_s$ systems, respectively. Decoherence effects may be properly treated in the Boltzmann equations through the prescription of \cite{Tulin:2012re}. Results are displayed on the left panel of Fig.~\ref{fig:alpha_beta_branching_ratio}, for which we assume that reheating occurs between $\sim 4 - 100 \text{MeV}$, that is, \emph{between BBN and hadronization}. To avoid excessive decoherence, baryogenesis favors reheating temperatures in the range $T_R \sim 15-50 \, \text{MeV}$. 

The behavior of the numerical results for $\alpha_{q} (T_R)$ can be understood as follows: at high reheat temperatures ($T_R \gtrsim 50 \text{MeV}$), the $B$ mesons formed from the decay of the $\Phi$ do not oscillate coherently and asymmetry production is therefore highly suppressed. As we consider lower values of $T_R$, the scattering rate for $e^{\pm} B \to e^{\pm} B$ drops such that the $B_s^0$ mesons are produced and oscillate coherently thereby producing a baryon asymmetry. At even lower temperatures, $B_d^0-\bar{B}_d^0$ coherent oscillations will also contribute to the asymmetry. The change of behavior of $\alpha_{s,d} (T_R)$ at very low $T_R$ is the result of dilution by a greater entropy density. 

Recall that for baryogenesis to proceed through $B$ meson oscillations and subsequent decay to the dark sector, a large positive net semileptonic-leptonic asymmetry is favored. As we saw above, the NP contributions to the neutral $B$ meson oscillating systems can enhance the semileptonic-leptonic asymmetry in $B_s^0$ decays to at most $a_{\mathrm{sl}}^s = 8.80 \times 10^{-4}$ while not being in conflict with measurements in $B$ physics. Meanwhile, $a^d_{\mathrm{sl}} < 0$ in all regions of parameter space, so that maximizing the produced baryon asymmetry favors the smallest possible value of this negative contribution, namely $a_{\mathrm{sl}}^d = -4 \times 10^{-4}$. At the same time, Fig.~\ref{fig:alpha_beta_branching_ratio} shows that the maximal values of $\alpha_s$ occurs for a reheat temperature of about $30$~MeV. Taking everything into account, we conclude that in order to reproduce the observed value of $Y=(8.718\pm 0.004)\times 10^{-11}$, a sizable inclusive branching ratio of $B$ mesons to baryons is needed. The required value for this branching ratio is displayed on the right panel of Fig.~\ref{fig:alpha_beta_branching_ratio} for extremal values of $a_{\mathrm{sl}}^{s,q}$, therefore representing a lower limit on the size of the branching fraction. The relatively large branching ratio is a robust prediction of the mechanism and provides a unique opportunity for high-energy experiments like LHCb \cite{Hicheur:2015oka} and Belle-II \cite{Kou:2018nap} to test this baryogenesis scenario. 

\newpage

\section{A Supersymmetric Realization}
\label{sec:Model}

The model independent analysis presented above allows to place constraints on the observables that play a direct role in the baryogenesis mechanism considered here. However, it is expected that other related observables can be used to indirectly test the scenario once the required minimal NP content is embedded in a consistent framework. With this aim, we construct a complete realization of the model in a supersymmetric context. The key ingredients of the theory which allow for baryogenesis to be realized are an unbroken $U(1)_R$ symmetry~\cite{Hall:1990hq, Randall:1992cq} and the presence of Dirac gauginos~\cite{Fayet:1974pd, Fayet:1975yi}. Supersymmetric models with Dirac gauginos and $U(1)_R$ symmetries are phenomenologically appealing~\cite{Fox:2002bu}. By forbidding Majorana gaugino masses and supersymmetric $a$-terms, the majority of collider constraints that push the MSSM into ever more fine tuned regions of parameter space are avoided. Their collider phenomenology has been greatly studied~\cite{Frugiuele:2012pe, Kalinowski:2015eca, Beauchesne:2017jou, Diessner:2017ske, Alvarado:2018rfl, Fox:2019ube}, along with flavor considerations~\cite{Kribs:2007ac}. Among other applications, these models can accommodate diverse baryogenesis scenarios~\cite{Fok:2012fb, Ipek:2016bpf}, sometimes at the cost of breaking the $R$-symmetry~\cite{Beauchesne:2017jou}. In this regard, realizing the mechanism of~\cite{Elor:2018twp} is a novel application of such ingredients and, as will be discussed, motivates a rather unstudied region of parameter space: a light (GeV mass scale) Dirac bino.

Let us briefly summarize the main ideas behind our supersymmetric theory for baryogenesis and DM from $B$ mesons. $R$-symmetries are known to arise naturally in supersymmetric constructions~\cite{Fayet:1974pd,Fayet:1975yi, Hall:1990hq,Randall:1992cq,Fox:2002bu,Kribs:2007ac}. Supercharges and their conjugates transform in opposite ways under an $R$-symmetry, allowing for particles within the same supermultiplet to carry different $R$-charges. Although $U(1)_R$ is usually taken to be broken at low energies, an interesting possibility is for it to remain exact\footnote{It is however difficult to argue against the existence of some amount of breaking due to supergravity, we will discuss this in Sec.~\ref{sec:Rsymbreaking}.} and be identified with baryon number. Following this avenue leads to quarks and squarks having different baryon number assignments, as shown in Tables~\ref{table:superfields} and \ref{table:fields}. In this way, a right-handed down-type squark $\tilde{d}_R$ has the required charge assignments to play the role of the heavy colored scalar in Eq.~\eqref{eq:Lag_psi}. We may also identify an adequate candidate for the neutral GeV scale Dirac baryon $\psi$ in the form of the Dirac bino. After integrating out the heavy squark, the effective four-fermion operator in Eq.~\eqref{eq:Lag_psi} arises from a combination of $R$-parity violating and gauge interactions. Furthermore, the Dirac partner of the bino is by definition a singlet under the SM gauge group, allowing it to naturally couple to a dark sector. The dark sector can be minimally accommodated by introducing a generic new singlet chiral multiplet. More interestingly, this new singlet can be identified with a right-handed neutrino multiplet. In this way, the bino mediates the decay of $B$ mesons into dark sector states. The complete set of superfield charge assignments is summarized in Table~\ref{table:superfields}, and the corresponding particle fields and their properties are given in Table~\ref{table:fields}. In what follows, we will further elaborate upon the ingredients of this model needed to realize baryogenesis.

\begin{table}[t]
\renewcommand{\arraystretch}{1.4}
\setlength{\arrayrulewidth}{.5mm}
\centering
\small
\setlength{\tabcolsep}{0.5 em}
\setlength{\arrayrulewidth}{.25mm}
\begin{tabular}{| c | c | c |}
    \hline
    Superfield  &   $R$-Charge (B no.)  & L no. \\ \hline\hline
    $\mathbf{U}^c, \mathbf{D}^c$ &   $2/3$  &   $0$  \\ \hline
    $\mathbf{Q}$ &   $4/3$ &   $0$   \\ \hline  
     $\mathbf{H}_u, \mathbf{H}_d$ &   $0$  &   $0$  \\ \hline
     $\mathbf{R}_u, \mathbf{R}_d$ &   $2$ &   $0$   \\ \hline
     $\mathbf{S}$, $\bT$, $\mathbf{O}$&   $0$  &   $0$  \\ \hline
     $\mathbf{L}$ &   $1$  &   $1$  \\ \hline
     $\mathbf{E}^c$, $\mathbf{N}_R^c$ &   $1$  &   $-1$  \\ \hline
\end{tabular}
\caption{\footnotesize{Summary of superfields with their $R$-charge (baryon number) and lepton number assignments. Each antichiral superfield has $R$-charge opposite to that of the corresponding chiral superfield.}}
\label{table:superfields}
\end{table}

\subsection{An Exact $U(1)_R$ Symmetry}

The model is similar to the one studied in \cite{Frugiuele:2012pe, Beauchesne:2017jou}. An exact $R$-symmetry requires a superpotential with $R$-charge of 2 (superspace derivatives $\mathcal{D}_\alpha$ have $R$-charge $-1$). Given the $U(1)_R$ charges of superfields (see Table~\ref{table:superfields}), only the following terms respect the $R$-symmetry: 
\begin{align}
\label{eq:WRPV}
\mathbf{W}  \,\,\,  \supset   \,\,\,   y_u \bQ \bH_u \bU^c - y_d \bQ \bH_d \bD^c - y_e \bL \bH_d \bE^c  + \frac{1}{2} \lambda_{ijk}^{''} \bU^c_i \bD^c_j \bD^c_k  +  \mu_u \bH_u \bR_d + \mu_d \bR_u \bH_d 
\end{align}
The first four terms of Eq.~\eqref{eq:WRPV} constitute the usual MSSM superpotential, with the addition of the ``$R$-parity violating" (RPV) term $\bU^c_i  \bD^c_j  \bD^c_k$ (the subscripts denote generation indices) which is now allowed\footnote{Note that although the $\lambda^{"}$ couplings are $R$-parity violating, they preserve the $U(1)_R$ symmetry (i.e. they do not break baryon number). As it is a very extended convection in the SUSY literature, we will henceforth refer to $\lambda^{"}$ as the RPV couplings. Other $R$-parity violating couplings, like the ones usually denoted by $\lambda$, $\lambda^{\prime}$ or $\mu^{\prime}$ (see e.g.~\cite{Martin:1997ns}) are not allowed by the $R$-symmetry.}. The rest describe the Higgs sector;  
the $\bR_{u,d}$ multiplets are added to generate $\mu$ terms which are forbidden in models with $R$-symmetry. Electroweak symmetry breaking proceeds as usual when the scalars of $\mathbf{H}_u$ and $\mathbf{H}_d$ acquire vacuum expectation values, while the VEVs of $\mathbf{R}_{u.d}$ remain zero.

Given the $U(1)_R$ charge assignments of the matter content (see Table~\ref{table:superfields}), $\bU^c$ and $\bD^c$ have an $R$-charge of $2/3$. If we explicitly write out the decomposition of the multiplets, e.g. $\bD^c = \tilde{d}_R^* + \sqrt{2} \theta d_{R}^\dagger + \theta^2 F_d $, we see that the d-type right-handed squarks have $R$-charge $-2/3$ while the quarks\footnote{\footnotesize{We use the notation of \cite{Martin:1997ns} to denote $d_R^\dagger$ as the Weyl spinor in the chiral superfield $\mathbf{D}^c$.}} have $R$-charge $1/3$ (recall that $\theta$ has $R$-charge 1). Since we identify baryon number with R-charge, $\tilde{d}_R$ has the appropriate charge assignments to be identified with the heavy colored scalar that mediates the effective interaction of Eq.~\eqref{eq:Lag_psi}.
At the quark/squark level, the $\bU^c \bD^c \bD^c$ term in the superpotential produces the interactions
\begin{align}
\label{eq:Lsquark}
\mathcal{L} \,\,\, \supset \,\,\, \lambda^{''}_{123} \left(  \tilde{u}_R^* s_R^\dagger b_R^\dagger +  \tilde{s}_R^* b_R^\dagger u_R^\dagger   + \tilde{b}_R^* u_R^\dagger s_R^\dagger \right)\,.
\end{align}
As in Sec.~\ref{sec:mechanism}, we have chosen a particular combination of quark flavor to keep the notation simple. In general, the role of $\tilde{d}_R$ can be played by any first or second generation up- or down-type squark, which means that any of the couplings $\lambda^{\prime\prime}_{ij3}$ with $i,j=1,2$ can be relevant for the baryogenesis mechanism.

\begin{table}[t]
\renewcommand{\arraystretch}{2}
\setlength{\arrayrulewidth}{.4mm}
\centering
\small
\setlength{\tabcolsep}{0.26 em}
\setlength{\arrayrulewidth}{.25mm}
\begin{tabular}{ |c  || c | c | c | c | c | c |}
    \hline
    Particle & Description  & Spin &  $Q_{EM}$ &   Baryon no.  &  Lepton no. &  Mass \\ \hline \hline
    $\, \Phi \,$ & ``Inflaton-like" scalar &    $0$  &   $0$ &    $0 $ & $0$ & $\, 11-100 \, \text{GeV} \, $ \\ \hline 
     $\, \tilde{q}_R \in \mathbf{D}^c \,, \mathbf{U}^c \, $ & Squark  &    $0$  &   $-1/3$ &    $ -2/3$  &  $0$ & $\, \mathcal{O}(\text{TeV}) \, $ \\ \hline 
 $\,  \begin{array}{c}\tilde{B} \in \mathbf{W}_\alpha \\ \lambda_s^\dagger \in \mathbf{S} \end{array}  $ & Dirac bino   &  $1/2$  &   $0$ &   $-1$  &  $0$ & $\,  \mathcal{O}(\text{GeV}) \,$\\ \hline  
     $\, \nu_R \in \mathbf{N}_R \,$ & Sterile neutrino &     $1/2$  &   $0$ &   $ 0$ & $1$  & $\, \mathcal{O}(\text{GeV})\, $ \\ \hline 
    $\, \tilde{\nu}_R \in \mathbf{N}_R \,$   & Sterile sneutrino DM  &  $0$  &   $0$ &    $-1$ & $1$ & $\, \mathcal{O}(\text{GeV})\, $\\ \hline 
\end{tabular}
\vspace{5mm}
\caption{\footnotesize{Summary of the component fields and their charge assignments, properties, and the superfield in which they are embedded. We work in terms of Weyl spinors so that the Dirac bino consists of two components $\psi = (\tilde{B}, \, \lambda_s^\dagger)$. Here, $\Phi$ is an ``inflaton-like" field that decays out of equilibrium producing $b$ and $\bar{b}$ quarks. The heavy colored squark $\tilde{q}_R$ is integrated out to mediate quark-bino four fermion interactions (note that various flavors of squarks may generate this interaction, as will be discussed further below). The Dirac bino mediates decays of hadrons into the dark sector. The dark sector is embedded into a right-handed neutrino supermultiplet where the sneutrino is stable and represents our DM candidate, while the sterile neutrino generally decays on time scales of interest for collider searches.}}
\label{table:fields}
\end{table}

%
%
\subsection{Dirac Gauginos}
In Wess-Zumino gauge, the field strength of a vector superfield can be written in the usual way as
\begin{align}\label{eq:field_strength_WZ}
\bW_\alpha^{\tilde{B}} = \tilde{B}_\alpha + \bigl[ D_1 \delta_\alpha^\beta+ \frac{i}{2} (\sigma^\mu \bar{\sigma}^\nu)^\beta_\alpha B_{\mu \nu} \bigr] \theta_\beta + i \theta^2   \sigma^\mu_{\alpha \dot{\beta}} \nabla_\mu \tilde{B}^{\dagger \dot{\beta}} \, ,
\end{align}
where fields are expressed in terms of the $y^\mu = x^\mu - i \bar{\theta} \sigma^\mu \theta$ coordinates.
The hypercharge $U(1)_Y$ field strength tensor is $B_{\mu \nu} = \partial_\mu B_\nu - \partial_\nu B_\mu$, and $\tilde{B}$ is the bino, that is, a left handed $2$-component Weyl spinor. One may construct a $4$-component Dirac gaugino by adding another $2$-component Weyl spinor to the theory. Concretely, let us introduce a new $\mathbf{S}$ superfield
\begin{align}
\label{eq:adjoint_singlet}
&\bS(y^\mu) = \phi_s + \sqrt{2} \lambda_{s}^\alpha  \theta_\alpha + \theta^\alpha \theta_\alpha F_s .
\end{align}
Gauge invariance requires the new field to transform in the adjoint representation, while the $R$-symmetry fixes its $R$-charge to vanish. 
Multiplets for the other SM gauge fields (a triplet $T$ and an octet $O$) are added in a similar way\footnote{The existence of these adjoint superfields can be motivated by imposing a $\mathcal{N}=2$ supersymmetry in the gauge sector, as was done in~\cite{Fox:2002bu}.}.

With the addition of the new superfields, Dirac gaugino mass terms can be generated~\cite{Fox:2002bu, Benakli:2011vb}. Consider the following superpotential terms: 
\begin{align}
\label{eq:DiracGauginoMass}
\mathcal{L}_{\mathrm{mass}} = \sqrt{2} \mathop{\int d^2 \theta}\, \bW^{' \alpha} \biggl[ \frac{c_1}{\Lambda_1} \bW_\alpha^{\tilde{B}} \bS  + \frac{c_2}{\Lambda_2} \bW_\alpha^{\tilde{W} i} \bT^i + \frac{c_3}{\Lambda_3} \bW_\alpha^{\tilde{g} a} \bO^a  \biggr] + \text{h.c.} \,.
\end{align}
SUSY breaking is introduced through a D-term spurion $\mathbf{W}^{' \alpha} = D \theta^\alpha$. Here, $c_i$ are $\mathcal{O}(1)$ coefficients, and in general we allow the SUSY breaking scale for each gauge group $\Lambda_i$ to be different. 
From  Eqs.~\eqref{eq:field_strength_WZ} and~\eqref{eq:adjoint_singlet} together with Eq.~\eqref{eq:DiracGauginoMass}, we see that upon SUSY breaking, Dirac mass terms
\begin{align}\label{eq:Dirac_Gaugino_mass_expanded}
\mathcal{L}_{\mathrm{mass}}^{\tilde{B}} \,\,\,  \supset \,\,\, \sqrt{2} c_1 \frac{D}{\Lambda_1} \tilde{B}^\alpha \lambda_{s, \alpha} + c_1\frac{D D_1}{\Lambda_1} \phi_s
\end{align}
are generated (the second term of \eqref{eq:Dirac_Gaugino_mass_expanded} vanishes on-shell when $D_1 = 0$). We denote the 4-component Dirac bino and its corresponding Dirac mass as
\begin{equation}
\label{eq:diracBinoMass}
\psi =\left( \begin{array}{c}  \tilde{B} \\ \lambda_s^\dagger \end{array} \right) \quad \text{and} \quad m_{\psi} = \sqrt{2} c_1 \frac{\langle D\rangle}{\Lambda_1} \, ,
\end{equation}
in consonance with Eq.~\eqref{eq:Lag_psi}. Similar expressions are obtained for the other gauginos, which acquire masses $m_{\psi_{2,3}} = \sqrt{2} c_{2,3}D/\Lambda_{2,3}$. In order to reproduce the spectrum required for baryogenesis, $m_\psi$ needs to be a few GeV. At the same time, collider~\cite{Cabrera:2011bi} and flavor~\cite{Gabbiani:1996hi,Bagger:1997gg} constraints generically push charginos and gluinos to lie above the TeV scale. If no other source of mass for gauginos is present, this implies that there must exist a hierarchy $(c_1/\Lambda_1) / (c_{2,3}/\Lambda_{2,3}) \lesssim 10^{-3}$. Such hierarchy can be explained either by invoking some tuning of the $c_i$ coefficients or by allowing for at least two separate sources of SUSY breaking~\cite{Cohen:2016kuf} so that $\Lambda_1 \ll \Lambda_{2,3}$. 

\subsection{Squark Masses and Couplings}
\label{subsec:squarkmassandcouplings}
Soft masses for squarks (and other superpartners) can be generated through an $F$-term SUSY breaking spurion $\mathbf{X}=\theta F$ ~\cite{Kribs:2007ac} at a scale $\Lambda_F$, 
which produces squark masses $m_{\tilde{q}} \sim F^* F / \Lambda_F$. Given the experimental constraints~\cite{Kribs:2007ac,Beauchesne:2017jou,Fox:2019ube} and depending on the scale of gluino and chargino masses, squarks as light as a few $100$ GeV can be feasible\footnote{Dirac gaugino masses produce a \emph{supersoft}~\cite{Fox:2002bu} scenario where radiative corrections to the squark masses are finite. This allows for squarks to be much lighter than gauginos, thus alleviating many of the collider and flavor constraints. For this reason, this paradigm is sometimes referred to as \emph{supersafe}. 
}.

As discussed above, the role of the colored scalar participating in the baryogenesis mechanism is played by a right-handed squark, which can be up- or down-type and belong to any of the first two generations. The squark mass is assumed to be in the TeV range so that it can be safely integrated out at the GeV scale where the baryogenesis dynamics happen. It is straightforward to see how the $\tilde{q}_R\, \bar{\psi}\, s$ term in Eq.~\eqref{eq:Lag_psi} is generated in the supersymmetric setup. Interacting chiral matter theories have gauge interactions of the form 
\begin{align}
\mathcal{L}_{\text{gauge}} \,\,\, \supset \,\,\, - \sqrt{2} g^\prime\,Q_{q_R} (\tilde{q}_R^*\, q_R\, \tilde{B}^\dagger) + \text{h.c.}\, ,
\label{eq:componentgaugeL}
\end{align}
where $Q_{q_R}$ and $g^\prime = e/\cos{\theta_W}$ denote the charge assignment and hypercharge gauge coupling (as customary, $e$ is the electromagnetic charge quantum and $\theta_W$ is the weak mixing angle). When $\tilde{s}_R$ is integrated out, the effective four-fermion interaction $ \tilde{B}\, u\, s\, b$ is recovered. Of course, other flavor combinations for the operator are possible: they are summarized in Table~\ref{table:flavor_structure} along with the resulting coefficient of the effective interaction. It is worth noting that the gauge coupling being fixed, the only unknown parameter that sets the strength of the effective interaction is $\lambda^{\prime\prime}_{ij3}$.
\begin{table}[t] 
\renewcommand{\arraystretch}{1.75}
\setlength{\arrayrulewidth}{.4mm}
\centering
\small
\setlength{\tabcolsep}{0.5 em}
\setlength{\arrayrulewidth}{.25mm}
\begin{tabular}{| c | c | c |}
    \hline
    four-fermion operator  &  Effective coupling  &  $\tilde{q}_R$ particle  \\ \hline \hline
    $\psi\, b\, u\, d$  &  $\lambda^{\prime\prime}_{113} \sqrt{2}\, g^\prime\, \left( \frac{Q_u}{m_{\tilde{u}}^2} + \frac{Q_d}{m_{\tilde{d}}^2} - \frac{Q_d}{m_{\tilde{b}}^2} \right)$  &  $\tilde{u}_R$, $\tilde{d}_R$, $\tilde{b}_R$  \\ \hline
    $\psi\, b\, u\, s$  &  $\lambda^{\prime\prime}_{123} \sqrt{2}\, g^\prime\, \left( \frac{Q_u}{m_{\tilde{u}}^2} + \frac{Q_d}{m_{\tilde{s}}^2} - \frac{Q_d}{m_{\tilde{b}}^2} \right)$  &  $\tilde{u}_R$, $\tilde{s}_R$, $\tilde{b}_R$  \\ \hline
    $\psi\, b\, c\, d$  &  $\lambda^{\prime\prime}_{213} \sqrt{2}\, g^\prime\, \left( \frac{Q_u}{m_{\tilde{c}}^2} + \frac{Q_d}{m_{\tilde{d}}^2} - \frac{Q_d}{m_{\tilde{b}}^2} \right)$  &  $\tilde{c}_R$, $\tilde{d}_R$, $\tilde{b}_R$  \\ \hline
    $\psi\, b\, c\, s$  &  $\lambda^{\prime\prime}_{223} \sqrt{2}\, g^\prime\, \left( \frac{Q_u}{m_{\tilde{c}}^2} + \frac{Q_d}{m_{\tilde{s}}^2} - \frac{Q_d}{m_{\tilde{b}}^2} \right)$  &  $\tilde{c}_R$, $\tilde{s}_R$, $\tilde{b}_R$  \\ \hline
\end{tabular}
\vspace{5mm}
\caption{\footnotesize{Exhaustive list of all the possible operators contribution to the baryogenesis mechanism in a $U(1)_R$ supersymmetric setup with Dirac gauginos.
}}
\label{table:flavor_structure}
\end{table}

\subsection{The Higgs Sector and Neutralino Mixing}
\label{sec:neutralino_mixing}
The number of neutral fermions in $R$-symmetric models is larger than in the MSSM due to the presence of the new $SU(2)$ doublets $R_{u,d}$ and the singlet and triplet fields $S$ and $T$. Here we follow~\cite{Fox:2019ube} to investigate the mixing between all these states after electroweak symmetry breaking. The relevant terms in the Higgs sector, allowed by gauge and $R$-Symmetries, are given by
\bea
\label{eq:HiggsW}
\mathbf{W} & \supset  \mu_u \bH_u \bR_u + \mu_d \bH_d \bR_d  + \bS \left( \lambda_u^{\tilde{B}} \bH_u R_u +  \lambda_d^{\tilde{B}} \bH_d \bR_d \right) + \bT \left( \lambda_u^{\tilde{W}} \bH_u \bR_u +  \lambda_d^{\tilde{W}} \bH_d \bR_d \right) .
\eea
Recall that $\bR_{u,d}$ are added to generate the otherwise $R$-symmetry forbidden $\mu$-terms. Electroweak symmetry breaking proceeds as usual when the scalars of $\mathbf{H}_u$ and $\mathbf{H}_d$ acquire vacuum expectation values $v_u$ and $v_d$ (the VEVs of $\mathbf{R}_{u.d}$ remain zero). The terms in~\eqref{eq:HiggsW}, together with the Dirac gaugino masses and the gauge potential, generate the Dirac mass mixing matrix
\bea
M_{\tilde{N}} =  \left( \begin{array}{cccc}
m_\psi & 0 & g^\prime v_u /\sqrt{2} & - g^\prime v_d / \sqrt{2}  \\
0 & m_{\tilde{W}} & -g v_u/\sqrt{2} & g v_d / \sqrt{2} \\ 
\lambda_u^{\tilde{B}} v_u/\sqrt{2}  & - \lambda_u^{\tilde{W}} v_u / \sqrt{2} & \mu_u & 0 \\
- \lambda_d^{\tilde{B}} v_d / \sqrt{2} & \lambda_d^{\tilde{W}} v_d / \sqrt{2} & 0 & \mu_d
 \end{array} \right) \, ,
\eea
which is written in the $(\tilde{B},\, \tilde{W},\, \tilde{R}_u,\, \tilde{R}_d)\times (\lambda_S,\, \lambda_T,\, \tilde{H}_u,\, \tilde{H}_d)$ basis. As usual, $g=e/\sin\theta_W$ and we require that $v_u^2 + v_d^2 = v^2/2 = (174\,\mathrm{GeV})^2$. In the large $\tan\beta \equiv v_u / v_d$ limit, the mass matrix simplifies to
\bea
M_{\tilde{N}} \simeq  \left( \begin{array}{cccc}
m_\psi & 0 & g^\prime v / \sqrt{2} & 0  \\
0 & m_{\tilde{W}} & -g v / \sqrt{2} & 0 \\ 
\lambda_u^{\tilde{B}} v / \sqrt{2}  & - \lambda_u^{\tilde{W}} v / \sqrt{2} & \mu_u & 0 \\
0 &0 & 0 & \mu_d
 \end{array} \right) \, ,
\eea
showing that one of the states with mass $\mu_d$ decouples. It is clear that if $\lambda_u^{\tilde{B}} = \lambda_u^{\tilde{W}} = 0$, there is no mixing between the bino, the neutral weakino and the higgsinos. Even if $\lambda_u^{\tilde{B},\tilde{W}}$ are nonzero, in the limit where $m_\psi \ll v \ll m_{\tilde{W}},\, \mu_u$, the lightest state is mostly bino with mass~$m_\psi$. This is the case of interest to us, as we are dealing with a GeV bino but other gauginos and higgsinos are at or above the TeV scale. In this limit, the mixings of the bino are approximately given by
\begin{align}\label{eq:bino_mass_mixing}
\theta_{\tilde{B}\tilde{W}} \simeq \frac{1}{2} \frac{\lambda_u^{\tilde{B}}\,g\,v^2}{m_{\tilde{W}}\,\mu} , \quad \quad
\theta_{\tilde{B}\tilde{H}_u} \simeq \frac{1}{\sqrt{2}} \frac{\lambda_u^{\tilde{B}}\,v}{\mu} .
\end{align}
Although not relevant for the baryogenesis dynamics, these mixings can play a role in constraining the model using flavor observables such as $\mu\rightarrow e\gamma$, which are discussed in Sec.~\ref{sec:deltaF1_observables}.

\subsection{A Dark Chiral Multiplet}
As was highlighted in Sec.~\ref{sec:mechanism}, the crucial step to generate a baryon asymmetry in the visible sector without incurring into global baryon number violation is to allow $\psi$ to decay into a dark sector where the corresponding antibaryon number is stored. In this way, a DM abundance is generated in parallel to the baryon asymmetry.

Thus far, the model we have introduced cannot accommodate DM. However, being uncharged under the SM gauge group, the singlet $S$ offers a natural portal to a dark sector. Given that the fermion in $S$ is the Dirac partner of the bino, such a coupling allows $\psi$ to decay to the dark sector. The minimal extension to take advantage of this portal includes a new chiral multiplet $\bphi$ where the $B=-1$ dark scalar $\phi$ and the $B=0$ Majorana spinor $\xi$ can both be embedded, 
\begin{equation}
\bphi = \phi^* + \sqrt{2} \theta^\alpha \xi_\alpha + \theta^2 F_\phi .
\end{equation}
The only superpotential terms allowed by gauge and $U(1)_R$ symmetries are precisely
\begin{equation}\label{eq:dark_matter_superpotential}
W \,\,\, \supset \,\, \,  \int \text{d}^2 \theta \,   \left(y_s \bS \bphi \bphi + m_\phi \bphi \bphi \right).
\end{equation}
The former generates the coupling $\lambda_s\, \phi^*\, \xi$ while the latter gives masses to the dark states. Note that the invariance of~\eqref{eq:dark_matter_superpotential} under $\bphi \leftrightarrow - \bphi$ acts to stabilize the DM (corresponding to the $\mathbb{Z}_2$ symmetry introduced in~\cite{Elor:2018twp}).

This most economic and generic extension contains all the ingredients to produce the baryon asymmetry and an asymmetric DM component. However, given that the dark states $\phi$ and $\xi$ and the mediator $\psi$ are rather light, a symmetric relic abundance tends to be overproduced \cite{Elor:2018twp} in this minimal setup. In order to partially deplete the symmetric abundance, new interactions have to be added to allow for the excess DM to annihilate\footnote{This problem is recurrent in asymmetric DM scenarios that identify the dark sector and baryon asymmetries. See~\cite{Davoudiasl:2010am} for other examples of how the symmetric DM component can be depleted.}. In our case, there exists an attractive solution to this problem: to identify the dark sector states with a right-handed neutrino supermultiplet.

\subsection{Sterile Sneutrino Dark Matter} \label{sec:sterile_sneutrino_DM}

In order to accommodate neutrino masses in the $U(1)_R$ symmetric model, the minimum requirement is the presence of a right-handed sterile neutrino multiplet $\mathbf{N}_R$ with $R$-charge $+1$,
\bea
\mathbf{N}_R^c = \tilde{\nu}_R^* + \sqrt{2} \theta \nu_R^\dagger + \theta^2 F_{\nu_R}^* \,.
\eea
Identifying $U(1)_R$ with $U(1)_B$ leads to the right-handed sneutrino $\tilde{\nu}_R$ carrying baryon number $-1$, while  the right-handed neutrino remains uncharged under baryon number.
At the same time, $\mathbf{N}_R^c$ has to carry $-1$ lepton number.
The following operators, allowed by all the symmetries (but including a $\Delta L = 2$ Majorana neutrino mass term), can be added to the superpotential: 
\bea
\label{eq:WNR}
\mathbf{W} \supset  \frac{\lambda_N}{4} \mathbf{S} \mathbf{N}_R^c \mathbf{N}_R^c + \mathbf{H}_u \mathbf{L}^I y_N^{I J} {\mathbf{N}_R^J}^c  + \frac{1}{2} \mathbf{N}_R^c M_M \mathbf{N}_R^c + \text{h.c.} \,.
\eea
To keep the discussion general, we allow for three flavors of sterile neutrinos ($I= e,\nu,\tau$; $J=1,2,3$). Note that the first and third terms are equivalent to the ones in~\eqref{eq:dark_matter_superpotential}, so the identification of $N_R^c$ as the DM multiplet is direct. The first term of \eqref{eq:WNR} may be expanded in component fields as
\bea
\label{eq:WLambdaN}
\frac{\lambda_N}{4}  \int \mathop{\mathrm{d}^2 \theta} \,  \mathbf{S} \mathbf{N}_R^c \mathbf{N}_R^c \supset \frac{1}{2} \lambda_N  \left( \lambda_s \nu_R^\dagger \tilde{\nu}_R^* + \phi_s \nu_R^\dagger \nu_R^\dagger \right) + \text{h.c.} \,.
\eea
This operator generates the three point interaction~\eqref{eq:DarkYukawa} that mediates the decay of the Dirac bino (through its right-handed partner $\lambda_s$) into the dark sector states $\nu_R$ and $\tilde{\nu}_R$. The second term in~\eqref{eq:WNR} generates Dirac neutrino masses. The $SU(2)_L$ multiplets can be expanded out as
\bea
\label{eq:Nmassandint}
\int \mathop{\mathrm{d}^2 \theta} \left(  \mathbf{H}_u^0  \mathbf{N}_L^I y_N^{I J} {\mathbf{N}_R^J}^c - \mathbf{H}_u^{+} \mathbf{E}_L^I   y_N^{I J} {\mathbf{N}_R^J}^c  \right) + \text{h.c.} = 2 y_N^{I J} \left( h_u^0 \nu_L^I {\nu_R^J}^{\dagger} + \tilde{h}_u^0 \tilde{\nu}_L^I {\nu_R^J}^{\dagger} + \tilde{h}_u^0 \nu_L^I {\tilde\nu^{J*}_R} \right)& \\ \nonumber
- 2 y_N^{I J} \left( h_u^{+} \ell_L^I {\nu_R^J}^{\dagger} + \tilde{h}_u^{+} \tilde{\ell}_L^I {\nu_R^J}^{\dagger} + \tilde{h}_u^{+} \ell_L^I {\tilde{\nu}_R^{J*}} \right)& + \text{h.c.}\, ,
\eea 
where $\ell^I= e, \, \mu, \, \tau \,$. Upon electroweak symmetry breaking, $\langle h_u^0 \rangle  = v_u$ produces a Dirac mass for the neutrino,
\begin{equation}
\label{eq:DiracNuMass}
\mathcal{L} \supset m_D \left( \nu_L^i {\nu_R^j}^\dagger + {\nu_L^i}^\dagger \nu_R   \right),
\end{equation}
with $m_D = 2 y_N v_u = 2 y_N v \sin \beta$. For simplicity, we have assumed a diagonal flavor structure and suppressed generation indices. This simple setup reproduces the type-I seesaw mechanism~\cite{Mohapatra:1979ia,GellMann:1980vs,Minkowski:1977sc,Schechter:1980gr}, whereby the light neutrinos obtain masses $m_\nu \simeq m_D^2 / M_M$, as long as $m_D\ll M_M$. For the baryogenesis mechanism to be viable, at least one of the right-handed neutrinos needs to have a mass in the GeV range so that $\psi$ can decay into it. In order to obtain a neutrino mass of $m_\nu \lesssim 0.1\,$eV, a small Yukawa $y_N\lesssim 10^{-8}$ is required, as will be further discussed in Sec.~\ref{subsec:oscSMnu}.

Aside from generating Dirac neutrino masses, the second term in~\eqref{eq:WNR} also provides the interactions necessary to avoid the overproduction of a symmetric DM component. The crucial difference to the generic case discussed in the previous section is the fact that the DM particles, being identified with sterile neutrino and sneutrino, now carry lepton number. The annihilation of right-handed sneutrinos into left-handed neutrinos can occur via $t$-channel exchange of a bino or a higgsino. We will study this in more detail in Sec.~\ref{sec:dark_matter}.

\subsection{$U(1)_R$ Symmetry Breaking from Supergravity} 
\label{sec:Rsymbreaking}
The model presented above relies on an exact global $U(1)_R$ symmetry which is identified with baryon number. While $R$-symmetries are motivated by top down constructions, we do generically expect them to be broken at low scales. Supersymmetry breaking effects coming from supergravity are generically always present (however see \cite{Luty:2002ff} for a possible alternative), and these generate a tiny Majorana mass for the gauginos through the conformal anomaly (as well as soft squark masses and $a$-terms). The order parameter is the gravitino mass $m_{3/2}$ so that gaugino Majorana masses 
\bea
M_1 \propto m_{3/2} = \frac{\langle F_\phi \rangle}{M_{\rm Pl}}
\eea
appear. Here, $M_{\rm Pl}$ denotes the Planck mass. 
In particular, Anomaly Mediated Symmetry Breaking (AMSB) \cite{Randall:1998uk,Giudice:1998xp} produces a small Majorana mass term for the bino
\bea
\label{eq:AMSBmajoranagaugino}
M_a =   \frac{\beta_{g_a}}{g_a} m_{3/2} \quad \Rightarrow \quad M_1 = \frac{{g^\prime}^2}{16 \pi^2} \frac{33}{5} m_{3/2} \,.
\eea
Additionally, soft squark masses $(m^2)^i_j = \frac{1}{2} |m_{3/2}|^2 \frac{d}{d t} \gamma^i_j$ and non-zero $a$-terms $a^{ijk} = - m_{3/2} \beta_{y^{ijk}}$ will be generated. Clearly, the introduction of (even tiny) $a$-terms, soft squark masses and Majorana gaugino masses from AMSB can spoil the phenomenological advantages of $R$-SUSY. Therefore, it is interesting to quantify the degree to which the $R$-symmetry must be exact, \emph{i.e.} the size of the gravitino mass and couplings $\lambda_{ijk}^{''}$ of Eq.~\eqref{eq:WRPV} that still accommodate a phenomenologically viable model. This will be further discussed in Sec.~\ref{sec:sugraconstraints}, where we will see that the most stringent constraints come from the AMBS Majorana gaugino masses that lead to dinucleon decay. 

\newpage

\section{Flavor Phenomenology}
\label{sec:Predictions}
After having outlined a complete model with a field content that realizes the mechanism of \cite{Elor:2018twp}, we may begin to preform calculations within this framework in order to identify the phenomenologically viable parameter space. As we will see, generating the observed baryon asymmetry and the DM relic abundance of the Universe requires a precise range of model parameters and observables, which are summarized in Table~\ref{tab:Parameters}. A key prediction of this scenario is the modification of some of the oscillation parameters of the neutral $B$ meson system with respect to the values expected in the SM. Because of this, sufficiently precise measurements of the $B$ meson system can \emph{directly} probe the baryogenesis dynamics. Any NP inducing such modifications is also expected to have an effect on other $\Delta F = 2$ observables, like oscillations in the $B$, $K$, and $D$ neutral meson systems; and on $\Delta F = 1$ processes such as $\mu \to e \gamma$ and $b \to s \gamma$. Making use of the supersymmetric model, these effects can be quantified within a consistent framework.

It is common lore that SUSY models with an $R$-Symmetry are not too severely constrained by flavor observables (see e.g.~\cite{Kribs:2007ac}). However, due to the presence of the $\mathcal{O}( \text{GeV} )$ bino and the $\lambdapp_{ijk}$ as the only RPV couplings, our mechanism highlights a different slice of parameter space from what has been previously considered. It is therefore important to re-evaluate the constraints and apply them to the relevant parameter space, as well to assess whether the model prediction for the semileptonic-leptonic asymmetry in the $B_s$ and $B_d$ systems lies within the range that can accommodate baryogenesis. 

The RPV couplings $\lambdapp_{ijk} \bU_i^c\bD_j^c\bD_k^c$ are an obvious source of flavor violation, and may also produce CP violation through complex phases. A comprehensive study of the phenomenological implications of RPV couplings was made in~\cite{Barbier:2004ez}. The main particularity of our setup is that the most constraining processes, \emph{i.e.} proton decay and nucleon-antinucleon oscillations, are forbidden as long as the $R-$symmetry remains exact (we discuss the implications of $R-$symmetry breaking in Sec.~\ref{sec:sugraconstraints}). As a consequence, the flavor violating observables described below give the strongest constraints for all the RPV couplings.

Soft masses for SM fermion superpartners can also be sources of flavor and CP violation. A particularity of the unbroken $U(1)_R$ is that left-right squark and slepton mixing is forbidden by the $R$-symmetry~\cite{Hall:1990hq}. The soft scalar mass matrices can be parametrized as usual by dimensionless ratios $\delta^{ij}_{LL}\equiv (m_{\tilde{q}}^2)^{ij} / | m^2_{\tilde{q}} | $ of the off-diagonal elements to the diagonal ones $m^2_{\tilde{q}}$, where $i,j$ are quark or lepton flavor indices (analogously for $\delta^{ij}_{RR}$). As we will see, CP violation can arise if these ratios have complex phases.

\subsection{$\Delta F = 2$ Meson Oscillations}
\label{sec:DF2osc}

In addition to the flavor violation already present in the SM, our setup allows for new processes contributing to neutral meson-antimeson oscillations, where flavor is violated by two units. For baryogenesis, a NP contribution in the $B^0$-$\bar{B^0}$ system is desirable. At the same time, experimental constraints, which are strongest in the $K^0$-$\bar{K^0}$ system, must be satisfied. Studies in general supersymmetric extensions of the SM~\cite{Gabbiani:1996hi,Bagger:1997gg}, in RPV contexts~\cite{Barbier:2004ez,Slavich:2000xm} and in $U(1)_R$ models with Dirac gauginos~\cite{Kribs:2007ac} have been performed. Here we extend and adapt those results to our scenario.

The physics of neutral meson oscillations, as well as the relevant observables, have been described in Section~\ref{sec:model_indep_asl} in a model independent way. We now link such observables to the NP model at hand. Firstly, the modification of the mass splitting between meson and antimeson can be computed as
\begin{equation}\label{eq:delta_M_bino}
\Delta M^{\rm NP} = 2\left( \sum_{i=1}^5 C_iM_i + \sum_{i=1}^3 \tilde{C}_i\tilde{M}_i \right).
\end{equation}
Here, $\ptwiddle{C}_i$ encode the model-specific partonic processes, which are discussed below. $\ptwiddle{M}_i$ correspond to the matrix elements of the resulting effective four-fermion operators $\ptwiddle{Q}_i$, that is, $\ptwiddle{M}_i = \braket{\bar{B}^0 | \ptwiddle{Q}_i | B^0}$ and similarly for $K^0$ and $B^0_s$. The operators are numbered following~\cite{Bagger:1997gg}. The ones relevant for us are $Q_1 = \overline{b}_L^\alpha \gamma_\mu d_L^\alpha\,\overline{b}_L^\beta \gamma_\mu d_L^\beta$, $\tilde{Q}_1 = \overline{b}_R^\alpha \gamma_\mu d_R^\alpha\,\overline{b}_R^\beta \gamma_\mu d_R^\beta$, $Q_4 = \overline{b}_R^\alpha d_L^\alpha\,\overline{b}_L^\beta d_R^\beta$ and $Q_5 = \overline{b}_R^\alpha d_L^\beta\,\overline{b}_L^\beta d_R^\alpha$, where $\alpha$ and $\beta$ are color indices. The corresponding matrix elements are
\begin{align}
M_1 &= \tilde{M}_1 = \frac{1}{3}f_B^2m_B B_1, \\ \nonumber
M_4 &= \left[\frac{1}{24} + \frac{1}{4}\left( \frac{m_B}{m_b+m_d} \right)^2 \right] f_B^2 m_B B_4 \, , \\ \nonumber
M_5 &= \left[\frac{1}{8} + \frac{1}{12}\left( \frac{m_B}{m_b+m_d} \right)^2 \right] f_B^2 m_B B_5 .
\end{align}
In our calculations, we use the values obtained in~\cite{Bazavov:2016nty} for the bag parameters $B_i$.

The decay rate difference $\Delta \Gamma$ is not modified in our setup, as there is no new common state to which both meson and antimeson can decay. Its value can therefore be set to the SM prediction as was done in Sec.~\ref{sec:ModelIndep}. As a consequence, the previous expression Eq.~\eqref{eq:delta_M_bino} also allows to calculate CP-violating observables, which depend on the complex phase of $\ptwiddle{C}_i$ with respect to the SM counterpart, as made explicit in Eq.~\eqref{eq:alpha_phi_alpha}. We are particularly interested in contributions to the semileptonic asymmetries in the $B^0_{s,d}$ systems, as these are the key observables that enter the baryogenesis mechanism.

At the partonic level, the meson oscillations arise from box diagrams. In the SM they are mediated by $W$ bosons, but in our supersymmetric scenario new contributions arise, which can be classified into two groups depending on the origin of the flavor violation.

\subsubsection*{Contributions from RPV couplings} 
The corresponding Feynman diagrams are shown in Fig.~\ref{fig:box_diagrams_RPV} in Appendix~\ref{sec:box_diagrams}. The first two contain four RPV vertices with $\lambdapp$ couplings, while the lower two diagrams involve mixed contributions with two RPV vertices and two gauge/Higgs vertices with the corresponding CKM factors. These diagrams were calculated in~\cite{Slavich:2000xm} for the (s)top couplings $\lambdapp_{3jk}$, here we generalize the computation to include also contributions from first two generation (s)quarks. The explicit analytic expressions are given in~\eqref{eq:box_diagrams_RPV}. Flavor changing squark mass mixing is neglected for this computation.
\renewcommand{\arraystretch}{1.3}
\begin{table*}[t]
\centering
\begin{subtable}{\textwidth}
\begin{center}
  \setlength{\arrayrulewidth}{.25mm}
\scalebox{0.93}{
\begin{tabular}{c | c c c}
  $K^0$  & $\lambdapps_{113}$ & $\lambdapps_{213}$ & $\lambdapps_{313}$ \\ \hline	
  $\lambdapp_{123}$ & $3.7\times 10^{-3}$ / $3.0\times 10^{-4}$ & -- / $9.0\times 10^{-2}$ & $2.5$ / $1.6\times 10^{-2}$ \\
  $\lambdapp_{223}$ & -- / $2.5$ & $3.7\times 10^{-3}$ / $3.0\times 10^{-4}$ & $0.25$ / $1.6\times 10^{-3}$ \\
  $\lambdapp_{323}$ & -- / $0.29$ & $0.22$ / $1.4\times 10^{-3}$ & $2.8\times 10^{-3}$ / $4.8\times 10^{-5}$
\end{tabular} }
\end{center}
\end{subtable}
\newline
\vspace*{0.5 cm}
\newline
\begin{subtable}{\textwidth}
\begin{center}
  \setlength{\arrayrulewidth}{.25mm}
\scalebox{0.93}{
\begin{tabular}{c | c c c}
  $B^0_d$  & $\lambdapps_{112}$ & $\lambdapps_{212}$ & $\lambdapps_{312}$ \\ \hline	
  $\lambdapp_{123}$ & $3.8\times 10^{-3}$ / $2.0\times 10^{-3}$ & -- / -- & $4.0$ / $1.1$ \\
  $\lambdapp_{223}$ & -- / -- & $3.8\times 10^{-3}$ / $2.0\times 10^{-3}$ & $0.40$ / $0.11$ \\
  $\lambdapp_{323}$ & -- / -- & -- / -- & $3.3\times 10^{-3}$ / $1.5\times 10^{-3}$
\end{tabular} }
\end{center}
\end{subtable}
\newline
\vspace*{0.5 cm}
\newline
\begin{subtable}{\textwidth}
\begin{center}
  \setlength{\arrayrulewidth}{.25mm}
\scalebox{0.93}{
\begin{tabular}{c | c c c}
  $B^0_s$  & $\lambdapps_{112}$ & $\lambdapps_{212}$ & $\lambdapps_{312}$ \\ \hline	
  $\lambdapp_{113}$ & $1.8\times10^{-2}$ / $1.0\times 10^{-2}$ & -- / -- & -- / -- \\
  $\lambdapp_{213}$ & -- / -- & $1.8\times 10^{-2}$ / $1.0\times10^{-2}$ & $2.3$ / $0.73$ \\
  $\lambdapp_{313}$ & -- / -- & -- / -- & $1.6\times 10^{-2}$ / $8.1\times 10^{-3}$
\end{tabular} }
\end{center}
\end{subtable}
\caption{\footnotesize{Constraints on products of RPV couplings from neutral meson oscillations. Each of the $K^0$, $B^0_d$ and $B^0_s$ systems constrain different combinations of couplings. For each product, the first limit assumes that the NP contribution is aligned with the SM one, while the second one assumes maximal CP-violating phase. A line indicates that the derived limit is above the unitarity bound on the couplings. The limits are computed assuming a universal squark mass $m_{\tilde{q}}=1\,\mathrm{TeV}$, and scale with $m_{\tilde{q}}$ (diagonal ones) or $m_{\tilde{q}}^2$ (off-diagonal ones). }}
\label{tab:DeltaF2_RPV}
\end{table*}

As can be seen in Eq.~\eqref{eq:box_diagrams_RPV}, one-loop diagrams involving RPV interactions generate the operators $\tilde{Q}_1$, $Q_4$ and $Q_5$. The former is associated with the box diagram involving four $\lambdapp$ couplings and therefore scales as $\tilde{C}_1\sim {\lambdapp}^4 / m_{\tilde{q}}^2$, where ${\lambdapp}^4$ symbolizes the appropriate product of four potentially different $\lambda^{\prime\prime (\star)}_{ijk}$ couplings (see Eq.~\eqref{eq:box_diagrams_RPV} for the explicit combinations that arise). For simplicity, we assume universal squark masses and compute the loop integrals in the limit where $m_{\tilde{q}}$ is much larger than any other mass\footnote{We do not take into account renormalization group running effects for the operators $\ptwiddle{Q}_i$ from the squark scale down to the hadronic scale. We estimate this approximation to be good within a factor of $\sim 2$ for the coefficients $\ptwiddle{C}_i$ and therefore do not expect a large effects on the limits presented in Table~\ref{tab:DeltaF2_RPV} and Fig.~\ref{fig:DeltaM_constraints}.}. The other two operators are associated with diagrams with only two RPV couplings, while the other two vertices correspond to weak interactions where flavor violation is due to CKM insertions. Therefore, they both scale as $C_{4,5}\sim {\lambdapp}^2 V^2 / m_{\tilde{q}}^2$, where $V^2$ represents the appropriate CKM factors (see Eq.~\eqref{eq:box_diagrams_RPV} for explicit expressions). These diagrams also carry a factor of $m_{u_i}m_{u_j} / m_W^2$, which introduces an extra suppression unless only top quarks run in the loop (\emph{i.e.} $i=j=3$). Note that the RPV couplings generically source CP violation in the neutral meson mixing because the $\lambdapp$ couplings can be complex.

Using the experimental measurements of the neutral meson oscillation parameters, $(\Delta M,\phi^{d,s})$ for $B^0_{d,s}$ and $(\Delta M,\epsilon_K)$ for $K^0$, together with the SM predictions for such observables\footnote{The SM predictions in the $K^0$ sector have large uncertainties and we therefore assume that the SM values saturate the experimental bounds to obtain conservative limits.}, we can obtain constraints on the products of RPV couplings. The resulting limits are compiled in Table~\ref{tab:DeltaF2_RPV}. For all of them, we assume a value of $m_{\tilde{q}}=1\,\mathrm{TeV}$. The limits on the diagonal of each table are dominated by the $\tilde{C}_1$ contribution with four $\lambdapp$ factors, which means that they scale with $m_{\tilde{q}}$. However, the off-diagonal entries in the tables, which rely on the $C_{4,5}$ contributions, scale with $m_{\tilde{q}}^2$ and are weaker due to the CKM and/or quark mass suppressions explained in the previous paragraph. The exception are the bottom right entries of the tables, which correspond to diagrams mediated by top quarks for which the contributions from $\tilde{C}_1$ and $C_{4,5}$ are of similar size for $m_{\tilde{q}}=1\,\mathrm{TeV}$. Each of the entries in the table quotes two limits: the left one corresponds to the case where the phase of the NP contribution to $M_{12}$ is aligned with the SM one; the limit in that case comes from measurements of $\Delta M$. The right one assumes that the NP phase is orthogonal to the SM one and therefore enhances CP violation, the limits in that case come from $\phi^{d,s}$ and $\epsilon_K$. For the general case where the phases are neither aligned nor orthogonal, the limit lies between the two extremal values quoted in Table~\ref{tab:DeltaF2_RPV}.

In the general case where the RPV couplings are complex and source CP violation, contributions to the semileptonic asymmetries in the $B^0_d$ and $B^0_s$ are expected. Indeed, as Fig.~\ref{fig:semileptonic_asymmetry_constraints} shows and was discussed in Sec.~\ref{sec:model_indep_asl}, NP effects can significantly enhance $a_{\mathrm{sl}}^{d,s}$ with respect to their SM value while satisfying all other constraints. For instance, the benchmark values used for the baryogenesis calculations are obtained if \emph{any} one product of complex RPV couplings saturates the corresponding limit in Table~\ref{tab:DeltaF2_RPV}.

\subsubsection*{Contributions from squark mass mixings} 
Off-diagonal terms in the squark mass matrix can also induce neutral meson oscillations. The bino-mediated diagrams are shown in Fig.~\ref{fig:box_diagrams} (App.~\ref{sec:box_diagrams}) for the $B^0_d$ system, the cases of $B^0_s$ and $K^0$ are analogous. For Dirac gauginos, the number of contributing diagrams is smaller than in the MSSM, as the Dirac partners do not couple to quarks and thus chirality flips in the internal gaugino propagators are not possible\footnote{For heavy gauginos that can be integrated out, effective dimension five operators like $\tilde{d}^\star_R\tilde{s}^\star_L\,d^\dagger_R s_L$ are absent, and the transition happens through dimension six operators like $\tilde{d}_L\partial_\mu \tilde{s}^\star_L\,d^\dagger_L\gamma^\mu s_L$. For a light bino this does not result in extra suppression, but the chirality structure of the diagrams is in any case different than for Majorana gauginos.}. Furthermore, the $U(1)_R$ symmetry prohibits the presence of $a$-terms, \emph{i.e.} no left-right squark mass mixing is allowed.

\begin{figure*}[t]
\centering
\begin{subfigure}[b]{\textwidth}
\label{fig:modelIndep}
\includegraphics[width=\linewidth]{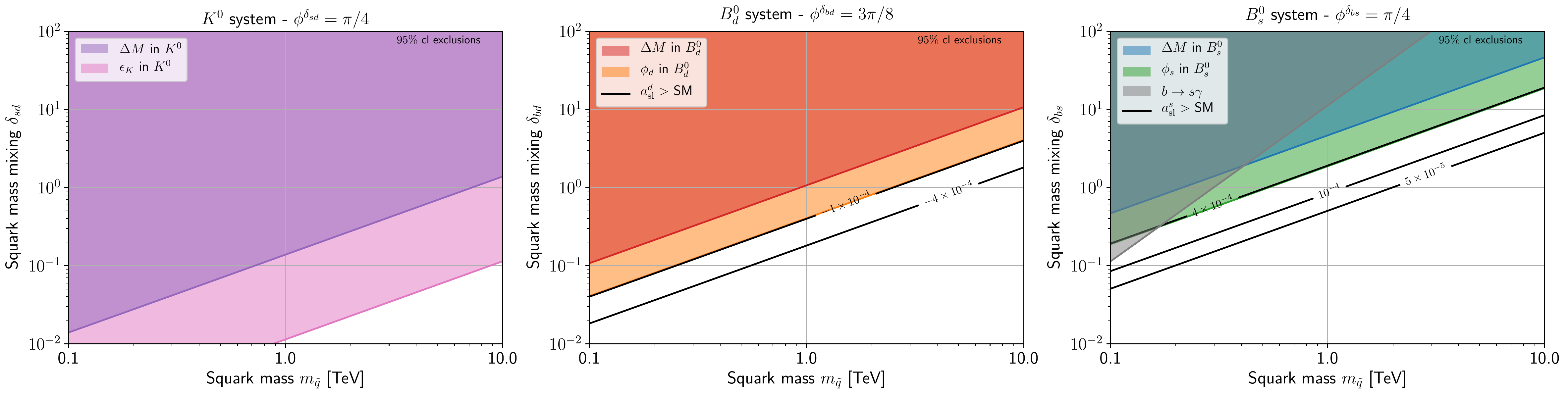}
\end{subfigure}
\begin{subfigure}[b]{\textwidth}
\includegraphics[width=\linewidth]{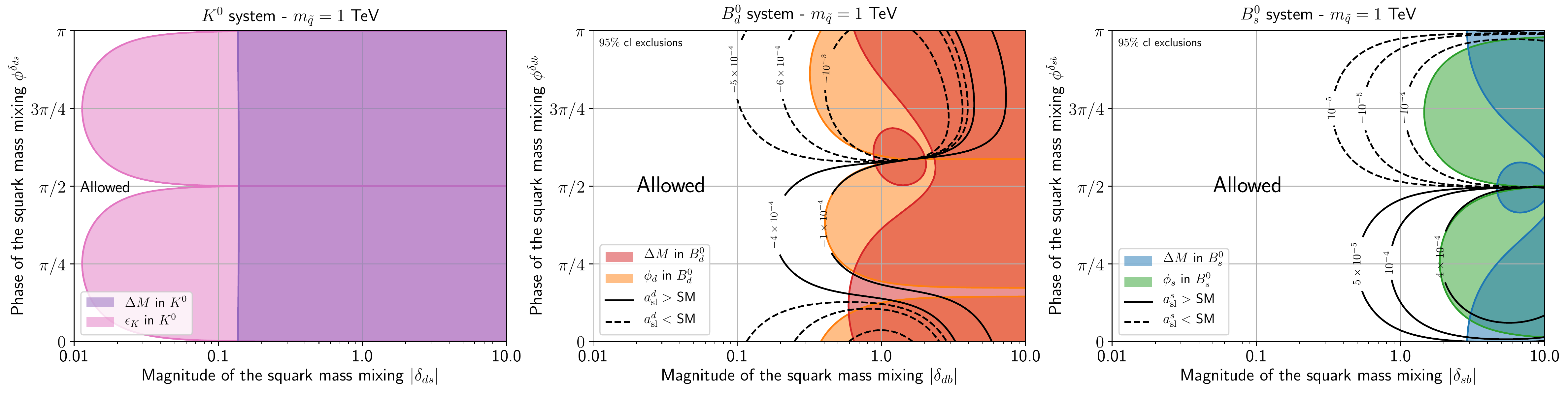}
\end{subfigure}
\caption{\footnotesize{Constraints on the soft squark masses coming from the bino contribution to mixing in neutral meson $K$, $B^0_d$ and $B^0_s$ systems. The shaded areas are excluded by the CP conserving and violating observables indicated in the labels, which are explained in Sec.~\ref{sec:ModelIndep}. For simplicity, we assume an approximately universal value $m_{\squark}$ for the diagonal elements of the mass matrix. The dependence on the bino mass drops out as long as $m_{\bino}\ll m_{\squark}$. The upper panels show constraints assuming maximal CP violation, while the lower ones show the dependence on the phases of the squark mass matrix elements. The contour lines in the $B^0_d$ and $B^0_s$ cases represent values of the semileptonic asymmetry in each system. As expected, the kaon sector measurements gives the strongest limits, showcasing the need for some flavor structure to produce sizeable contributions to the semileptonic asymmetries.}}
\label{fig:DeltaM_constraints}
\end{figure*}

The authors of ~\cite{Kribs:2007ac} considered the $\Delta F = 2$ flavor constraints  arising from box diagrams with gluinos, which due to the stronger coupling give the largest contribution if the masses of all gauginos are assumed to be similar. This is not necessarily the case in the regime where $m_\psi\sim 1\,$GeV is much lighter than gluinos and winos. The analytic expressions for the contribution of a light bino to the oscillation parameters defined in Sec.~\ref{sec:ModelIndep} are given in App.~\ref{sec:box_diagrams}. Parametrically, the contribution of a light GeV bino to $\Delta M$ is of the same order as that of a gluino which is about $10$ times heavier than the squarks. 

CP violation arises when $\Delta M^{\rm NP}$ is a complex quantity. This happens if the off-diagonal elements of the squark mixing matrix are complex, and thus so are $\delta_{LL}$ and $\delta_{RR}$. The CP-violating observables defined in Sec.~\ref{sec:ModelIndep} can thus receive NP contributions. With this, the experimental constraints detailed in Sec.~\ref{sec:ModelIndep}, together with the analogous ones for the kaon system ($\Delta M_K$ and $\epsilon_K$, see \cite{Artuso:2015swg}) can be used to constraint squark mixing. Figure~\ref{fig:DeltaM_constraints} shows the excluded ranges of values for the diagonal and (generally complex) off-diagonal elements $m_{\squark}$ and $\delta$. The non-trivial dependence of the constraints on the phase of $\delta$ comes from interference with the SM contributions, \emph{i.e.} the alignment between the NP effects and their SM counterparts. 

The values of the semileptonic asymmetries as a function of the model parameters are shown as contour lines in Fig.~\ref{fig:DeltaM_constraints}. For squark masses in the TeV scale and $\mathcal{O}(1)$ absolute values of the mixings $\delta_{sb}$ and $\delta_{bd}$, a broad interval of phases lead to semileptonic asymmetries compatible with baryogenesis. At the same time, the stringent bounds in the kaon sector restrict $\delta_{ds}$ to be $\lesssim 10^{-2}$ unless its phase is close to zero (or $\pi/2$). This is compatible with baryogenesis, as $\delta_{ds}$ does not play a role in $B$ meson oscillations, but points towards some flavor structure that particularly suppresses any flavor violation which does not involve the third generation; see for instance~ \cite{Barbieri:1995uv,Kuzmin:1996sy} for constructions where such flavor structures arise.  

\subsection{$\Delta F = 1$ Observables}
\label{sec:deltaF1_observables}
Processes in which flavor is violated by one unit can also probe model parameters relevant for baryogenesis. We limit the discussion to two observables, $b \rightarrow s \gamma$ and $\mu \rightarrow e \gamma$, as they offer the best experimental prospects. Although the latter is the best tested experimentally, the former is directly related to parameters relevant for the baryogenesis mechanism. For this study we mainly adapt the results of~\cite{Kribs:2007ac} to our setup.

\subsubsection*{Contributions from RPV couplings} 
The diagrams contributing to the $b\rightarrow s\gamma$ decay are shown in Fig.~\ref{fig:DeltaF1_diagram_RPV} in Appendix~\ref{sec:box_diagrams}, along with the full analytic expression for the decay rate, given in Eq.~\eqref{eq:Br_b_sgamma_RPV}. Assuming a universal squark mass $m_{\tilde{q}}$, we obtain
\begin{equation}
\text{Br} (b\rightarrow s\gamma) \simeq 4.2\cdot 10^{-7} \left( \frac{\rm TeV}{m_{\tilde{q}}} \right)^4 \left| \sum_{i=1}^3 \lambdapps_{i12}\lambdapp_{i13} \right|^2.
\end{equation}
Comparing with the experimental measurement~\cite{Amhis:2012bh} $\text{Br}  ({\bar{B}\rightarrow X_s\gamma} )= \left( 3.55 \pm 0.24 \pm 0.09 \right)\times 10^{-4}$ and the SM prediction~\cite{Misiak:2010tk} $\text{Br}  ({\bar{B}\rightarrow X_s\gamma}) = \left( 3.15 \pm 0.23 \right)\times 10^{-4}$, we see that this process does not place any meaningful constraint on the RPV couplings for squark masses around or above the TeV scale.

\subsubsection*{Contributions from squark and slepton mass mixings} 
Flavor violating decays can also be triggered by off-diagonal elements in the mass matrix of squarks (for $b\rightarrow s\gamma$) or sleptons (for $\mu\rightarrow e\gamma$). Given the particularities of the $R$-symmetric model, \emph{i.e.} the absence of $L-R$ sfermion mixing and the inert nature of the Dirac partner of the gauginos, the bino can only mediate two kind of diagrams contributing to this process. They are depicted in Fig.~\ref{fig:DeltaF1_diagram_bino} for $b\rightarrow s\gamma$, the ones for $\mu\rightarrow e\gamma$ are analogous. The first diagram requires an external chirality flip while the second one involves mixing between the appropriate Dirac higgsino and the bino, as discussed in Sec.~\ref{sec:neutralino_mixing}.

The flavor-violating decay of the $b$ quark is most important because it directly probes the couplings relevant for baryogenesis. Evaluating Eq.~\eqref{eq:b_sgamma_bino} in the small $\tan\beta$ limit, we obtain a branching ratio
\begin{equation}
\text{Br} (b\rightarrow s\gamma) = \Gamma(b\rightarrow s\gamma) \tau_{B} \,\,  \simeq \,\, 8.4\times 10^{-8} \left( \frac{\rm TeV}{m_{\tilde{q}}} \right)^4 \left( \left(\frac{7}{8}\right)^2 \delta_{LL}^2 + \delta_{RR}^2 \right).
\end{equation}
This should be compared with the experimental measurement~\cite{Amhis:2012bh} $\text{Br}  ({\bar{B}\rightarrow X_s\gamma} )= \left( 3.55 \pm 0.24 \pm 0.09 \right)\times 10^{-4}$, which is compatible with the SM prediction~\cite{Misiak:2010tk} $\text{Br}  ({\bar{B}\rightarrow X_s\gamma}) = \left( 3.15 \pm 0.23 \right)\times 10^{-4}$. The corresponding exclusion on the $b-s$ squark mixing is shown in the top right panel of Fig.~\ref{fig:DeltaM_constraints}.

The computation of the flavor-violating muon decay is analogous to the $b\rightarrow s\gamma$ one with the obvious substitutions, see Eq.~\eqref{eq:mu_egamma_bino} for the corresponding expression. Evaluating it again in the small $\tan\beta$ limit, we find
\begin{equation}
\text{Br} (\mu\rightarrow e\gamma)  \simeq 3.4\times10^{-8} \left( \frac{\rm TeV}{m_{\tilde{\ell}}} \right)^4 \left( \left(\frac{7}{8}\right)^2 \delta_{LL}^2 + \delta_{RR}^2 \right).
\end{equation}
This process is extremely suppressed in the SM, so this contribution is constrained by the stringent experimental upper limit $\text{Br}  ({\mu^+\rightarrow e^+\gamma} )< 4.2\times 10^{-13}$ ($90\%$ cl) set by the MEG experiment~\cite{TheMEG:2016wtm}. For sleptons at the TeV scale, the mass mixing in the $e-\mu$ sector is bound to be $\delta_{e\mu}< 4\times 10^{-3}$, which again indicates that any flavor violation that only involves the first two generations should be suppressed.

\section{Theoretical and Phenomenological Considerations}
\label{sec:BandDMconst}
We now discuss constraints on the model parameters that arise from the requirement of achieving the measured matter-antimatter asymmetry and DM abundance. The model independent results of Sec.~\ref{sec:ModelIndep} are applied to the $R$-SUSY model of Sec.~\ref{sec:Model}, as is the analysis of flavor observables performed in Sec.~\ref{sec:Predictions}. With this, we find that successful baryogenesis requires squark masses in the scale of $1-4$ TeV and at least one of the $\lambdapp_{ij3}$ RPV couplings to be $\mathcal{O}(1)$. Both these requirements stem from the necessity for a large branching fraction of (visible) baryon number violating decays of neutral $B$ mesons.
The observed DM abundance is easily accounted for with order one couplings of $\psi$ to the sterile neutrino multiplet. 
We further discuss constraints that apply to parameters that do not directly control baryogenesis. Such effects are rather a byproduct of realizing the mechanism in the $R$-SUSY model. At the phenomenological level, we discuss bounds coming from neutrino physics and astrophysics. On the more theoretical side, supergravity considerations on the size of $R$-symmetry breaking constrain the gravitino mass and some RPV couplings.

\subsection{Baryogenesis}
\label{subsec:BranchingFrac}
As was highlighted in Sec.~\ref{sec:model_indep_br}, successful baryogenesis requires a relatively large branching ratio of $B$ mesons into final states containing a single baryon and dark sector particles. In the supersymmetric setup, such decays can be mediated by any of the four-fermion operators listed in Table~\ref{table:flavor_structure}. Assuming that the decay is dominated by a single one of them yields a branching ratio
\begin{equation}
\label{eq:Brfraction}
\Br \sim 10^{-3}\left (\frac{\Delta m}{ 2\,\mathrm{GeV}}\right)^4 \left((1.9\,\mathrm{TeV})^2  \lambda^{\prime\prime}_{ij3} \sqrt{2}\, g^\prime\, \left( \frac{Q_u}{m_{\tilde{u_i}}^2} + \frac{Q_d}{m_{\tilde{d_j}}^2} - \frac{Q_d}{m_{\tilde{b}}^2} \right)  \right)^2,
\end{equation}
where $\Delta m = m_{B^0} - m_{\mathcal{B}} - m_X$ is the mass splitting between $B$ meson and final states, and $X$ represents the dark particles together with any other mesons. Note that Eq.~\eqref{eq:Brfraction} is computed using a spectator approximation \cite{Aitken:2017wie}, which breaks down when the mass splitting $\Delta m \lesssim 2 \, \text{GeV}$. In general, given the hadron spectrum of interest here, we expect $2 \, \text{GeV} \lesssim \Delta m \lesssim 3 \, \text{GeV}$ for the operators of interest. For final states involving charm quarks, $\Delta m$ may be too small to reliably use the spectator quark approximation and an analytic estimation of the branching ratio is not feasible.

For hypercharge coupling $g^\prime (m_b) \sim  0.34$ and assuming degenerate squark masses, achieving a large enough branching fraction for baryogenesis
\bea
\label{eq:Brdeg}
\Br \gtrsim 10^{-3} \quad \text{requires} \quad m_{\text{squark}} \lesssim 1 \,  \text{TeV} \sqrt{ \lambda^{\prime\prime}_{ij3} } \frac{\Delta m}{2 \text{GeV}}  \,,
\eea
where we have used the maximum positive value of $(a^s_{\mathrm{sl}})_{\rm max} = 4.1\times 10^{-4}$ and the minimum negative of $(a^d_{\mathrm{sl}})_{\rm min} =- 9.0\times 10^{-5} $ to maximize the asymmetry. It is also important to take into account the lower bound on squark masses of $\sim1$~TeV, which is discussed in more detail in Sect.~\ref{sec:LHC_colored_scalars}. 
Because of this, it is clear that attaining a large enough branching fraction to accommodate the observed baryon asymmetry favors at least one of the RPV couplings being large, together with (at least one) squark mass in the TeV scale\footnote{\footnotesize{The parameter space for squark masses can be opened up a little by increasing $ \lambda^{\prime\prime}_{ij3}$ up to where perturbativity begins to break down (for instance \cite{Barbier:2004ez} computes the renormalization group running in $R$-SUSY models and obtains $\lambda^{''}_{123}<1.25$).}}, that is
\begin{align}
\label{eq:BaryoConst}
\text{Max} \left[ \lambda^{''}_{ij3} \right] \sim \mathcal{O}(1) \,, \quad \text{for} \quad i,\, j = 1,2 \quad \text{and} \quad m_{\tilde{q}_R} = 1 - 4 \text{TeV} \,.
\end{align}
Here the range in the allowed squark mass parameter space takes into account enhancement to the coupling from the renormalization group running of the anomaly dimension of the 4 fermi operator\footnote{\footnotesize{A detailed RGE study is beyond the scope of the current work.}}.

\subsection{Dark Matter Relic Abundance}
\label{sec:dark_matter}
The mechanism of \cite{Elor:2018twp} tends to generically overproduce a symmetric DM component, therefore requiring additional dark sector interactions to deplete its abundance. This is the case when the DM is constituted by the lightest dark sector particle. However, if the dark sector states are identified with a right-handed neutrino multiplet, the sneutrino is stable on cosmological time scales even if it is slightly heavier than the right-handed neutrino. This is due to the fact that the $R$-symmetry is exact and baryon number remains conserved. The sterile sneutrino carries baryon number $-1$ while the right-handed neutrino has zero baryon number, so as long as the mass difference $m_{\nu_R}-m_{\tilde{\nu}_R}$ is smaller than the neutron mass, sneutrinos cannot kinematically decay to right-handed neutrinos and remain cosmologically stable.

In the situation where $m_{\tilde{\nu}_R} > m_{\nu_R}$, a sneutrino-antisneutrino pair can annihilate into right-handed neutrinos. After that, the $\nu_R$ decays as usual into SM leptons, thus contributing to the depletion of the symmetric component of DM.
The annihilation cross section for $\tilde{\nu}_R \tilde{\nu}^\star_R \rightarrow \nu_R \nu_R$ is dominated by the exchange of light binos. The thermally averaged cross section is given by
\begin{equation}
\langle \sigma v \rangle = \frac{\lambda_N ^4 m_{\nu_R}^2\sqrt{m_{\tilde{\nu}_R}^2-m_{\nu_R}^2}}{32\pi m_{\tilde{\nu_R}} (m_{\tilde{\nu_R}}^2-m_{\nu_R}^2+m_\psi^2)^2}\,.
\end{equation}
With this, we can easily leverage the results of \cite{Elor:2018twp} to find that the observed DM relic abundance is reproduced for values of the superpotential coupling $\lambda_{N}$ of the order
\begin{align}
\label{eq:constDMrelicabundance}
\lambda_{N} \sim \mathcal{O}(10^{-1}) \quad \Rightarrow \quad \Omega_{\rm DM} h^2 = 0.12 \,.
\end{align}
In what follows we take this as the benchmark value for $\lambda_N$. It is worth emphasizing once again that this requires the sneutrino to be heavier than right-handed neutrino. Otherwise, the kinematic suppression of the annihilation rate by a factor of $e^{-\Delta m/T}$, where $\Delta m=m_{\nu_R}-m_{\tilde{\nu_R}}$ makes this annihilation channel inefficient. Therefore, if right-handed neutrinos are heavier than sneutrinos, additional dark sector interactions are generally needed to reproduce the correct relic abundance.

\subsection{Generation of Neutrino Masses}
\label{subsec:oscSMnu}
As was discussed in Sec.~\ref{sec:sterile_sneutrino_DM}, the SM neutrinos $\nu$ acquire a tiny mass through their Dirac mass mixing with the Majorana right-handed neutrinos $\nu_R$. This is nothing but the well-known type-I seesaw mechanism~\cite{Mohapatra:1979ia,GellMann:1980vs,Minkowski:1977sc,Schechter:1980gr}. As usual, diagonalization of the neutrino mass matrix with terms from Eq.~\eqref{eq:DiracNuMass} and Eq.~\eqref{eq:Nmassandint} leads to a large eigenvalue $m_{\nu_R}\sim M_M$ and a small one $m_\nu =  4 \, y_N^2 \, v^2 \,  \sin^2 \beta / m_{\nu_R}$, as long as the Yukawa $y_N \sin{\beta}$ is sufficiently small. Successful baryogenesis requires the existence of at least one right-handed neutrino with a mass $m_{\nu _R} \sim \mathrm{GeV}$. At the same time, current experimental data constrain the mass of any of the SM neutrinos to be $m_\nu \lesssim 0.12\, \mathrm{eV}$ \cite{Drewes:2013gca,Anelli:2015pba,Abazajian:2012ys}. Fulfilling these two conditions bounds $\nu_R$'s contribution to the seesaw, constraining the sterile neutrino Yukawa coupling to be\footnote{\footnotesize{Baryogenesis only requires the existence of one GeV-scale right-handed neutrino, but additional sterile neutrinos may be present and play a role in the generation of SM neutrino masses. We are therefore unable to make more precise statements regarding the spectrum of neutrino masses and Yukawa couplings. 
}}
\bea
\label{eq:NuOscConst}
 y_N \, \sin \beta  \lesssim 2\cdot 10^{-8} \sqrt{ \frac{m_{\nu_R}}{1 \, \mathrm{GeV}} } \,.
\eea
This translates into an upper limit for the mixing between the active and the sterile neutrino, which is given as usual by the ratio of the Dirac to Majorana mass terms, corresponding to $\left| U \right| \equiv y_N \, v \, \sin \beta / m_{\nu_R}$ in terms of our model parameters. We find
\begin{equation}
    \left|U\right|^2 \lesssim 10^{-10} \frac{m_{\nu_R}}{1 \, \mathrm{GeV}}\,.
\end{equation}

The usual phenomenological constraints for type-I seesaw models with GeV-scale right-handed neutrinos apply to our scenario. The strongest limits come from a combination of beam dump, LEP and $B$ factory searches (see~\cite{Alekhin:2015byh} for a recent review). Meaningful bounds can also be obtained from displaced vertices searches at LHC~\cite{Boiarska:2019jcw}. In the most simplistic scenario which we discuss here, the mixing between sterile and active neutrino flavors is expected to be small, $\left| U \right|^2\sim 10^{-11}$, which is beyond the reach of current experimental setups. 

As a particular feature, our model predicts that a relatively large number of sterile neutrinos are produced in $B$ meson decays. As was discussed in Sec.~\ref{subsec:BranchingFrac}, a new decay channel for $B$ mesons with branching ratio of $\gtrsim 10^{-3}$ is required for baryogenesis. The bino produced in this process decays predominantly to a right-handed neutrino and sneutrino pair, constituting a large new production cross section for sterile neutrinos. Experiments like LHCb and other $B$ factories and SHiP~\cite{Anelli:2015pba}, where $\sim 7\cdot 10^{13}$ $B$ mesons are expected to be produced, are best posed to take advantage of this production channel, potentially allowing to place meaningful constraints in the parameters of the model. We leave a more detailed investigation of this interesting possibility for future work.

\subsection{Dark Matter Stability}
\label{subsec:DMstability}
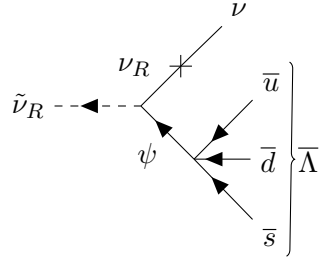
\begin{figure}[t]
\centering
\begin{tikzpicture} 
\begin{feynman}
\vertex (a) {\(\tilde{\nu}_{R}\)}; 
\vertex [right=of a] (b); 
\vertex [above right=0.75cm of b] (d);
\vertex [above right=0.75cm of d] (f1) {\( \nu \)};
\vertex [below right=1cm of b] (c);
\vertex [above right=1.1cm of c] (f2) {\(\overline u\)}; 
\vertex [below right=1.1cm of c] (f3) {\(\overline s\)};
\vertex [right=0.75cm of c] (f4) {\(\overline d\)};
\diagram* {
(a) -- [anti charged scalar] (b) -- [edge label=\(\nu_R\), insertion=1.] (d) -- (f1), 
(b) -- [anti fermion, edge label'=\(\psi\)] (c), 
(c) -- [anti fermion] (f2),
(c) -- [anti fermion] (f3),
(c) -- [anti fermion] (f4),
}; 
\draw [decoration={brace}, decorate] (f2.north east) -- (f3.south east) node [pos=0.5, right] {\(\overline{\Lambda}\)};
\end{feynman}
\end{tikzpicture}
\caption{\footnotesize{Decay of the right-handed sneutrino DM particle into three light (anti)quarks (which hadronize into i.e. a $\bar{\Lambda}$ baryon) and a neutrino. The process is mediated by an off-shell bino and right-handed neutrino, as well as the RPV coupling $\lambdapp_{112}$. Propagator arrows indicate the flow of baryon number. }}
\label{fig:DM_decay}
\end{figure}
The stability of the right-handed sneutrino DM is a consequence of kinematic relations and the action of a stabilizing symmetry which is a combination of baryon and lepton number. 
This stability is however not absolute, as decays into three light (anti)quarks and a SM neutrino are possible via a higher dimensional operator. The diagram for the process is shown in Fig.~\ref{fig:DM_decay}. The first vertex corresponds to the coupling of the Dirac bino to the dark sector sterile neutrino multiplet through Eq.~\eqref{eq:WLambdaN}. The second is an effective four-fermion operator analogous to the ones shown in Table~\ref{table:flavor_structure}, but with a different flavor combination, namely
\begin{align}
\mathcal{O} = \sqrt{2} g^\prime \lambdapp_{112}  \left( \frac{Q_u}{m_{\tilde{u}}^2} + \frac{Q_d}{m_{\tilde{d}}^2} - \frac{Q_d}{m_{\tilde{s}}^2} \right) \psi \, u \, d \, s \, .
\label{eq:lightquarkcoupling}
\end{align}
The decay is thus induced by the $\lambdapp_{112}$ RPV coupling, which is not directly relevant for baryogenesis. The full process $\tilde{\nu}_R \rightarrow \psi^{*} + \nu_R^{(*)} \rightarrow \bar{u} + \bar{d} +  \bar{s}  + \nu_{\rm SM}$ is mediated by an off-shell bino and is further suppressed by the small mixing between the active and the sterile neutrino, so the decay rate is expected to be very small. Let us nevertheless estimate it to obtain a constraint on the RPV coupling to light quarks from the requirement of DM stability.

The hadronized final state must contain a baryon and a strange quark, so the lowest mass final state in the decay contains a $\Lambda$ baryon, which has a mass $m_\Lambda \simeq 1.1$ GeV. As the DM mass is constrained to be in the range $1.2-2.7$~GeV, the maximum momentum for any final state particle is around $1.1$~GeV.  This is too small for the simple approximation of considering 3 free light quarks as decay products to be good. However, we can leverage the fact that the available energy is low enough so that final states with fewer particles, perhaps only one hadron, should dominate the width. We therefore compute the rate for the decay  $\tilde{\nu}_R\rightarrow \overline{\Lambda} +\nu$. A rough estimate can be obtained by using dimensional analysis to match $uds \leftrightarrow 4\pi f_\pi^3\,\Lambda$. The lifetime for the two-body decay is then
\begin{align}
\label{eq:DMdecay}
\tau_{\tilde{\nu}_R \rightarrow \bar{\Lambda} \, \nu} \,\,\, \sim \,\,\, 4.3 \times 10^{17} \text{sec} \,\, &\frac{10^{-11}}{\left| U \right|^2} \left( \frac{4.2 \times 10^{-5}}{ \lambda^{''}_{112}}\right)^2  \left(\frac{0.1}{\lambda_N}\right)^2 \, \frac{\left(\frac{m_{\psi} }{3 \, \text{GeV}} \right)^2 \left(\frac{m_{\tilde{\nu}_R}}{2\, \text{GeV}} \right)  \left(\frac{3 \, \text{GeV}^2}{\Delta m^2} \right)   }{  \left(\text{TeV}^2 \left( \frac{Q_u}{m_{\tilde{u}}^2} + \frac{Q_d}{m_{\tilde{d}}^2} - \frac{Q_d}{m_{\tilde{s}}^2}\right) \right)^{2} }\, \, ,
\end{align}
where $\Delta m^2 = m_{\tilde{\nu}_R}^2 - m_\Lambda^2$ and the rest of the notation is as described in previous sections. A most conservative bound can be obtained by requiring that this lifetime is larger than the age of the Universe, $4.3 \times 10^{17} \, \text{sec}$. Using the benchmark parameter values indicated in Eq.~\eqref{eq:DMdecay}, we obtain a constraint on the RPV coupling,
\begin{equation}\label{eq:lambda112_constraint}
	 \lambdapp_{112}\lesssim 10^{-5} \,.
\end{equation}
The operator of Eq.~\eqref{eq:lightquarkcoupling} also leads to a (visible) baryon number non-conserving decay of the sterile neutrino, $\nu_R\rightarrow \tilde{\nu}_R\, \bar{u}\,\bar{d}\,\bar{s}$. As one sterile neutrino is produced for each DM antibaryon during baryogenesis, such a process could washout the generated baryon asymmetry if its rate is too large. However, the width of this decay channel is set by the same RPV coupling which is bound to be small by DM stability in Eq. \Eq{eq:DMdecay}). Because of this, its rate is negligible compared to the usual baryon number preserving SM right-handed neutrino decays.

Even if the bound Eq.~\eqref{eq:lambda112_constraint} is satisfied and the DM is cosmologically stable, the sneutrino may have a non-negligible decay width into antiquarks and a neutrino. As discussed above, the most likely decay mode is expected to be into $\overline{\Lambda}+\nu$. If such process occurs inside a large volume detector such as as Super-Kamiokande \cite{Fukuda:2002uc}, the pions and the annihilation of the $\bar{p}$ and $\bar{n}$ from the $\overline{\Lambda} $ decay could produce a detectable signal. Furthermore, observations of the antiproton flux in cosmic rays can be used to the constrain the rate of DM annihilation or decay into antimatter~\cite{Bergstrom:1999jc,Garny:2012vt,Ibarra:2013cra,Fornengo:2013xda,Cirelli:2013hv}. However, the antiprotons produced in our setup have very low energies, which makes modelling their propagation in the galactic medium a difficult task. On top of that, the background contribution to the low-energy antiproton flux from astrophysical sources is hard to model~\cite{evoli,Cowsik:2016bwg}. A careful analysis would therefore be needed to apply any such constraint to our scenario.

 \subsection{How Exact Must the $R$-Symmetry Be?}
 \label{sec:sugraconstraints}
As discussed in Sec.~\ref{sec:Rsymbreaking}, AMSB \cite{Randall:1998uk,Giudice:1998xp} effects coming from supergravity generate small Majorana gaugino Masses, soft squark masses and $a$-terms (trilinear scalar couplings), all of order the gravitino mass $m_{3/2}$. While other supergravity scenarios exist in which these contributions are suppressed (see e.g. \cite{Luty:2002ff}), it is interesting to study generic constraints on the gravitino mass and associated bounds on the RPV couplings. 

The Majorana mass term in Eq.~\eqref{eq:AMSBmajoranagaugino} allows for two neutrons to decay into 2 kaons (dinucleon decay) or neutron-antineutron oscillations via a Majorana mass insertion. In a nuclear environment, the  antineutrons annihilate with other nucleons leading to the decay of the nucleus. This process is severely constrained by the lack of observation of such decays. Flavor structure may also severely suppress this process  \cite{Aitken:2017wie}, with the most constrained operators involving only up, down and strange quarks.  Such operators are generated via the combination of the $\lambdapp_{112}$ operator and a Majorana mass for the bino, or by any of the other $\lambdapp_{ijk}$  in combination with a Majorana bino mass and  flavor violation from the weak interactions or squark mixing parameters. Among the RPV couplings, $\lambdapp_{112}$ is the most constrained as it can lead to dinucleon decay without any squark mixing or weak interaction insertion. 
Adapting the results of \cite{Aitken:2017wie}, where the dinucleon decay bound is translated into the $\Lambda-\overline{\Lambda}$ mixing rate $\delta_{\Lambda\Lambda}$, to the $R$-SUSY setup presented here, we find
\bea
\label{eq:nnbound}
\delta_{\Lambda\Lambda} \simeq (2\times 10^{-2}\, \text{GeV}^3)^2  \frac{M_1}{ m_\psi^2-m_n^2 } \left(\frac{ g^\prime\, \lambdapp_{112} }{m_{\tilde{q}_R}^2} \right)^2 \lesssim    10^{-30} \, \text{GeV}\,,
\eea
where $m_\psi$ is the Dirac mass of the bino given in Eq.~\eqref{eq:diracBinoMass}, $\lambdapp_{112}$ is the RPV coupling as defined in Eq.~\eqref{eq:WRPV} and $M_1$ is the bino Majorana mass, e.g. generated by AMSB (see Eq.~\eqref{eq:AMSBmajoranagaugino}). Given that in our model the $R$-symmetry is identified with baryon number, the gravitino is a baryon and therefore subject to the constraints coming from neutron stars as discussed in Sec.~\ref{sec:neutron_stars}, \emph{i.e.} it must satisfy $m_{3/2} \gtrsim 1.2 \, \text{GeV}$. Taking this into account and using the vanilla AMSB scenario without any additional suppression,  Eq.~\eqref{eq:nnbound} implies the bound
\bea
\label{eq:L111bound}
\lambdapp_{112} \, \lesssim 5 \times 10^{-6}
\eea
for $\mathcal{O}(1\, \text{TeV})$ squark masses. Note that this is similar to the bound  from DM stability discussed in Sec.~\ref{subsec:DMstability}. 
Additionally, the squark mass matrix receive a contribution from AMSB as discussed in Sec.~\ref{sec:Rsymbreaking}. This effect should comply with the limits obtained in Sec.~\ref{subsec:squarkmassandcouplings}. Finally, AMSB generates non-zero $a$-terms which can also contribute to di-nucleon decay. This effect is however sub-leading to the one discussed above arising from the Majorana bino mass. All in all, in the scenario where the $R$-symmetry is slightly broken by supergravity effects, avoiding a too large dinucleon decay rate places strong constraints on some combinations of RPV and flavor violating couplings.

\subsection{Dark Matter Capture in Neutron Stars}
\label{sec:neutron_stars}
The extremely dense environment found inside neutron stars has been utilized as a laboratory for constraining DM interactions and properties. Neutron stars, the remnants of a collapsed core of a giant star, are prevented from further gravitational collapse into a black hole by Fermi pressure of neutrons and electrons. DM that interacts with neutrons and electrons can upset this balance leading to destabilization of the star. Many studies have been conducted to investigate such effects and the constraints they imply~\cite{Gould:1989gw,deLavallaz:2010wp,McDermott:2011jp,Bell:2013xk,Bramante:2013hn,Gresham:2018rqo,Garani:2018kkd}. As an example of an application, in Sec.~\ref{sec:mechanism} we have used the results of \cite{McKeen:2018xwc} to restrict our parameter space to dark baryon masses greater than $\sim 1.2$~GeV, as lighter dark baryons would lead to a process that would deplete the neutrons (and therefore the Fermi pressure) via conversion into DM within the star.

We now examine constraints on our model arising from the possibility that DM is captured inside the core of a neutron star. Scattering processes of DM on electrons or neutrons within the neutron star imparts the DM's kinetic energy on the target such that the emitted DM particle may not have enough energy to escape the star. DM that accumulates within the core of a neutron star due to this loss of kinetic energy will begin to self gravitate, eventually overcoming the Fermi pressure, thereby destabilizing the star and leading to gravitational collapse~\cite{Guver:2012ba,Bertone:2007ae}. 

In our framework, DM-neutron scattering proceeds through two higher dimensional four-fermion operators coupling to light quarks, and as such has a negligible rate.  However, DM-electron scattering $e^-\tilde{\nu}_R \rightarrow e^-\tilde{\nu}_R$ occurs via a one loop diagram involving a bino, W boson and sterile neutrinos. The cross section for this process is given approximately by
\begin{align}\label{eq:scattering_DM_electrons}
\sigma_{e \, \tilde{\nu}_R}  \sim \frac{1}{16\pi}\left( \frac{g^2\,m_{\tilde{\nu_R}}^2\, \lambda_{N}^2\, y_N^2\, v^2\, \sin^2 \beta}{m_{\psi}\, m_{\nu_R}^2\, m_{W}^2 }   \right)^2  \sim 10^{-64}\,\text{cm}^2 \left(\frac{y_N \sin \beta}{10^{-8}}\right)^4 \,,
\end{align}
where on the right hand side we have fixed GeV mass dark sector particles and TeV scale squarks along with other benchmark values as discussed above. Note that since we consider bosonic DM there is no contribution to the Fermi pressure from the captured DM, and we simply recast the bound from~\cite{Bertoni:2013bsa} on the scattering cross section that would result in enough DM being captured to cause it to self gravitate. To avoid this, the DM-electron cross-section needs to be smaller than $\sim 10^{-58}\, \text{cm}^2$ (for 1 GeV DM mass). Therefore, black hole formation is avoided for small enough sterile neutrino oscillation parameters,
\bea
\label{eq:sigmaenuNS}
\sigma_{e \, \tilde{\nu}_R} \lesssim 10^{-64} \text{cm}^2  \quad \Rightarrow \quad y_N \sin \beta \,\, \lesssim \,\, 3 \, \times 10^{-7} \,.
\eea
Recall from Eq.~\eqref{eq:NuOscConst} that in a type-I seesaw framework,  $y_N \sin \beta$ should not exceed $\sim 10^{-8}$ in order to accommodate constraints on the maximal size of SM neutrino masses. Therefore, in the simplest type-I seesaw framework DM capture in neutron stars proceeds at a low enough rate and does not induce the collapse of the star.

The accumulation of DM in the core of the neutron star is even less constrained in our framework, as the captured DM carries antibaryon number and can therefore annihilate with neutrons within the star. Such annihilation process may proceed through several different channels in our model. Let us consider as an example the annihilation to a sterile neutrino and a kaon, which is kinetically favored and proceeds through t-channel exchange of a Dirac bino. We estimate the cross section for the parton level process $\tilde{\nu}_R+udd\rightarrow \nu_R+d\bar{s}$ as 
\begin{align}
\label{eq:sigmaAnnNS}
\sigma_{\mathrm{annh}} &\sim\frac{1}{8\pi}\frac{(4\pi f_{\pi}^2)^2(m_{\tilde{\nu_R}}^2+m_n^2-m_{K^0}^2-m_{\nu_R}^2)}{m^2_{\psi}(m_{\tilde{\nu_R}}^2+m_n^2)}\lambda_N^2 \left(\lambda^{\prime\prime}_{112} \sqrt{2}\, g^{\prime}\, \left( \frac{Q_u}{m_{\tilde{u}}^2} + \frac{Q_d}{m_{\tilde{d}}^2} - \frac{Q_d}{m_{\tilde{b}}^2} \right)\right)^2 \\ \nonumber  &\sim 10^{-48}\, \text{cm}^2\, {\lambdapp_{112}}^2 ,
\end{align}
where we have used our usual benchmark values. For given annihilation and capture rates $C_a$ and $C_c$, which measure the probability of a single particle annihilating or being captured over a unit time interval, the number $N_{\tilde{\nu_R}}$ of DM particles captured by a neutron star evolves according to
\begin{equation}
\frac{dN_{\tilde{\nu_R}}}{dt}=C_c-C_a N_{\tilde{\nu_R}} \, .
\end{equation}
The capture rate $C_c$ depends on the scattering rate of DM with neutron star constituents. As shown above, in our case the scattering with electrons dominates over the one with neutrons. Using Eq.~\eqref{eq:scattering_DM_electrons} and the results of~\cite{Guver:2012ba}, the capture rate is estimated to be
\begin{equation}
   C_c \, \sim  \, 10^{23} \, \text{yr}^{-1} \, \left(\frac{\sigma_{e \, \tilde{\nu}_R} }{10^{-55}\, \mathrm{cm}^2} \right) .
\end{equation}
Assuming that DM particles at the core of the neutron star are relativistic and that the density of surrounding neutrons is similar to nuclear density $n_s$, the annihilation rate is simply given by $C_a\sim\sigma^{\mathrm{inc}}_{\mathrm{annh}} \, n_s$, 
where $\sigma^{\mathrm{inc}}_{\mathrm{annh}}$ is the inclusive cross section for DM annihilation with neutrons, which receives a large contribution from~\Eq{eq:sigmaAnnNS}. The number of DM particles that is required for bosonic DM to form a black hole~\cite{Guver:2012ba} is about $N_{\tilde{\nu_R}}\approx 10^{38}$.
To avoid the accumulation of too many DM particles inside the neutron star, we can simply require $C_a N_{\tilde{\nu_R}}>C_c$. This condition translates into a lower bound on the annihilation cross section, which in our case reads
\begin{equation}
 \sigma^{\mathrm{inc}}_{\mathrm{annh}}\gtrsim 10^{-80}\,\text{cm}^2 \, \left(\frac{y_N\, \text{sin}\beta}{10^{-8}} \right)^4 \,.
\end{equation}
This condition is easily satisfied by the annihilation process described above.

\section{Signals at Colliders and $B$ Factories}
\label{sec:Signals}
In this section we identify the main features of the model that are likely to be tested in current and upcoming terrestrial experiments, with focus on colliders and $B$ factories. As we will see, a rich phenomenology including exotic $B$ decays and long lived particle searches provides an opportunity to fully test this baryogenesis scenario. 
More precisely, the large $\sim 10^{-3}$ branching ratio of $B$ mesons to baryons and missing transverse energy ($\MET$) Eq.~\eqref{eq:Brfraction}, along with the requirement of $1-4$ TeV squark masses  (Eqs.~\eqref{eq:Brdeg} and \eqref{eq:BaryoConst}), result in multiple correlated signals arising from processes occurring at different vertices within LHC detectors. 

\subsection{Semileptonic Asymmetries}
An avenue to probe the baryogenesis scenario is through the semileptonic asymmetries. As was discussed in Sec.~\ref{sec:model_indep_asl}, an enhancement of $a_{\mathrm{sl}}^s$ and/or $a_{\mathrm{sl}}^d$ with respect to their SM values is required to reproduce the observed baryon asymmetry. This amplification of the CP violation in the $B$ meson system can be tested as the still relatively loose experimental bounds tighten up on the SM predictions. Given stronger constraints on the semileptonic asymmetries, larger values of the branching ratio of $B$ to a baryon and missing energy than the ones quoted in Eq.~\eqref{eq:Brdeg} would be required for the viability of the baryogenesis mechanism, strengthening the signals of the searches described below. Therefore, experimental efforts in this direction complement the searches proposed below in testing the dynamics of the model.

\subsection{Exotic $B$ Meson Decays at $B$ Factories}
As was discussed in Sec.~\ref{subsec:BranchingFrac}, successful baryogenesis requires a new decay mode of $B$ mesons with a relatively large branching ratio of $\sim 10^{-3}$. The decay products include a single baryon and dark sector states (a sneutrino DM particle and a sterile neutrino) carrying baryon number, potentially along with some mesons. This apparently baryon number violating decay can be looked for in $B$ factories like BABAR, Belle~\cite{Bevan:2014iga} and Belle~II~\cite{Kou:2018nap}. Depending on what precise combination of coupling and light squark dominates the process (see Eq.~\eqref{eq:Brfraction}), different baryons and mesons are expected to appear as decay products. Generically, protons and strange or charmed baryons  will be produced.  A more   experimentally challenging situation would arise if a large fraction of decays occurs into a neutron and dark particles, as such a virtually invisible final state would be extremely difficult to detect. However, all of the possible operators contributing to the decay contain either an up quark or a charm quark, and either a down quark or a strange quark. The only operator that would not produce either charmed or strange particles in the final state is isospin symmetric, so averaging over both the $B^0$ and $B^+$ decays it is valid to assume an equal number of decays containing protons and neutrons.

A dedicated search for exotic $B$ decays with a final state containing a baryon and missing energy has not been performed to this date. However, an inclusive branching ratio $\mathrm{Br}\left(B\rightarrow \mathcal{B}+\mathrm{anything}\right) = 6.8\pm 0.6 \%$ was reported in~\cite{Albrecht:1992he} (see also~\cite{Tanabashi:2018oca}), which is large compared to the known exclusive modes. A loose limit $\mathrm{Br}\left(B\rightarrow \mathcal{B}+X\right) \lesssim 1-2 \, \%$ can be derived assuming that the (also reported~\cite{Albrecht:1988sj,Crawford:1991at}) modes $\mathrm{Br}\left(B\rightarrow p\bar{p}+\mathrm{anything}\right) = 2.47\pm 0.23 \%$ and $\mathrm{Br}\left(B\rightarrow \Lambda\bar{p} / \bar{\Lambda}p+\mathrm{anything}\right) = 2.5\pm 0.4 \, \%$ do not contain any invisible fermion in the final state. A dedicated search in BABAR or Belle II is highly desirable and has the potential to probe the full parameter space, given the sensitivity\footnote{We thank S. Robertson for his input on this matter.} of similar searches like $\mathrm{Br}\left(B\rightarrow \Lambda\bar{p} \nu \bar{\nu} \right) < 3\cdot10^{-5}$ ($90\%$ cl) using BABAR data \cite{babarrare}. An additional signal of interest arises in the case when the mixing of the sterile neutrino with SM neutrinos is large enough for it to decay into SM states within the detection volume. This possibility is discussed in Sec.~\ref{subsec:LHCLLD}.

\subsection{LHC Searches for Heavy Colored Scalars}\label{sec:LHC_colored_scalars}
As highlighted in Eq.~\eqref{eq:BaryoConst}, the mass of at least one of the squarks must to be $\mathcal{O} (\rm TeV)$, in order for it to mediate a large enough $B^0 \rightarrow \mathcal{B} + X$ branching ratio. In this mass range, such colored scalars can potentially be produced at the LHC. Let us therefore briefly discuss the existing constraints on the mass of the squarks arising from collider searches. Our discussion is based on that of~\cite{Aitken:2017wie}. Studies in similar setups have been performed in~\cite{Beauchesne:2017jou, Chalons:2018gez, Fox:2019ube}, but are not applicable to our setup as they focus on binos which are significantly heavier than the ones of interest here.

If kinematically possible, squarks can be pair produced through the usual QCD processes. Additionally, resonant production of a single squark can occur through the $\lambda_{ijk}^{''} \bU^c_i \bD^c_j \bD^c_k$ operator. The latter channel can constrain larger masses due to the lower kinematic threshold of the single production process. Once produced, a squark $\tilde{q}_R$ can decay either to a pair of quarks (through the aforementioned coupling) or to a quark and a bino (through the gauge coupling in Eq.~\eqref{eq:componentgaugeL}). The bino decays mostly into dark sector states and therefore contributes to missing energy. 

Searches for dijet resonances and monojet events containing missing energy are best suited to place constrains on the mass of $\tilde{q}_R$. The limits were computed in~\cite{Aitken:2017wie}; although the detailed limits depend on the values and flavor structure of the couplings $\lambda_{ijk}^{''}$, the reach of the searches does not extend beyond $1-1.2\,$TeV. In this work we therefore conservatively assume that squarks lie above this scale, in which case no limits from LHC searches for colored scalars exist. Note however that this is not strictly necessary, as lighter squarks are allowed providing $\lambda_{ijk}^{''}$ are small enough. A detailed study of the flavor structure and comparison with the requirement in Eq.~\eqref{eq:BaryoConst} is left for future work.

\subsection{Exotic Decays of $b$ Hadrons at the LHC}
\label{sec:QCDbprod}
\begin{figure}[t]
\centering
\begin{subfigure}[b]{0.48\textwidth}
\centering
\begin{tikzpicture} \begin{feynman}
\vertex (a1); 
\vertex [right=1cm of a1] (b);
\vertex [below right= of b] (f1) {\( \overline{b} \)};
\vertex [above right= of b] (c); 
\vertex [right=of c] (d);
\vertex [above right= of c] (f2) {\( \overline{u} \)};
\vertex [below right= of c] (f3) {\( \overline{d} \)};
\vertex [above right=1cm of d] (f4) {\( \nu_R \)};
\vertex [below right=1cm of d] (f5) {\( \tilde{\nu}_R \)};
\diagram* {
(a1) -- [gluon] (b), 
(b) -- [anti fermion] (f1),
(b) -- [fermion, edge label = \(b\)] (c), 
(c) -- [fermion, edge label=\( \psi \)] (d),
(c) -- [anti fermion] (f2),
(c) -- [anti fermion] (f3),
(d) -- (f4),
(d) -- [charged scalar] (f5),
}; 
\end{feynman}
\end{tikzpicture}
\end{subfigure}
\begin{subfigure}[b]{0.48\textwidth}
\centering
\begin{tikzpicture} \begin{feynman}
\vertex (a1) {\( u \)}; 
\vertex [below right=of a1] (b);
\vertex [below left=of b] (a2) {\( d \)};
\vertex [above right=of b] (c); 
\vertex [below right=of b] (f1) {\( \overline{b} \)};
\vertex [above right=1cm of c] (f2) {\(\nu_R\)};
\vertex [below right=1cm of c] (f3) {\(\tilde{\nu}_R\)};
\diagram* {
(a1) -- [fermion] (b) -- [anti fermion] (a2), 
(b) -- [anti fermion] (f1), 
(b) -- [fermion, edge label'=\( \psi \)] (c),
(c) -- (f2),
(c) -- [charged scalar] (f3),
}; 
\end{feynman}
\end{tikzpicture}
\end{subfigure}
\caption{\footnotesize{Production channels of the bino at the LHC, also showing its decay into a neutrino/sneutrino pair. The left diagram schematically shows the decay of a b quark in a $b\,\bar{b}$ pair through the effective four-fermion operator $\psi\, u\, d\, b$ (this is a parton level diagram, the $b$ quark would hadronize before decaying).
The right one corresponds to direct production of the bino through the same effective four-fermion operator. Propagator arrows indicate the flow of baryon number.}}
\label{fig:bino_production_LHC}
\end{figure}
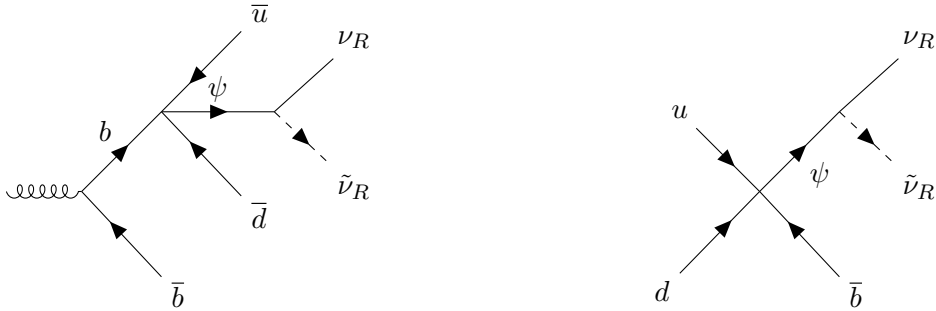

At the LHC, $b \, \bar{b}$ pairs are copiously produced via strong interactions. These $b$-quarks subsequently hadronize and decay. Through the effective four-fermion operators in Table~\ref{table:flavor_structure}, one of the $b$-quarks may undergo the visible baryon number violating decay $b\rightarrow \bar{u} \bar{d} \psi$ (or $b\rightarrow \bar{u} \bar{s} \psi$). Recall that $\psi$ is the Dirac bino (Eq.~\eqref{eq:diracBinoMass}), which subsequently decays into two dark sector states: a stable sneutrino DM particle and a sterile neutrino. If the sterile neutrino is sufficiently long-lived, both particles leave the detector contributing to missing transverse energy of the event. The rate for this exotic $b$ decay is directly related to the baryogenesis dynamics (see Eq.~\eqref{eq:Brfraction}) and is expected to be significant, corresponding to a branching fraction of $\sim 10^{-3}$.

At the partonic level, the process of interest is shown in Fig.~\ref{fig:bino_production_LHC} (\emph{left}) and can be summarized as
\bea
\bar{b} \, b \rightarrow  \bar{b} \, \bar{u}  \, \bar{q} \, \psi \,, \quad \text{where} \quad q = s \, , b \, ,
\eea
and $\psi$ subsequently decays into a DM particle and a sterile neutrino, both of which are assumed to be stable at detector scales (the possibility of the sterile neutrino decaying before leaving the detector is discussed in Sec.~\ref{subsec:LHCLLD}).
The corresponding signature at the LHC is an event with a single jet along with significant missing transverse energy ($\MET$) \cite{Abe:2018bpo}. The fact that the single jet is produced from a $b$-quark allows for the use of $b$-tagging \cite{Scodellaro:2017wli}, which provides a further distinctive feature of the process.

\subsection{LHC Searches for $j + \MET$}
\label{sec:Bviolbprod}
A similar final state to the one discussed in Sec.~\ref{sec:QCDbprod} can be generated through $2\rightarrow2$ processes in $pp$ collisions. Assuming that the squark mediator is too heavy to be produced on-shell\footnote{Close to the squark threshold deviations from Eq.~\eqref{eq:bino_production_cross_section} are expected, but the order of magnitude estimate remains valid. The careful analysis required to obtain a more precise expression lies beyond the scope of this work.}, we may calculate the rate using the effective four-fermion operators given in Table~\ref{table:flavor_structure}.   
As an example, we consider the process $u\,d\rightarrow \psi\, \bar{b}$, which involves couplings relevant for baryogenesis. The corresponding diagram is shown in Fig.~\ref{fig:bino_production_LHC} (\emph{right}). At the partonic level, we can estimate the production cross section as 
\begin{equation}
\label{eq:bino_production_cross_section}
\sigma_{ud\rightarrow\psi\bar{b}}\sim \mathrm{pb}\,{\lambdapp_{113}}^2\left( \frac{1\,\mathrm{TeV}}{m_{\tilde{q}}} \right)^4\left( \frac{\sqrt{s_{ud}}}{1\,\mathrm{TeV}} \right)^2,
\end{equation}
where $\sqrt{s_{ud}}$ is the partonic center-of-mass energy and we have neglected all fermion masses. Similarly to the exotic $b$-hadron decay, this process results in a distinctive final state consisting of a single $b$-jet plus missing transverse energy corresponding to the decay of the bino into dark sector particles. The generalization to the other RPV operators $\lambdapp_{ijk}$ is straightforward and motivates more general mono-jet searches~\cite{Abe:2018bpo}.

\subsection{LHC Searches for Long Lived Particles} 
\label{subsec:LHCLLD}
In the discussion of the previous section we have assumed the Dirac bino to decay into dark sector states which leave the detector as missing transverse energy, but this might not always be the case. In fact, in the model presented above the dark sector particles are identified with a sneutrino and a right-handed neutrino. The sneutrino is sufficiently stable to be a good DM candidate and therefore invisible in any high energy experiment. The sterile neutrino $\nu_R$ is however unstable and expected to decay back into SM particles. Interestingly, for the range of parameters of interest for baryogenesis, the sterile neutrino can decay with a lifetime that is long on collider time scales, thereby providing exciting implications for long lived particle searches at MATHUSLA \cite{Lubatti:2019vkf}, FASER \cite{Feng:2017uoz}, CODEX-b\cite{Gligorov:2017nwh}, and the ATLAS muon tracker \cite{Aaboud:2018aqj}. 

\begin{figure}[t]
\centering
\begin{subfigure}[b]{0.32\textwidth}
\centering
\begin{tikzpicture} \begin{feynman}
\vertex (a) {\(\nu_{R}\)}; 
\vertex [right=of a] (b); 
\vertex [above right=of b] (f1) {\(\tilde{\nu}_{R}\)};
\vertex [below right=of b] (c);
\vertex [above right=of c] (f2) {\(u\)}; 
\vertex [below right=of c] (f3) {\(d\)};
\vertex [right=of c] (f4) {\(s\)};
\diagram* {
(a) -- (b) -- [anti charged scalar] (f1), 
(b) -- [fermion, edge label'=\(\psi\)] (c), 
(c) -- [fermion] (f2),
(c) -- [fermion] (f3),
(c) -- [fermion] (f4),
}; 
\end{feynman}
\end{tikzpicture}
\end{subfigure}
\begin{subfigure}[b]{0.32\textwidth}
\centering
\begin{tikzpicture} \begin{feynman}
\vertex (a) {\(\nu_{R}\)}; 
\vertex [right= 1cm of a] (a1); 
\vertex [right= 1cm of a1] (b); 
\vertex [above right=of b] (f1) {\(\ell\)};
\vertex [below right=of b] (c);
\vertex [above right=of c] (f2) {\(q^\prime\)}; 
\vertex [below right=of c] (f3) {\(\overline{q}\)};
\diagram* {
(a) -- [insertion=1] (a1) -- [edge label'=\(\nu_{\rm SM}\)] (b) -- (f1), 
(b) -- [boson, edge label'=\(W^{\pm}\)] (c), 
(c) -- [fermion] (f2),
(c) -- [anti fermion] (f3),
}; 
\end{feynman}
\end{tikzpicture}
\end{subfigure}
\begin{subfigure}[b]{0.32\textwidth}
\centering
\begin{tikzpicture} \begin{feynman}
\vertex (a) {\(\nu_{R}\)}; 
\vertex [right=1cm of a] (a1); 
\vertex [right=1cm of a1] (b); 
\vertex [above right=of b] (f1) {\(\nu_{\rm SM}\)};
\vertex [below right=of b] (c);
\vertex [above right=of c] (f2) {\(\ell^-\)}; 
\vertex [below right=of c] (f3) {\(\ell^+\)};
\diagram* {
(a) -- [insertion=1] (a1) -- [edge label'=\(\nu_{\rm SM}\)] (b) -- (f1), 
(b) -- [boson, edge label'=\(Z^0\)] (c), 
(c) -- [fermion] (f2),
(c) -- [anti fermion] (f3),
}; 
\end{feynman}
\end{tikzpicture}
\end{subfigure}
\caption{\footnotesize{Decays of the right-handed neutrino through an RPV coupling (left) and charged (center) and neutral (center) current interactions after a Dirac mass insertion. Propagator arrows indicate the flow of baryon number. }}
\label{fig:RHN_decays}
\end{figure}
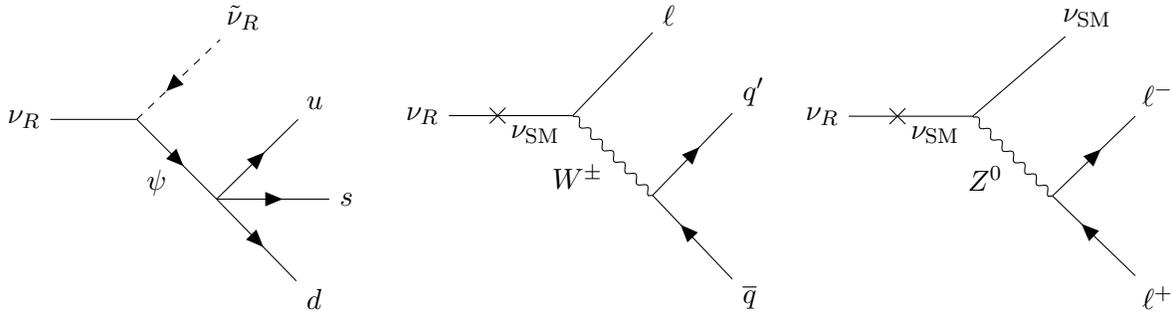
The right-handed neutrino can in principle decay into DM and three light quarks through an off-shell bino and an RPV coupling as shown in Fig.~\ref{fig:RHN_decays} (\emph{left}). However, this process is related to the same coupling that is required to be small by DM stability (as discussed in Sec.~\ref{sec:dark_matter}). As a consequence, this decay channel is constrained to have a rate that is too small for detection purposes\footnote{\footnotesize{Additional frameworks of low-scale baryogenesis through meson and baryon oscillations contain decays which are long lived on collider time scales and can produce three light quarks.
For instance, in the set-up of \cite{Aitken:2017wie}, baryogenesis proceeds through the oscillation of charmed and beautiful baryons which creates an asymmetry in a Majorana fermion $\chi$. $\chi$ can be produced at the LHC, and would be relatively long lived before decaying into light quarks $\chi \rightarrow u \, d \, s$. Note that this scenario does however \emph{not} accommodate DM.}}.
Nevertheless, the neutral and charged current decay of the sterile neutrinos through a mass mixing to SM neutrinos may in general be long lived\footnote{\footnotesize{Constructing models which produce collider signals from long-lived sterile neutrino decays is a subject of much study \cite{Atre:2009rg}}.}. In particular, the processes $\nu_R \rightarrow q^\prime \, \bar{q} \, l$ and $\nu_R \rightarrow l^+ \, l^{-} \, \nu$ (see the center and left diagrams in Fig.~\ref{fig:RHN_decays}) are long lived with decay lengths of order  $1 - 10^6$ meters for values of the Yukawa couplings that lie within the requisite range to accommodate SM neutrino masses. Therefore, depending on the exact parameters and experimental setup, the sterile neutrino might either leave the detector as missing energy or decay in a displaced vertex, motivating the use of long lived particle detection techniques.

In the framework presented here, the prompt production and subsequent long lived decay of the sterile neutrino would produce novel signals at the LHC. Let us exemplify the particular case involving $b$ quarks, which is the interesting one for baryogenesis. As discuss in the previous sections, two different production channels are possible, each one of them producing a signature with some distinctive features.

Firstly, the creation of a $b$ quark and a Dirac bino from a $pp$ collision can proceed as in Fig.~\ref{fig:bino_production_LHC} (\emph{right}) through the $\psi \, u \, d\, b$ operator in Table~\ref{table:flavor_structure}. The cross section estimated in Eq.~\eqref{eq:bino_production_cross_section}. The bino then promptly decays into a sterile neutrino and DM through the operator in Eq.~\eqref{eq:WLambdaN}. This is followed by the long lived decay of the sterile neutrino as discussed above, resulting in a lepton jet (for the CC case). The parton-level process is therefore
\begin{equation}
u \,d \rightarrow b \, \psi \, \rightarrow b \, \nu_R \, \tilde{\nu}_R \,\rightarrow b \, q^\prime \, \bar{q} \,  l \, \tilde{\nu}_R  \,\,\, , \text{or}  \,\,\, b \, l^+ \, l^{-} \,  \nu \, \tilde{\nu}_R \, \,.
\end{equation}
In this case, the long lived particle is centrally produced. A potential study in, for instance, the ATLAS detector would involve a mono-jet signal (as discussed in Sec.~\ref{sec:Bviolbprod}) and an associated long lived decay leading to a lepton jet \cite{Aad:2015sms}. Note that our signal is novel in that it contains a single lepton within the jet.

Secondly, QCD production of a $b\bar{b}$ pair can be followed by the decay of one of the quarks through the same four-fermion operator $\psi \, u \, d\, b$, resulting in a forward production of quarks, DM and a sterile neutrino as shown in Fig.~\ref{fig:bino_production_LHC} (\emph{left}). The prompt $b$ decay signal can be searched for as discussed in Sec.~\ref{sec:QCDbprod} by triggering on the associately produced $b$ quark. The long lived sterile neutrino decays through the charged or neutral current channels just discussed. At the partonic level, this corresponds to the process
\begin{align}
b \rightarrow \bar{u} \, \bar{d}\, \psi  \rightarrow \bar{u} \, \bar{d}\,  \tilde{\nu_R} \, \nu_R  \rightarrow  \bar{u} \, \bar{d} \,  \tilde{\nu_R} \, q \, \bar{q} \,  l  \,\,\, , \text{or} \,\,\,  \bar{u} \, \bar{d}\,  \tilde{\nu_R} \,l^+ \, l^{-}  \,  \nu \,.
\end{align}
The kinematics of these decays would be primed for forward searches such as FASER.
We leave a detailed analysis of all these interesting collider signals to future work.

\section{Summary and Outlook}
\label{sec:Outlook}

In this work we have constructed a supersymmetric realization of the mechanism proposed in~\cite{Elor:2018twp}, where the baryon asymmetry and the DM abundance of the Universe are produced from neutral $B$ meson oscillation and subsequent decay into a dark sector. We have introduced a model of $R$-SUSY in which an exact $U(1)_R$ symmetry is identified with $U(1)_B$ baryon number (such that particles within the same multiplet are charged differently under baryon number). Heavy squarks of mass $1-2$ TeV are integrated out to generate baryon number conserving four-fermion operators coupling SM quarks to a light Dirac bino (about 2-4 GeV in mass) which carries baryon number. 

The existence of an exact $R$-symmetry forbids the existence of Majorana masses for gauginos, which are instead of Dirac type. This is a crucial feature of the model, as the Majorana gauginos and the absence of $R$-parity allow for neutron-antineutron oscillations and dinucleon decays, both of which processes are very tightly constrained experimentally. That said, it is well known that supergravity effects generically break the $R$-symmetry and generate Majorana gaugino masses (along with soft squark masses and $a$-terms). Taking this into account, we have quantified the degree to which the $R$-symmetry needs to be exact in the face of the experimental constraints.

The baryogenesis mechanism starts when $b$-quarks and antiquarks are produced from the late decay of an ``inflaton-like" TeV scale particle at temperatures $T_R \sim 10-50 \, \text{MeV}$. The quarks then hadronize into neutral $B_{s,d}^0$ mesons which oscillate and quickly decay into a baryon and a Dirac bino via the aforementioned operator. The Dirac bino further decays into dark sector particles, which are identified with a right-handed sterile neutrino and its baryon number carrying superpartner \textemdash a sterile sneutrino which is stable on cosmological time scales and and therefore constitutes the DM. 

\renewcommand{\arraystretch}{1.3}
\begin{table*}[t]
\begin{center}
  \setlength{\arrayrulewidth}{.25mm}
\scalebox{0.88}{
\begin{tabular}{|cccc|}
\hline
  Parameter  & Description & Achieve Baryogenesis and DM &   Reference \\ \hline	
  $m_\Phi$ & $\Phi$ mass & 	  $11-100$ GeV   & - \\
  $T_R$  & reheat temperature  & $10 - 50 \, \text{MeV}$ & Fig.~\ref{fig:alpha_beta_branching_ratio}, Sec.~\ref{sec:model_indep_br}\\
  $m_{\tilde{q}_R}$ & squark mass & $1-4 \, \text{TeV}$ &  Fig.~\ref{fig:alpha_beta_branching_ratio}, Sec.~\ref{subsec:BranchingFrac}, Sec.~\ref{sec:LHC_colored_scalars} \\
  $m_{\psi}$ & Dirac bino mass& $1.2 -  4.2  \, \text{GeV} $  & Sec.~\ref{sec:mechanism}, see also \cite{Elor:2018twp} \\
  $m_{\nu_R}$&  Majorana neutrino mass  &   $1.2 \, -  2.7 \, \text{GeV} $ &  Sec.~\ref{sec:mechanism}, see also \cite{Elor:2018twp}  \\
  $m_{\tilde{\nu_R}}$ &  sneutrino mass  &  $ 1.2 \, -  2.7 \, \text{GeV}$   &	Sec.~\ref{sec:mechanism}, see also \cite{Elor:2018twp}  \\
   $ \text{Br}$&  Br of $B \to \nu \tilde{\nu} + \text{Baryon}$ & $4\times10^{-4}-0.1$	 & Fig.~\ref{fig:alpha_beta_branching_ratio}, Sec.~\ref{sec:model_indep_br}, Sec.~\ref{subsec:BranchingFrac}	  \\
      $a_{\mathrm{sl}}^d$&  asymmetry in $B_d^0$ &   $\bigl[ -8.9\times 10^{-4},\,-9.0\times 10^{-5}\bigr]$  & Fig.~\ref{fig:semileptonic_asymmetry_constraints}, Sec.~\ref{sec:model_indep_asl},	Fig.~\ref{fig:DeltaM_constraints}, Sec.~\ref{sec:DF2osc} \\
   $a_{\mathrm{sl}}^s$&  asymmetry in $B_s^0$ &  $\bigl[-2.1\times 10^{-4},\, + 4.1\times 10^{-4}\bigr]$ & 	 Fig.~\ref{fig:semileptonic_asymmetry_constraints}, Sec.~\ref{sec:model_indep_asl}, Fig.~\ref{fig:DeltaM_constraints}, Sec.~\ref{sec:DF2osc}  \\ 
   $\text{Max} \bigl[ \lambda^{\prime\prime}_{ij3} \bigr]$&  RPV coupling &  $\gtrsim 1$ & Sec.~\ref{subsec:BranchingFrac}  \\ 
    $\lambda_{112}^{''} $ & RPV coupling $ u \, d\, s$	 & $  \lesssim 10^{-5}$   &  Sec.~\ref{subsec:DMstability}	  \\
   $\lambda_N $&  coupling for $\lambda_s\,\tilde{\nu}_R \, \nu_R$	 & $\mathcal{O}(0.1)$   &  Sec.~\ref{sec:dark_matter}	  \\
   \hline
\end{tabular} }
\end{center}
\caption{\footnotesize{
Ranges of parameters relevant for generating the observed baryon asymmetry ($Y_B = 8.7\times 10^{-11}$) and DM abundance ($\Omega_{\rm DM} h^2 = 0.12$) of the Universe and which are consistent with all experimental constraints. For convenience, in the third column we provide a link to the relevant section in this paper where each result is discussed. 
}}
\label{tab:Parameters}
\end{table*}

We have shown that the $R$-SUSY model presented here reproduces the measured baryon asymmetry and DM abundance of the Universe while being consistent with experimental constraints from colliders and flavor observables. In particular, this mechanisms motivates a study of a yet unexplored region of the $R$-SUSY model parameter space \textemdash that of a light Dirac bino. Indeed, this work represents the first excursion into the exploration of this slice of parameter space. 

Focusing first on the flavor observables that are relevant for the baryogenesis dynamics, we have shown that the NP modifications to the semileptonic-leptonic asymmetries $a_{\mathrm{sl}}^{s,d}$ in the neutral $B_{s,d}^0 - \bar{B}^0_{s,d}$ systems can be large enough to generate a sufficient baryon asymmetry while being allowed by current experimental constraints. In our model, the extra CP violation that enhances the semileptonic asymmetries can be traced back to complex phases in the RPV couplings $\lambdapp_{ijk}$ or in the off-diagonal elements of the squark mass matrix. Another robust prediction of the mechanism is the existence of a new decay channel of $B$ mesons into a baryon and dark sector states, which needs to have a large branching fraction $\gtrsim 10^{-3}$. In the supersymmetric setup, this is only possible if at least one of the squarks has a mass close to $1$~TeV while the RPV couplings $\lambdapp_{ij3}$ have a flavor structure that allows at least one of the combinations $  11,\,12,\,21\,\mathrm{or}\,22$ to be $\mathcal{O}(1)$. The $R$-SUSY NP modifications to additional flavor observables (most notably $K^0-\bar{K}^0$ oscillations and the branching ratio of $b\rightarrow s \gamma$ and $\mu \rightarrow e \gamma$) further constrain couplings that, while not directly relevant for baryogenesis, give a hint towards the flavor structure of the model. The values of the parameters for which $R$-SUSY successfully realizes baryogenesis are summarized in Table.~\ref{tab:Parameters}. 

As a new feature of the supersymmetric model, the dark sector particle content may be elegantly realized as a sterile neutrino supermultiplet. The Dirac bino then decays predominantly into the sterile neutrino and the cosmologically stable sterile (scalar) sneutrino, which carries baryon number $-1$ and constitutes the DM. In this way, equal and opposite matter-antimatter asymmetries are generated in the visible and dark sectors without violating the global baryon number of the Universe. Dark sector interactions between the sneutrino and neutrino act to deplete the otherwise overproduced symmetric abundance of the sneutrino DM. These partially invisible decays of the bino (and, by extension, of $b$-flavored hadrons) can be looked for at $B$-factories like Babar and Belle II or at the LHC, by exploiting mono-jet searches and $b$-tagging techniques. In addition to that, the decays of the sterile neutrino into SM fermions may be long lived on collider time scales and can be searched for at experiments like SHiP, MATHUSLA, FASER, CODEX-b, and the ATLAS and CMS muon trackers. Additional constraints on model parameters are derived from measurements of neutrino oscillations and by exploiting astrophysical observations, most importantly the existence of stable neutron stars.

To conclude, we have presented a supersymmetric model that can achieve low-scale baryogenesis and DM production without baryon number violation. In contrast to the standard lore of high scale baryogenesis, our setup enjoys a plethora of possible experimental signatures ranging from measurements of flavor observables to long lived decays at colliders. Such exciting possibilities leave the door open to fully test this baryogenesis scenario in the not too distant future.

\acknowledgments

We thank Shih-Chieh Hsu and Henry Lubatti for useful discussions regarding long lived decay searches at the LHC. We also thank  Shih-Chieh Hsu for useful comments on the draft.
GA thanks the University of Washington for kind hospitality during the early stages of the project.
GA receives support from the Fundaci\'on ``la Caixa" via a ``la Caixa" postgraduate fellowship.
AEN is supported in part by the Kenneth K. Young Chair in Physics. AEN, GE and HX are supported in part by the U.S. Department of Energy Award DE-SC0011637.
This project has received support from the European Union's Horizon 2020 research and innovation programme under the Marie Sklodowska-Curie grant agreement No 674896.

\newpage

\appendices

\section{Analytic Expressions for Flavor Violating Observables} 
\label{sec:box_diagrams}
Here we give the analytic expressions and some more details on the evaluation of the NP contributions to the flavor violating observables discussed in Sec.~\ref{sec:Predictions}. We distinguish two kinds of contributions depending on whether the light bino plays a role in the process or not.

\subsection{Contributions from RPV Couplings}
\label{sec:box_diagrams_RPV}
The $R-$parity violating couplings $\lambda^{\prime\prime}_{ijk}$ can mediate flavor changing transitions and induce CP-violating effects through their complex phases. A comprehensive study of RPV SUSY phenomenology was done in~\cite{Barbier:2004ez}, to which we refer for any further details not discussed here.

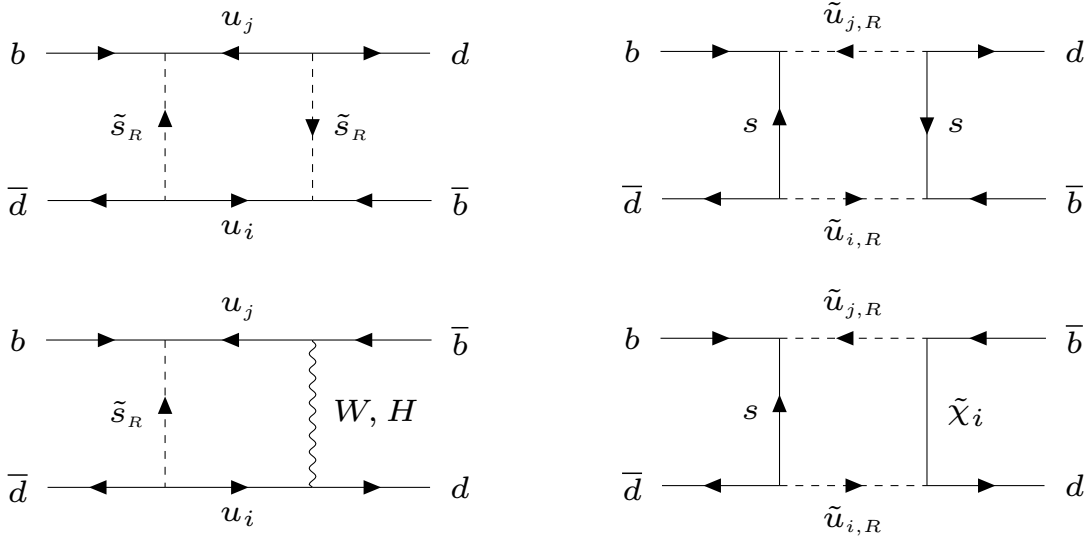
\begin{figure}[t]
\centering
\begin{subfigure}[c]{0.49\textwidth}
\centering
\feynmandiagram[scale=1.4,transform shape, horizontal=a to b] { 
i1 [particle={\tiny \(b\)}]
-- [fermion] a
-- [anti fermion, edge label={\tiny \( u_{\scaleto{j}{3pt}}\)}] b
-- [fermion] f1 [particle={\tiny\( d \)}],
i2 [particle={\tiny\( \overline d\)}]
-- [anti fermion] c
-- [fermion, edge label'={\tiny \( u_{\scaleto{i}{3pt}} \)}] d
-- [anti fermion] f2 [particle={\tiny\( \overline b \)}],
{ [same layer] a -- [anti charged scalar, edge label'={\tiny \( \tilde{s}_{\scaleto{R}{2pt}} \)}] c },
{ [same layer] b -- [charged scalar, edge label={\tiny \( \tilde{s}_{\scaleto{R}{2pt}} \)}] d},
};
\end{subfigure}
\begin{subfigure}[c]{0.49\textwidth}
\centering
\feynmandiagram[scale=1.4,transform shape, horizontal=a to b] { 
i1 [particle={\tiny \( b\)}]
-- [fermion] a 
-- [anti charged scalar, edge label={\tiny \( \tilde u_{\scaleto{j, R}{3pt}}\)}] b
-- [fermion] f1 [particle={\tiny\( d \)}],
i2 [particle={\tiny\( \overline d\)}]
-- [anti fermion] c
-- [charged scalar, edge label'={\tiny \( \tilde u_{\scaleto{i, R}{3pt}} \)}] d
-- [anti fermion] f2 [particle={\tiny\( \overline b \)}], 
{ [same layer] a -- [anti fermion, edge label'={\tiny \( s \)}] c },
{ [same layer] b -- [fermion, edge label={\tiny \( s \)}] d},
};
\end{subfigure}
\begin{subfigure}[c]{0.49\textwidth}
\centering
\feynmandiagram[scale=1.4,transform shape, horizontal=a to b] { 
i1 [particle={\tiny \( b\)}]
-- [fermion] a
-- [anti fermion, edge label={\tiny \( u_{\scaleto{j}{3pt}}\)}] b
-- [anti fermion] f1 [particle={\tiny\( \overline b \)}],
i2 [particle={\tiny\( \overline d \)}]
-- [anti fermion] c
-- [fermion, edge label'={\tiny \( u_{\scaleto{i}{3pt}} \)}] d
-- [fermion] f2 [particle={\tiny\( d \)}],
{ [same layer] a -- [anti charged scalar, edge label'={\tiny \( \tilde{s}_{\scaleto{R}{2pt}} \)}] c },
{ [same layer] b -- [boson, edge label={\tiny \( W,H \)}] d},
};
\end{subfigure}
\begin{subfigure}[c]{0.49\textwidth}
\centering
\feynmandiagram[scale=1.4,transform shape, horizontal=a to b] { 
i1 [particle={\tiny \( b\)}]
-- [fermion] a
-- [anti charged scalar, edge label={\tiny \( \tilde u_{\scaleto{j, R}{3pt}}\)}] b
-- [anti fermion] f1 [particle={\tiny\( \overline b \)}],
i2 [particle={\tiny\( \overline d\)}]
-- [anti fermion] c
-- [charged scalar, edge label'={\tiny \( \tilde u_{\scaleto{i, R}{3pt}} \)}] d
-- [fermion] f2 [particle={\tiny\( d \)}],
{ [same layer] a -- [anti fermion, edge label'={\tiny \( s \)}] c },
{ [same layer] b -- [edge label={\tiny \( \tilde \chi_i \)}] d},
};
\end{subfigure}
\caption{\footnotesize{Diagrams involving the RPV couplings $\lambda^{\prime\prime}_{ijk}$ and contributing to the neutral meson oscillations. The case of the $B_d^0$ system is shown as an example, the $K^0$ and $B^0_s$ are analogous with the obvious substitutions. Propagator arrows indicate the flow of baryon number.}}
\label{fig:box_diagrams_RPV}
\end{figure}

\paragraph{$\Delta F=2$ processes.} Neutral meson oscillation parameters are modified by the existence of diagrams like the ones depicted in Fig.~\ref{fig:box_diagrams_RPV} for the case of the $B^0_d$ system. The corresponding box diagrams were computed in~\cite{Slavich:2000xm} for the (s)top couplings, here we generalize the results including the two lighter generations. One important difference in our setup is the absence of left-right squark mass mixing, which is forbidden by the $R$-symmetry. Taking this into account, we find
\begin{align}\label{eq:box_diagrams_RPV}
\tilde{C}_1 &= \sum_{i,j=1}^3 \frac{1}{4\pi^2}\, \lambdapps_{i12}\lambdapp_{i23}\lambdapps_{j12}\lambdapp_{j23}\, \left[ I_4 \left( m_s^2, m_s^2, m_{\tilde{u}_{i,R}}^2, m_{\tilde{u}_{j,R}}^2\right) + I_4 \left( m_{u_i}^2, m_{u_j}^2, m_{\tilde{s}_{R}}^2, m_{\tilde{s}_{R}}^2\right) \right] ,\\ \nonumber
C_4 &= -C_5, \\ \nonumber
C_5 &= \sum_{i,j=1}^3 \frac{g^2}{4\pi^2}\, \lambdapps_{i12}\lambdapp_{j23}\, V_{i1}V_{j3}^\star\, m_{u_{i}} m_{u_{j}} \bigg [ I_2 \left( m_{\tilde{s}_R}^2, m_W^2, m_{u_{i}}^2, m_{u_{j}}^2\right) + \frac{1}{4m_W^2} I_4 \left( m_{\tilde{s}_R}^2, m_W^2, m_{u_{i}}^2, m_{u_{j}}^2\right) \\ \nonumber
&\qquad\qquad\qquad\qquad\qquad\qquad\qquad\qquad + \frac{1}{4m_W^2 \tan^2\beta} I_4 \left( m_{\tilde{s}_{R}}^2, m_{H^{+}}^2, m_{u_{i}}^2, m_{u_{j}}^2 \right) \bigg ] \\ \nonumber
&+ \sum_{i,j=1}^3 \sum_{r=1}^2 \frac{g^2}{8\pi^2}\, \lambdapps_{i12}\lambdapp_{j23}\, V_{i1}V_{j3}^\star\, \frac{m_{u_i}m_{u_j}}{2 m_W^2 \sin^2\beta} \left| C_{r2} \right|^2 \, I_4 \left( m_b^2, m_{\tilde{\chi}_r^+}^2, m_{\tilde{u}_{i}}^2, m_{\tilde{u}_{j}}^2\right) .
\end{align}
Here, $V_{ij}$ denotes the corresponding CKM matrix element and $C_{rs}$ refers to the mixing matrix of positive charginos. The loop integrals are given by
\begin{equation}
I_n (m_1^2, m_2^2, m_3^2, m_4^2) = \int_0^\infty \frac{k^n \diff k^2}{\left(k^2+m_1^2\right) \left(k^2+m_2^2\right) \left(k^2+m_3^2\right) \left(k^2+m_4^2\right)} .
\end{equation}
In the limit where one of the masses $M$ is much larger than the others, we have $I_4\sim 1/M^2$ and $I_2\sim 1/(M\mu)^2$, where $\mu$ is some combination of the remaining masses. Additionally, logarithmic factors with quotients of masses can appear in the expressions. Therefore, squark masses dominate the integrals and the coefficients $C_i$ are always suppressed by $1/m_{\tilde q}^2$.

\paragraph{$\Delta F=1$ processes.}
\begin{figure}[t]
\centering
\begin{subfigure}[b]{0.49\textwidth}
\centering
\begin{tikzpicture} \begin{feynman}
\vertex (a0) {$b_L$}; 
\vertex[right=1cm of a0] (a1); 
\vertex[right=1cm of a1] (a2);
\vertex[right=2cm of a2] (a3);
\vertex[right=1.5cm of a3] (a4) {$s_R$};
\vertex at ($(a2)!0.5!(a3) + (0, 1.0cm)$) (d); 
\vertex at ($(a3) + (0.5cm, 0.5cm)$) (c1); 
\vertex at ($(a3) + (1.25cm, 1.25cm)$) (c2); 
\diagram* {
(a4) -- [anti fermion] (a3) -- [fermion, edge label=$d_R$] (a2) -- [anti fermion, insertion=0.9, edge label=$b_R$] (a1) -- [anti fermion] (a0),
(a2) -- [anti charged scalar, quarter left, edge label=$\tilde{u}_{R,i}$] (d) -- [anti charged scalar, quarter left] (a3),
(c1) -- [boson, edge label=$\gamma$] (c2),
};
\end{feynman} 
\end{tikzpicture}
\end{subfigure}
\begin{subfigure}[b]{0.49\textwidth}
\centering
\begin{tikzpicture} \begin{feynman}
\vertex (a0) {$b_L$}; 
\vertex[right=1cm of a0] (a1); 
\vertex[right=1cm of a1] (a2);
\vertex[right=2cm of a2] (a3);
\vertex[right=1.5cm of a3] (a4) {$s_R$};
\vertex at ($(a2)!0.5!(a3) + (0, 1.0cm)$) (d); 
\vertex at ($(a3) + (0.5cm, 0.5cm)$) (c1); 
\vertex at ($(a3) + (1.25cm, 1.25cm)$) (c2); 
\diagram* {
(a4) -- [anti fermion] (a3) -- [fermion, edge label=$u_{R,i}$] (a2) -- [anti fermion, insertion=0.9, edge label=$b_R$] (a1) -- [anti fermion] (a0),
(a2) -- [anti charged scalar, quarter left, edge label=$\tilde{d}_{R}$] (d) -- [anti charged scalar, quarter left] (a3),
(c1) -- [boson, edge label=$\gamma$] (c2),
};
\end{feynman} 
\end{tikzpicture}
\end{subfigure}
\caption{\footnotesize{Diagrams for the flavor-violating decay $b\rightarrow s \gamma$ mediated by RPV interactions. The photon can be attached to any charged line. Propagator arrows indicate the flow of baryon number.}}
\label{fig:DeltaF1_diagram_RPV}
\end{figure}
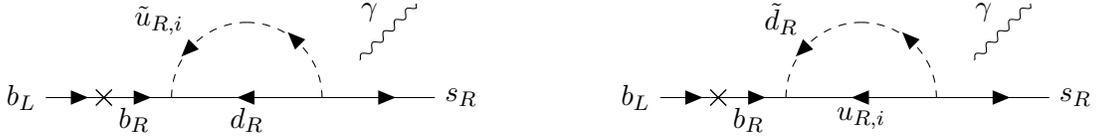
The RPV couplings alone can induce the flavor-violating decay $b\rightarrow s\gamma$ through the diagrams shown in Fig.~\ref{fig:DeltaF1_diagram_RPV}. Summing all contributions, the resulting branching ratio is
\begin{equation}\label{eq:Br_b_sgamma_RPV}
\text{Br} (b\rightarrow s\gamma) = \frac{2N_{C}\,\alpha_{\rm em}\,m_b^5}{4}\cdot \tau_B \left( \left| \sum_{i=1}^3 B^{\mathrm{RPV}}_{R1} + \tilde{B}^{\mathrm{RPV}}_{R1} \right|^2 \right),
\end{equation}
where
\begin{align}
B^{\mathrm{RPV}}_{R1} &= - \frac{\lambdapps_{i12}\lambdapp_{i13}}{864\pi^2}\frac{1}{m_{\tilde{u}_{R,i}}^2} G_1\left( x_{d} \right) , \\ \nonumber
\tilde{B}^{\mathrm{RPV}}_{R1} &= - \frac{\lambdapps_{i12}\lambdapp_{i13}}{864\pi^2}\frac{1}{m_{\tilde{d}_R}^2} G_1\left( x_{u_i} \right),
\end{align}
The loop function $G_1$ is defined in Eq.~\eqref{eq:G1_G2_definition} and is evaluated at $x_d=m^2_{d}/m^2_{\tilde{u}_{R,i}}$ or $x_{u_i}=m^2_{u_i}/m^2_{\tilde{d}_{R}}$.
For heavy squarks, we have $G_1(x\ll 1)\simeq 1/2$.

\subsection{Contributions from Squark Mass Mixing}
\label{sec:box_diagrams_bino}

\begin{figure}[t]
\centering
\begin{subfigure}[b]{0.49\textwidth}
\centering
\feynmandiagram[scale=1.4,transform shape, horizontal=a to b] { 
i1 [particle={\tiny \( b\)}]
-- [fermion] a
-- [scalar, with reversed arrow=0.3, with reversed arrow=0.8, insertion={[size=5pt]0.5}, edge label={\tiny \(\tilde b_{\scaleto{L/R}{3pt}} \quad \tilde d_{\scaleto{L/R}{3pt}}\)}] b
-- [fermion] f1 [particle={\tiny\( d \)}],
i2 [particle={\tiny\( \overline d\)}]
-- [anti fermion] c
-- [scalar, with arrow=0.25, with arrow=0.75, insertion={[size=5pt]0.5}, edge label'={\tiny \(\tilde d_{\scaleto{L/R}{3pt}} \quad \tilde b_{\scaleto{L/R}{3pt}}\)}] d
-- [anti fermion] f2 [particle={\tiny\( \overline b\)}],
{ [same layer] a -- [fermion, boson, edge label'={\tiny \( \tilde{B} \)}] c },
{ [same layer] b -- [anti fermion, boson, edge label={\tiny \( \tilde{B} \)}] d},
};
\end{subfigure}
\begin{subfigure}[b]{0.49\textwidth}
\centering
\feynmandiagram[scale=1.4,transform shape, horizontal=a to b] { 
i1 [particle={\tiny \( b\)}]
-- [fermion] a
-- [scalar, with reversed arrow=0.3, with reversed arrow=0.8, insertion={[size=5pt]0.5}, edge label={\tiny \(\tilde b_{\scaleto{L/R}{3pt}} \quad \tilde d_{\scaleto{L/R}{3pt}}\)}] b
-- [fermion] f1 [particle={\tiny\( d \)}],
i2 [particle={\tiny\( \overline d^{\scaleto{c}{2pt}} \)}]
-- [fermion] c
-- [scalar, with reversed arrow=0.3, with reversed arrow=0.8, insertion={[size=5pt]0.5}, edge label'={\tiny \(\tilde d_{\scaleto{R/L}{3pt}} \quad \tilde b_{\scaleto{R/L}{3pt}}\)}] d
-- [fermion] f2 [particle={\tiny\( \overline b^{\scaleto{c}{2pt}} \)}],
{ [same layer] a -- d },
{ [same layer] a -- [boson, with arrow = 0.25, edge label={\tiny \( \quad \tilde{B} \)}] d },
{ [same layer] b -- c },
{ [same layer] b -- [boson, with reversed arrow = 0.25, edge label'={\tiny \( \tilde{B} \quad \)}] c},
};
\end{subfigure}
\caption{\footnotesize{Diagrams mediated by Dirac binos and involving squark mass mixing which contribute to neutral meson oscillations. The case of the $B_d^0$ system is shown as an example, the $K^0$ and $B^0_s$ are analogous with the obvious substitutions. Note that only $L-L$ or $R-R$ (and not $L-R$) squark mixing is allowed by the $R$-symmetry. Propagator arrows indicate the flow of baryon number.}}
\label{fig:box_diagrams}
\end{figure}
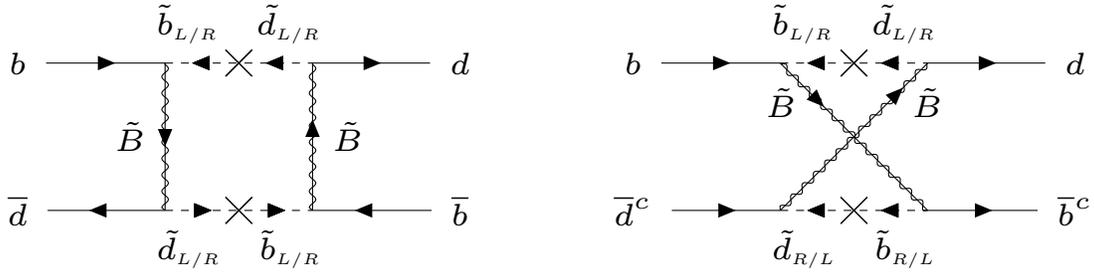

Here we compute the contribution of a light Dirac bino to the meson oscillation parameters, following the spirit of~\cite{Kribs:2007ac}. The flavor violation is generated in this case by off diagonal elements in the squark soft mass mixing matrix. To be specific, we give the expressions for consider $B_d^0$, but the computations for $B_s^0$ and $K$ are analogous. The relevant diagrams are shown in Fig.~\ref{fig:box_diagrams}. With the notation of Eq.~\eqref{eq:delta_M_bino}, we obtain
\begin{align}
 C_1 &= \left(\frac{1}{6}\right)^4 \frac{\alpha_Y^2}{m_\squark^2} \delta_{LL}^2 \tilde{f}_6(\lambda_\bino), \\
 \tilde{C}_1 &= \left(\frac{1}{6}\right)^4 \frac{\alpha_Y^2}{m_\squark^2} \delta_{RR}^2 \tilde{f}_6(\lambda_\bino), \\
 C_4 &= \left(\frac{1}{6}\right)^4 \frac{\alpha_Y^2}{m_\squark^2}\, 4\delta_{LL}\delta_{RR}\, \tilde{f}_6(\lambda_\bino), 
\end{align}
Comparing with~\cite{Kribs:2007ac}, we see that binos give rise to fewer operators than gluinos due to the more restricted color structure of the diagrams. The loop function depends on the ratio $\lambda_\bino = m_{\bino}^2 / m_{\squark}^2$ and is given by
\begin{equation}
\tilde{f}_6(\lambda_\bino) = \frac{6\lambda_\bino(1+\lambda_\bino)\log\lambda_\bino - \lambda_\bino^3-9\lambda_\bino^2+9\lambda + 1 }{3(1-\lambda_\bino)^5}\,,
\end{equation}
which approaches $\tilde{f}_6\rightarrow 1/3$ when $m_{\bino}\ll m_{\squark}$. To get this result we have neglected the momentum of the external quarks and applied Fierz rearrangements to obtain the desired color structures.
In general, $\Delta M^{\rm NP}$ is a complex quantity because the off-diagonal elements of the squark mixing matrix can be complex, and thus so are $\delta_{LL}$ and $\delta_{RR}$. Effects in the CP-violating observables defined in Sec.~\ref{sec:ModelIndep} are therefore expected. Note however that $C_1$ is always real because it only depends on the absolute value of the squark mixing parameters $\delta^2=\delta\cdot\delta^\star$. Analogous formulas, with the obvious substitutions, hold for the $B^0_s$ and $K^0$ systems.

\begin{figure}[t]
\centering
\begin{subfigure}[b]{0.49\textwidth}
\centering
\begin{tikzpicture} \begin{feynman}
\vertex (a0) {$b_R$}; 
\vertex[right=1cm of a0] (a1); 
\vertex[right=1cm of a1] (a2); 
\vertex[right=2cm of a2] (a3);
\vertex[right=1.5cm of a3] (a4) {$s_L$};
\vertex at ($(a2)!0.5!(a3) + (0, 1.0cm)$) (d); 
\vertex at ($(a3) + (0.5cm, 0.5cm)$) (c1); 
\vertex at ($(a3) + (1.25cm, 1.25cm)$) (c2); 
\diagram* {
(a4) -- [anti fermion] (a3) -- [anti fermion, boson, edge label=$\tilde{B}$] (a2) -- [anti fermion, insertion=0.9, edge label=$b_L$] (a1) -- [anti fermion] (a0),
(a2) -- [anti charged scalar, quarter left, edge label=$\tilde{b}_{L}$, insertion=0.99] (d) -- [anti charged scalar, quarter left, edge label=$\tilde{s}_L$] (a3),
(c1) -- [boson, edge label=$\gamma$] (c2),
};
\end{feynman} 
\end{tikzpicture}
\end{subfigure}
\begin{subfigure}[b]{0.49\textwidth}
\centering
\begin{tikzpicture} \begin{feynman}
\vertex (a1) {$b_R$}; 
\vertex[right=1.5cm of a1] (a2);
\vertex[right=1cm of a2] (a21); 
\vertex[right=1cm of a21] (a3); 
\vertex[right=1.5cm of a3] (a4) {$s_L$};
\vertex at ($(a2)!0.5!(a3) + (0, 1.0cm)$) (d); 
\vertex at ($(a3) + (0.5cm, 0.5cm)$) (c1); 
\vertex at ($(a3) + (1.25cm, 1.25cm)$) (c2); 
\diagram* {
(a4) -- [anti fermion] (a3) -- [anti fermion, boson, edge label=$\tilde{B}$, insertion=0.99] (a21) -- [anti fermion, edge label=$\tilde{H}_d$] (a2) --  [anti fermion] (a1),
(a2) -- [anti charged scalar, quarter left, edge label=$\tilde{b}_{L}$, insertion=0.99] (d) -- [anti charged scalar, quarter left, edge label=$\tilde{s}_L$] (a3),
(c1) -- [boson, edge label=$\gamma$] (c2),
};
\end{feynman} 
\end{tikzpicture}
\end{subfigure}
\caption{\footnotesize{Diagrams for the flavor-violating decay $b\rightarrow s \gamma$ mediated by a light bino, with external (left) and internal (right) chirality flip. The flavor violation arises through off-diagonal terms in the squark mass matrix. The photon can be attached to any charged line. Propagator arrows indicate the flow of baryon number.}}
\label{fig:DeltaF1_diagram_bino}
\end{figure}
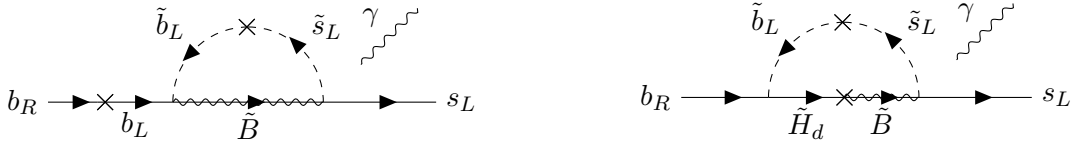

\paragraph{$\Delta F=1$ processes.}
The calculation of these processes are done following the conventions of~\cite{Kribs:2007ac}. The diagrams for $b\rightarrow s\gamma$ are shown in Fig.~\ref{fig:DeltaF1_diagram_bino}. Summing contributions from both diagrams, the decay width of a $b$ quark to a $s$ quark and a photon mediated by a light bino can be written as
\begin{equation}\label{eq:b_sgamma_bino}
\text{Br} (b\rightarrow s\gamma) = \frac{N_c\,\alpha_{\rm em}\,m_b^5}{4}\cdot \tau_B \left( \left| B_{L1} + B_{L2} \right|^2 + \left| B_{R1} + B_{R2} \right|^2 \right),
\end{equation}
where $B_{L1,R1}$ correspond to diagrams with an external chirality flip,
\begin{align}
B_{L1} &= - \frac{\alpha_Y}{432\pi}\frac{\delta_{LL}}{m_{\tilde{q}}^2} G_1(x_{\tilde{B}}) , \\ \nonumber
B_{R1} &= - \frac{\alpha_Y}{108\pi}\frac{\delta_{RR}}{m_{\tilde{q}}^2} G_1(x_{\tilde{B}}),
\end{align}
while $B_{L2,R2}$ come from the diagrams with internal bino-higgsino mixing,
\begin{align}
B_{L2} &= -\frac{\alpha_Y}{144\pi} \cos^2\beta \frac{\delta_{LL}}{m_{\tilde{q}}^2} G_2(x_{\tilde{B}}) ,\\ \nonumber
B_{R2} &= \frac{\alpha_Y}{72\pi} \cos^2\beta \frac{\delta_{RR}}{m_{\tilde{q}}^2} G_2(x_{\tilde{B}}).
\end{align}
The loop functions $G_1$ and $G_2$, which are to be evaluated at $x=m^2_{\tilde{B}}/m^2_{\tilde{q}}$, are given by
\begin{align}\label{eq:G1_G2_definition}
G_1(x) &= \frac{17x^3-9x^2-9x+1-6x^2(x+3)\log x}{2(1-x)^5}, \\ \nonumber
G_2(x) &= \frac{-5x^2+4x+1+2x(x+2)\log x}{(1-x)^4}.
\end{align}
In the case where the bino is much lighter than the squarks, we have $G_1(x_{\tilde{B}})\simeq 1/2$ and $G_2(x_{\tilde{B}})\simeq 1$.

The computation of the flavor-violating muon decay $\mu\rightarrow e \gamma$ is very similar to the $b\rightarrow s\gamma$ transition, with the obvious substitutions. We therefore directly give the result
\begin{equation}
\text{Br} (\mu\rightarrow e\gamma) = \frac{48\alpha_{\rm em}\pi^3}{G_F^2} \left( \left| A_{L1} + A_{L2} \right|^2 + \left| A_{R1} + A_{R2} \right|^2 \right),
\end{equation}
where
\begin{align}\label{eq:mu_egamma_bino}
A_{L1} &= - \frac{\alpha_Y}{48\pi}\frac{\delta_{LL}}{m_{\tilde{\ell}}^2} G_1(x_{\tilde{B}}) ,\\ \nonumber
A_{R1} &= - \frac{\alpha_Y}{12\pi}\frac{\delta_{RR}}{m_{\tilde{\ell}}^2} G_1(x_{\tilde{B}}),\\ \nonumber
A_{L2} &= -\frac{\alpha_Y}{16\pi} \cos^2\beta \frac{\delta_{LL}}{m_{\tilde{\ell}}^2} G_2(x_{\tilde{B}}) ,\\ \nonumber
A_{R2} &= \frac{\alpha_Y}{8\pi} \cos^2\beta \frac{\delta_{RR}}{m_{\tilde{\ell}}^2} G_2(x_{\tilde{B}}).
\end{align}
The loop functions as defined above should now be evaluated at $x=m^2_{\tilde{B}}/m^2_{\tilde{\ell}}$.

\begin{spacing}{1.1}
\bibliography{Refs}
\bibliographystyle{utphys}
\end{spacing}

\end{document}